\documentclass{aa}

\input psfig.tex

\usepackage{graphicx}
\usepackage{natbib}
\usepackage{array}
\bibpunct{(}{)}{;}{a}{}{,}

\usepackage{rotating}    

\input psfig.tex
\usepackage{latexsym}
\usepackage{natbib}
\usepackage{amssymb}
\usepackage{amsmath}

\usepackage{graphicx}
\usepackage{graphics}
\usepackage{fancyhdr}
\usepackage{morefloats}
\begin{document}

\title{EvoL: The new Padova T-SPH parallel code for cosmological simulations - I. Basic code: gravity and hydrodynamics}
\subtitle{}
     \author{Emiliano Merlin \inst{1}, Umberto Buonomo \inst{1}, Tommaso Grassi \inst{1}, Lorenzo Piovan \inst{1} \& Cesare Chiosi \inst{1}}
     \offprints{E. Merlin }
     \institute{$^1 $ Department of Astronomy, University of Padova,
                Vicolo dell'Osservatorio 3, 35122 Padova, Italy  \\
                \email{emiliano.merlin@unipd.it; cesare.chiosi@unipd.it;
                umberto.buonomo@unipd.it; tommaso.grassi@unipd.it; lorenzo.piovan@unipd.it }
      }
     \date{Received: November 2009; Accepted:  }


\abstract {We present \textsc{EvoL}, the new release of the Padova
N-body code for cosmological simulations of galaxy formation and
evolution. In this paper, the basic Tree + SPH code is
presented and analysed, together with an overview on the software
architectures.} {\textsc{EvoL} is a flexible parallel Fortran95
code, specifically designed for simulations of cosmological
structure formation on cluster, galactic and sub-galactic scales.}
{\textsc{EvoL} is a fully Lagrangian self-adaptive code, based on
the classical Oct-tree by Barnes \& Hut (1986) and on the Smoothed
Particle Hydrodynamics algorithm (SPH, Lucy 1977). It includes
special features such as adaptive softening lengths with correcting
extra-terms, and modern formulations of SPH and artificial
viscosity. It is designed to be run in parallel on multiple CPUs to
optimize the performance and save computational time.} {We describe
the code in detail, and present the results of a number of standard
hydrodynamical tests.} {}

\titlerunning{EvoL I}
\authorrunning{E. Merlin et al.}
\maketitle


\section{Introduction} \label{intro}

The importance of numerical simulations in modern science is
constantly growing, because of the complexity, the multi-scaling
properties, and the non-linearity of many physical phenomena. When
analytical predictions are not possible, we are forced to compute
the evolution of physical systems numerically. Typical examples in
astrophysical context are the problems of structure and galaxy
formation and evolution. Over the past two decades, thanks to highly
sophisticated cosmological and galaxy-sized numerical simulations  a
number of issues concerning the formation of cosmic systems and
their evolution along the history of the Universe have been
clarified. However, an equivalent number of new questions have been
raised, so that complete understanding of how galaxies and clusters
formed and evolved along the Hubble time is still out of our reach.
This is especially true at the light of many recent observations
that often appear at odds with well established theories (see for
instance the long debated question of the monolithic versus
hierarchical mechanism of galaxy formation and their many variants)
and to require new theoretical scenarios able to match the
observational data.

To this aim, more and more accurate and detailed numerical
simulations are still needed. They indeed are the best tool to our
disposal to investigate such complex phenomena as the formation and
evolution of galaxies within a consistent cosmological context. A
number of astrophysical codes for galaxy-sized and cosmological
simulations are freely and publicly available. They display a huge
range of options and functions; each of them is best suited for a
set of particular problems and experiments, and may suffer  one or
more drawbacks. Among the best known, we recall  \textsc{Flash}
\citep{Fryxell2000}, \textsc{Gadget} \citep{Springel2001} and its
second release \textsc{Gadget2} \citep{Springel2005},
\textsc{Gasoline} \citep{Wadsley2004}, \textsc{Hydra}
\citep{Couchman2005}, \textsc{Enzo} \citep{Norman2007}, and
\textsc{Vine} \citep{Wetzstein2008}.

\textsc{EvoL} is the new release of the Padova N-body code 
\citep[\textsc{Pd-Tsph}, ][]{Carraro1998,Lia2002,Merlin2007}.
It is a flexible, fully Lagrangian, parallel, and self-adaptive N-body code,
written in Fortran95 in a straightforward and user-friendly format.

\textsc{EvoL} describes the dynamical evolution of a system of
interacting discrete masses (particles) moving under the mutual
gravitational forces and/or the gravitational action of external
bodies, plus, when appropriate, under mutual hydrodynamical interactions. 
Such particles can represent real discrete bodies or a fluid
contained in  volume elements of suitable size. A numerical
simulation of a physical system follows the temporal evolution of it
using small but finite \textit{time-steps} to approximate the
equations of motion to a finite-differences problem. In the
Lagrangian description, no reference grid is superposed to the
volume under consideration, while the particles move under their
mutual (or external) interactions. Each particles carries a mass, a
position, a velocity, plus (when necessary) a list of physical
features such as density, temperature, chemical composition, etc.

To simulate the dynamical evolution of a region of the Universe, one
has to properly model the main different material components, namely
Dark Matter, Gas, and Stars), representing them with different
species of particles. Moreover, a physically robust model for the
fundamental interactions (in addition to gravity) is required, at
the various scales of interest. Finally, a suitable cosmological
framework is necessary, together with a coherent setting of the
boundary conditions and with an efficient algorithm to follow the
temporal evolution of the system. \textsc{EvoL} is designed to
respond to all of these requirements, improving upon the previous
versions of the Padova Tree-SPH code under many
aspects.\footnote{The core of the old \textsc{Pd-Tsph} code was
written during the '90s by C. Chiosi, G. Carraro, C. Lia and C.
Dalla Vecchia \citep{Carraro1998, Lia2002}. Over the years, many
researchers added their contribution to the development of the code.
In its original release, the \textsc{Pd-Tsph} was a basic Tree + SPH
code, written in Fortran90, conceptually similar to
\textsc{Tree-SPH} by \citet{Hernquist1989}. Schematically,
\textsc{Pd-Tsph} used an early formulation of SPH \citep{Benz1990}
to solve the equations of motion for the gas component, and the
\citet{Barnes1986} Tree algorithm to compute the gravitational
interactions.} In the following sections, the main features of
\textsc{EvoL} are presented in detail, together with  a review of
some general considerations about the adopted algorithms whenever
appropriate.

This paper presents the basic features of the code; namely, the
Tree-SPH (i.e. Oct-Tree plus Smoothing Particle Hydrodynamics)
formalism, its implementation in the parallel architecture of the
code, and the results of a number of classic hydrodynamic tests. The
non-standard algorithms (e.g. radiative cooling functions, chemical
evolution, star formation, energy feedback) will be presented in a
following companion paper (Merlin et al. 2009, in preparation).

The plan of the paper is as follows. In Sect. \ref{gravity}
we describe the aspects of the code related to the gravitational
interaction. In  Sect. \ref{description_SPH} we deal with the
hydrodynamical treatment of fluids and its approximation in the SPH language.
Sect. \ref{description_integration} illustrates some aspects related to the
integration technique, such as the time steps, the periodic boundary conditions,
and the parallelization of the code. Sect. \ref{test} contains and discusses
numerous tests aimed at assessing the performance of the code. 
Sect.\ref{Conclusions} summarizes some concluding remarks. 
Finally, Appendices A and B contains some technical
details related the N-dimensional spline kernel (Appendix A) and
the equations of motion and energy conservation (Appendix B).

\section{Description of the code: Gravity} \label{gravity}

Gravity is the leading force behind the formation of cosmic
structures, on many scales of interest, from galaxy clusters down to
individual stars and planets. Classical gravitation is a well
understood interaction. As long as General Relativity, Particle
Physics or exotic treatments such as Modified Dynamics are not
considered
\footnote{A relativistic formulation of both gravitational
and hydrodynamical interactions is possible \citep[for a general
summary see e.g.][]{Rosswog2009}, but is generally unessential in
problems of cosmological structure formation.}, it is described by
Newton's law of universal gravitation:
\begin{eqnarray}
\vec{F}_{ij} = G\frac{m_i m_j}{|\vec{r}_j-\vec{r}_i|^3}(\vec{r}_j-\vec{r}_i),
\label{newton}
\end{eqnarray}

\noindent where $G$ is the gravitational constant. Given a density
field $\rho(\vec{r})$, the gravitational potential $\Phi$ is
obtained via Poisson equation:
\begin{eqnarray}
\nabla^2 \Phi(\vec{r}) = 4 \pi G \rho(\vec{r}),
\label{Poisson}
\end{eqnarray}

\noindent and $\vec{F}_{ij} = \nabla \Phi(\vec{r})$.

The gravitational force exerted on a given body (i.e., particle) by
the whole system of bodies within a simulation can be obtained by
the vectorial summation of all the particles' contributions (without
considering, for the moment, the action of the infinite region
external to the computational volume; this can indeed be taken into
account using periodic boundary conditions, see Sect.
\ref{periodism}). This is simply the straightforward application of
Eq. \ref{newton}. Anyway, in practice this approach is not efficient,
and may also lead to artificial divergences, as explained in the following
Sections.

\subsection{Softening of the gravitational force}  \label{soft}

Close encounters between particles can cause numerical errors,
essentially because of the time and mass resolution limits.
Moreover, when dealing with continuous fluids rather than with single, isolated objects,
one tries to solve Eq. \ref{Poisson} rather than Eq. \ref{newton},
and must therefore try to model a \textit{smooth} density field given a distribution 
of particles with mass. In addition, dense clumps of particles may also steal large amounts of
computational time. To cope with all these issues, it is common practice to
\textit{soften} the gravitational force between close pairs of
bodies: if the distance between two particles becomes smaller than a
suitable \textit{softening length} $\epsilon$, the force exerted on
each body is corrected and progressively reduced to zero with
decreasing distance.

Different forms of the force softening  function can be used. A
possible expression is given by the following formula:

\begin{eqnarray}
&& \frac{F(\vec{r},\epsilon)}{G m_i} = m_j \frac{\vec{r}}{|\vec{r}|} \times \\
&& \begin{cases}
1 / \epsilon^2 [\frac{4}{3}u-\frac{6}{5}u^3+\frac{1}{2}u^4] & \text{if $0 \leq u < 1$,} \\
1 / \epsilon^2 [\frac{8}{3}u-3u^2+\frac{6}{5}u^3-\frac{1}{6}u^4-\frac{1}{15u^2}] & \text{if $1 \leq u < 2$,} \\
1/|\vec{r}|^2 & \text{if $u \geq 2$,}
\end{cases}\nonumber
\label{softforce}
\end{eqnarray}

\noindent where $\vec{r}$ is the distance between particles $i$ and
$j$, and $u \equiv|\vec{r}|/\epsilon$. This expression for the force
softening corresponds (via the Poisson equation) to a density
distribution \textit{kernel} function proposed by
\citet{Monaghan1985} and widely adopted in Smoothed Particles
Hydrodynamics algorithms (see Sect. \ref{description_SPH}) \footnote{Throughout
the paper, we only use the 3-dimensional formulation of kernels. See
Appendix A for the 1-D and 2-D forms.}:

\begin{eqnarray}
&& W(r,\epsilon) = \frac{1}{\pi h^3} \times
\begin{cases}
1 - \frac{3}{2}u^2+\frac{3}{4}u^3 & \text{if $0 \leq u < 1$,} \\
\frac{1}{4}(2-u)^3 & \text{if $1 \leq u < 2$,} \\
0 & \text{if $u \geq 2$.}
\end{cases}
\label{spline}
\end{eqnarray}

\noindent Note that in Eq. \ref{softforce} the softening length
$\epsilon$ is assumed to be the same for the two particles $i$ and
$j$. In the more general situation in which each particle carries
its own $\epsilon$, a symmetrisation is needed to ensure energy and
momentum conservation: this can be achieved either using $\epsilon =
\bar{\epsilon_{ij}}=(\epsilon_i+\epsilon_j)/2$, or by symmetrizing
the softened force after computing it with the two different values
$\epsilon_i$ and $\epsilon_j$.

The softening lengths can be fixed in space and time, or may be let
vary with local conditions. If the softening length is kept
constant, the choice of its numerical value is of primary
importance: for too small a softening length it will result in noisy
force estimates, while for too large a value it will systematically
bias the force in an unphysical manner 
\citep[][ and see also Sect. \ref{bosstest}]{Merritt1996,
Romeo1998, Athanassoula2000, Price2007}. Unfortunately, the
``optimal'' value for the softening length depends on the particular
system being simulated, and in general it is  not  known a priori. A
standard solution is to assign to each particle a softening length
proportional to its mass, keeping it fixed throughout the whole
simulation\footnote{Recently, \citet{Shirokov2007} pointed out that
since particles are of unequal mass, and hence unequal softening
lengths, one should actually compute the pairwise gravitational
force and potential by solving a double integral over the particle
volumes. Therefore, the computation of the interactions using the
classic approach is likely  not accurate. Given the practical
difficulty in evaluating those integrals, they also provide a
numerical approximation.}, or letting it vary
with time, or redshift, if a cosmological background is included. 
A clear advantage of keeping $\epsilon$ fixed in space is that energy is 
naturally conserved, but on the other hand the smallest resolvable length 
scale is fixed at the beginning of the simulation and remains the same 
in the whole spatial domain independently of the
real evolution of the density field. A collapsing or expanding body may 
quickly reach a size where the flow is dominated by the softening in 
one region, while in another the flow may become unphysically point-like.

Obviously, the accuracy can be greatly improved if $\epsilon$ is let
vary according to the local particle number density \citep[see e.g.
][]{Dehnen2001}. Moreover, in principle, if particles in a
simulation are used to sample a continuous fluid (whose physics is
determined by the Navier-Stokes or the Boltzmann equations) the
properties of such points should always change accordingly to the
local properties of the fluid they are sampling, to optimize the
self-adaptive nature of Lagrangian methods.  On the other hand, if
particles represent discrete objects (single stars, or galaxies in a
cluster, etc.), their softening lengths might perhaps be considered
an intrinsic property of such objects and may be kept constant,
depending on their mass and not on the local density of particles.
In cosmological and galaxy-sized simulations, gas and Dark Matter
particles are point-masses sampling the underlying density field,
and stellar particles represent clusters of thousands of stars prone
to the gravitational action of the nearby distribution of matter;
thus a fully adaptive approach seems to be adequate to describe the
evolution of these fluids. However, it can be easily shown that
simply letting $\epsilon$ change freely would result in a poor
conservation of global energy and momentum, even if in some cases
the errors could be small \citep[see e.g.][]{Price2007,
Wetzstein2008}.

To cope with this, \textsc{EvoL} allows for the adaptive evolution
of individual softening lengths, but includes in the equations of motion
of particles self-consistent correcting additional terms.
Such terms can be derived \citep[see][]{Price2007} starting from
an analysis of the Lagrangian describing a self-gravitating, softened,
collisionless system of $N$ particles, which is given by\footnote{
A clear advantage of using the Lagrangian to derive the equations of
motion is that, provided the Lagrangian is appropriately
symmetrised, momentum and energy conservation are guaranteed. Note
that this derivation closely matches the one described
in Sect. \ref{description_SPH} to derive the variational formulation of the
SPH formalism and the so-called $\nabla h$ correcting terms.}

\begin{eqnarray}
L=\sum_{i=1}^N m_i \left( \frac{1}{2}v_i^2 - \Phi_i \right),
\label{Lagrangian}
\end{eqnarray}

\noindent where $\Phi$ is the gravitational potential of the $i$-th particle,
\begin{eqnarray}
\Phi(\vec{r}_i) = - G \sum_{j=1}^N m_j \phi (|\vec{r}_{ij}|,\epsilon_i),
\label{potential}
\end{eqnarray}

\noindent and $\phi$ is a softening kernel ($\vec{r}_{ij}=\vec{r}_i-\vec{r}_j$).
Assuming a density distribution described by the standard spline kernel Eq.
\ref{spline}, the latter becomes

\begin{eqnarray}
&& \phi(\vec{r},\epsilon) = \\
&& \begin{cases}
1 / \epsilon [\frac{2}{3}u^2-\frac{3}{10}u^4+\frac{1}{10}u^5-\frac{7}{5}] & \text{if $0 \leq u < 1$,} \\
1 / \epsilon [\frac{4}{3}u^2-u^3+\frac{3}{10}u^4-\frac{1}{30}u^5-\frac{8}{5}+\frac{1}{15u}] & \text{if $1 \leq u < 2$,} \\
-1/|\vec{r}| & \text{if $u \geq 2$,}
\end{cases}\nonumber
\label{softphi}
\end{eqnarray}

\noindent where again $u \equiv |\vec{r}|/\epsilon$. Note that the force softening, Eq.
\ref{softforce}, is obtained taking the spatial derivative of Eq. \ref{softphi}.

The equations of motion are obtained from the Euler-Lagrange
equations,
\begin{eqnarray}
\frac{d}{dt} \left(\frac{\partial L}{\partial \vec{v}_i} \right) - \frac{\partial L}{\partial \vec{r}_i} = 0,
\label{EulerLagrange}
\end{eqnarray}

\noindent which give
\begin{eqnarray}
m_i \frac{d\vec{v}_i}{dt} = \frac{\partial L}{\partial \vec{r}_i}.
\end{eqnarray}

The gravitational part of the Lagrangian is
\begin{eqnarray}
L_{grav} & = & - \sum_{j=1}^N m_j \Phi_j = \\
& & - \frac{G}{2}\sum_j\sum_k m_j m_k \phi_{jk}(\epsilon_j) \nonumber
\label{Lgrav}
\end{eqnarray}

\noindent where for the sake of clarity $\phi_{ij}(\epsilon_i) =
\phi(|\vec{r}_{ij}|,\epsilon_i)$; swapping indices, the partial
derivative $\partial L_{grav} / \partial \vec{r}_i$ results

\begin{eqnarray}
& & \frac{\partial L_{grav}}{\partial \vec{r}_i} = \\ \nonumber
& & -\frac{G}{2}\sum_j\sum_k m_j m_k \left[ \frac{\partial \phi_{jk}(\epsilon_j)}{\partial|r_{jk}|}|_{\epsilon} \frac{\partial|r_{jk}|}{\partial \vec{r}_i} + \frac{\partial \phi_{jk}(\epsilon_j)}{\partial \epsilon_j}|_{r} \frac{\partial \epsilon_j}{\partial \vec{r}_i} \right],
\label{110}
\end{eqnarray}

\noindent where
\begin{eqnarray}
\frac{\partial|r_{jk}|}{\partial \vec{r}_i} =
\frac{\vec{r}_j-\vec{r}_k}{|\vec{r}_j-\vec{r}_k|}(\delta_{ji}-\delta_{ki}).
\end{eqnarray}

To obtain the required terms in the second addend on the right part
of Eq. \ref{110}, the softening length must be related to the
particle coordinates so that $\epsilon=\epsilon(n)$, where $n$ is
the number density of particles at particle $i$ location. To this
aim, one can start from the general interpolation rule that any
function of coordinates $A(\vec{r})$ can be expressed in terms of
its values at a disordered set of points, and the integral
interpolating  the function $A(\vec{r})$ can be defined by
\begin{eqnarray}
A_{int}(\vec{r})=\int{A(\vec{r}')W(\vec{r}-\vec{r}',\epsilon)\vec{dr'}},
\label{Aint}
\end{eqnarray}

\noindent where $W$ is the density kernel Eq. \ref{spline}. This is
the same rule at the basis of the SPH scheme (see Sect. \ref{description_SPH})
\footnote{Note that in principle $W$ should be defined over the
whole space, as in the first SPH calculations by \citet{Gingold1977}
where a Gaussian-shaped kernel was adopted. For practical purposes,
anyway, its domain is made compact putting $W=0$ for distances
greater than $\eta \epsilon$, with $\eta>0$, from the position of
the active particle.}.

Approximating the integral to a summation over particles for practice
purposes\footnote{\citet{Price2007} use mass the density in place of number
density to achieve the desired solutions. The formulation in terms of number
density is equivalent and can be advantageous when dealing with particles
of different masses. Also note that this is a reformulation of the well
known SPH density summation, see Sect. \ref{description_SPH}.}, one obtains
\begin{eqnarray}
A_{sum}(\vec{r}_i) = A_i = \sum_j{\frac{A_j}{n_j} W(\vec{r}_{ij},\epsilon_i)},
\label{Asum}
\end{eqnarray}

\noindent where $i$ marks the ``active'' particle that lays at the
position at which the function is being evaluated, and $j$ are its
neighbouring particles (or simply \textit{neighbours}). The error by 
doing so depends on the disorder of particles and is in general $O(h^2)$
\citep{Monaghan1992}. Thus,

\begin{eqnarray}
n_i = \sum_j W(|\vec{r}_{ij}|,\epsilon_i),
\label{rho0}
\end{eqnarray}

\noindent where it has been put $A = n$; so, there is no need to
know in advance the value of the density of neighbouring particles
to compute the density of the active one.

To link $\epsilon$ and $n$, a safe prescription is to use

\begin{eqnarray}
\epsilon_i = \eta \left( \frac{1}{n_i} \right)^{1/3},
\label{etan}
\end{eqnarray}

\noindent where $\eta$ is a dimensionless parameter. With this law,
the weighted number of particles within a softening sphere is
tentatively held constant, i.e.

\begin{eqnarray}
\frac{4}{3}(2 \epsilon_i)^3 n_i = N_{nei,id},
\end{eqnarray}

\noindent with $N_{nei,id} \def \frac{4}{3} \pi (2 \eta)^3$. Numerical experiments
have shown that a safe choice is $1.2 \leq \eta \leq 1.5$, corresponding to
$60 \leq N_{nei,id} \leq 110$ \citep{Price2007}.

The term needed in Eq. \ref{110} is

\begin{eqnarray}
\frac{\partial \epsilon_j}{\partial \vec{r}_i} = \frac{\partial
\epsilon_j}{\partial n_j} \frac{\partial n_j}{\partial \vec{r}_i}.
\end{eqnarray}

\noindent The first factor is given by the derivative of Eq.
\ref{etan},

\begin{eqnarray}
\frac{\partial \epsilon_j}{\partial n_j} = - \frac{\epsilon_j}{3 n_j},
\end{eqnarray}

\noindent whereas the second factor is the spatial derivative of Eq.
\ref{rho0}, which results:
\begin{eqnarray}
\frac{\partial n_j}{\partial \vec{r}_i} = \frac{1}{\Upsilon_j}\sum_p \frac{\partial W_{jp}(\epsilon_j)}{\partial \vec{r}_i} (\delta_{ij}-\delta_{ip}),
\end{eqnarray}

\noindent where it was defined
\begin{eqnarray}
\Upsilon_i = \left[ 1 - \frac{\partial \epsilon_i}{\partial n_i} \sum_j \frac{\partial W_{ij}(\epsilon_i)}{\partial \epsilon_i}\right].
\end{eqnarray}

\noindent Rearranging and putting $G\equiv1$, one finally finds
\begin{eqnarray}
\frac{\partial L_{grav}}{\partial \vec{r}_i} & = & \\
& & -m_i \sum_j m_j \left[\frac{\phi_{ij}'(\epsilon_i)+\phi_{ij}'(\epsilon_j)}{2}\right]\frac{\vec{r}_i-\vec{r}_j}{|\vec{r}_i-\vec{r}_j|} \nonumber\\
& & - m_i \sum_j m_j \frac{1}{2} \left[ \frac{\xi_i}{\Upsilon_i} \frac{\partial W_{ij}(\epsilon_i)}{\partial \vec{r}_i} + \frac{\xi_j}{\Upsilon_j} \frac{\partial W_{ij}(\epsilon_i)}{\partial \vec{r}_j} \right], \nonumber
\label{dLgravdr}
\end{eqnarray}

\noindent where
\begin{eqnarray}
\xi_i = \frac{\partial \epsilon_i}{\partial n_i} \sum_j \frac{\partial \phi_{ij}(\epsilon_i)}{\partial \epsilon_i}.
\label{zeta}
\end{eqnarray}

\noindent The $\partial \phi / \partial \epsilon$ terms used in Eq.
\ref{zeta} can be tabulated together with  the other kernel
functions:

\begin{eqnarray}
&& \frac {\partial \phi}{\partial \epsilon} = \\
&& \begin{cases}
1 / \epsilon^2 [-2 u^2 + \frac{3}
{2} u^4 - \frac{3}{5} u^5 + \frac{7}{5}] & \text{if $0 \leq u < 1$,} \\
1 / \epsilon^2 [-4 u^2 + 4 u^3 + \frac{3}{2} u^4  + \frac{1}{5} u^5
+\frac{8}{5}] &\text{if $1 \leq u < 2$,} \\
0 &\text{if $u \geq 2$,}
\end{cases}\nonumber
\label{dphidh}
\end{eqnarray}

\noindent where, as usual, $u \equiv|\vec{r}|/\epsilon$. Finally, the
$\partial W /\partial \vec{r}$ and $\partial W / \partial \epsilon$
terms are easily obtained deriving the explicit expression of
$W(|\vec{r}|,\epsilon)$.

In Eq. 22 (and Eq. 12), the first term is the
classical gravitational interaction, whereas the second term is a
new, extra-term which restores energy conservation for varying
softening lengths. Since $\xi$ is defined as a negative  quantity
for positive kernels, this term acts as a negative pressure,
\textit{increasing} the gravitational force. To obtain the $\xi$ and
$\Upsilon$ correcting terms, each particle $i$ must perform a loop
over the other particles, summing the contributions from the ones
that satisfy the criterion $|\vec{r}_{ij}| \leq 2 \times
\max(\epsilon_i, \epsilon_j)$ \footnote{Thus, \textit{all}
particles, and not only SPH particles, need to find their
neighbours to compute these terms (see Sect. \ref{tree}).
While the $\Omega$ term described in Sect. \ref{description_SPH} is only needed
for the SPH algorithm, and therefore only SPH particles concur to compute
it (considering only SPH neighbours in turn), the $\xi$ and
$\Upsilon$ gravitational terms are required for any kind of
gravitating particle, and must be computed looping on \textit{all}
neighbouring particles, both SPH and not. The time lost in the
neighbour searching routine for non-gaseous particles can be at
least partially balanced adopting the individual time-stepping
scheme (see Sect. \ref{inddt}). In this case, large softening lengths
in low-density regions result in larger individual time-steps for
particles belonging to those regions.}.

The relation between $\epsilon$ and $n$, defined by Eq. \ref{etan},
leads to a non-linear equation which can be solved self-consistently
for each particle. \citet{Price2007} proposed an iterative method in
which, beginning from a starting guess for both $n$ and $\epsilon$,
the solution of the equation

\begin{eqnarray}
f(\epsilon) = |n_i(\epsilon_i) - n_{sum,i}(\epsilon_i)| = 0,
\label{iter}
\end{eqnarray}

\noindent where $n_{sum,i}(\epsilon_i)$ is the mass density computed
via summation over neighbouring particles (Eq. \ref{rho0}) and
$n_i(\epsilon_i)$ is the density obtained from Eq. \ref{etan}, is
searched by means of an iterative procedure, that can be solved
adopting a classical bisection scheme, or more efficient routines
such as the Newton-Raphson method. In this case, a loop is iterated
until $|\epsilon_{i,new}-\epsilon_i|/\epsilon_{i,init} <
\gamma_{TOL}$, where $\gamma_{TOL}$ is a tolerance parameter $\sim
10^{-2} - 10^{-3}$, $\epsilon_{i,init}$ is the value of $\epsilon$
for the particle $i$ at the beginning of the procedure, $\epsilon_i$
is its current value, and
$\epsilon_{i,new}=\epsilon_i-f(\epsilon_i)/f'(\epsilon_i)$; the
derivative of Eq. \ref{iter} is given by
\begin{eqnarray}
f'(\epsilon_i) = \frac{\partial n_i}{\partial \epsilon_i} - \sum_j m_j \frac{\partial W_{ij}(\epsilon_i)}{\partial \epsilon_i} = - \frac{3n_i}{\epsilon_i}\Upsilon_i.
\end{eqnarray}

\noindent To increase the efficiency of this process, a predicted
value for both $\epsilon$ and $n$ can be obtained at the beginning
of each time-step using a discretized formulation of the Lagrangian
continuity equation,

\begin{eqnarray}
\frac{dn}{dt}= - n(\nabla\cdot\vec{v}).
\label{continuity}
\end{eqnarray}

%

\noindent Such formulation can be obtained taking the time
derivative of Eq. \ref{rho0}, which results
\begin{eqnarray}
\frac{dn_i}{dt} = \frac{1}{\Upsilon_i} \sum_j (\vec{v}_i-\vec{v}_j) \cdot \nabla_i W(\vec{r}_{ij},\epsilon),
\label{143}
\end{eqnarray}

\noindent while
\begin{eqnarray}
\frac{d\epsilon_i}{dt} = \frac{\partial \epsilon_i}{\partial n_i} \frac{dn_i}{dt}.
\end{eqnarray}

Combining Eq. \ref{143} with Eq. \ref{continuity}, one can also see
that the velocity divergence at  $i$ particle location is given by

\begin{eqnarray}
\nabla \cdot \vec{v}_i = -\frac{1}{n_i \Upsilon_i}
\sum_j(\vec{v}_i-\vec{v}_j) \cdot \nabla_i W(\vec{r}_{ij},\epsilon).
\end{eqnarray}

\noindent The adoption of this \textit{adaptive softening length}
formalism, with the correcting extra-terms in the equation of
motion, results in small errors, always comparable in magnitude to
the ones found with the ``optimal'' $\epsilon$ \citep[see][their Figs.
2, 3 and 4]{Price2007}.

At the beginning of a simulation, the user can select whether he/she
prefers to adopt  the constant or the adaptive softening lengths
formalism, switching on or off a suitable flag.

\subsection{Hierarchical oct-tree structure} \label{tree}

A direct summation of the contribution of all particles should in
principle be performed for each particle at each time-step to
correctly compute the  gravitational interaction, leading to a
computational cost increasing with $N^2$ ($N$ being the number of
particles). A convenient alternative to this unpractical approach
are the so-called \textit{tree} structures, in which particles are
arranged in a hierarchy of groups or ``cells''. In this way, when the
force on a particle is computed, the contribution by distant groups
of bodies can be approximated by their lowest multipole moments, 
reducing the computational cost to evaluate the total force 
to $O(N logN)$. In the classical \citet{Barnes1986} scheme,
the computational spatial domain is hierarchically partitioned into
a sequence of cubes, where each cube contains eight siblings, each
with half the side-length of the parent cube. These cubes form the
nodes of an oct-tree structure. The tree is constructed such that
each node (cube) contains either a desired number of particles
(usually one, or one per particles type - Dark Matter, gas, stars), or is
progenitor of further nodes, in which case the node carries the
monopole and quadrupole moments of all the particles that lie inside
the cube. The force computation then proceeds by walking along the
tree, and summing up the appropriate force contributions from tree
nodes. In the standard tree walk, the global gravitational force
exerted by a node of linear size $l$ on a particle $i$ is considered
only if $r > l/\theta$, where $r$ is the distance of the node from
the active particle and $\theta$ is an \textit{accuracy parameter}
(usually $\leq 1$): if a node fulfills the criterion, the tree walk
along this branch can be terminated, otherwise it is ``opened'', and
the walk is continued with all its siblings. The contribution from
individual particles is thus considered only when they are
sufficiently close to the particle $i$.

To cope with some problems pointed out by \citet{Salmon1994} about
the worst-case behaviour of this standard criterion for commonly
employed opening angles, \citet{Dubinski1996} introduced the
simple modification
\begin{eqnarray}
r > l/\theta + \delta,
\label{opcrit}
\end{eqnarray}

\noindent where the $\delta$ is the distance of the geometric center
of a cell to its center of mass. This provides protection against
pathological cases where the center of mass lies close to an edge of
a cell.

%

In practice, this whole scheme only includes the monopole moment of
the force exerted by distant groups of particles. Higher orders can
be included, and the common practice is to include the quadrupole
moment corrections (\textsc{EvoL} indeed includes them). Still
higher multipole orders would result in a worst performance without
significant gains in computational accuracy \citep{McMillan1993}.

Note that if one has the total mass, the center of mass, and the
weighted average softening length of a node of the oct-tree structure,
the softened expression of the gravitational force can be straightforwardly computed
treating the cell as a single particle if the opening criterion is
fulfilled but the cell is still sufficiently near for the force to
need to be softened.

As first suggested by \citet{Hernquist1989}, the tree structure can
be also used to obtain the individual lists of neighbours
for each body. At each time step each (active) particle can build its
list of interacting neighbouring particles while walking the tree and
opening only sufficiently nearby cells, until nearby bodies are
reached and linked.

\section{Description of the code: Hydrodynamics} \label{description_SPH}

Astrophysical gaseous plasmas can generally be reasonably
approximated to highly compressible, unviscous fluids, in which
anyway a fundamental role is played by violent phenomena such as
strong shocks, high energy point-like explosions and/or supersonic
turbulent motions. \textsc{EvoL} follows the basic gas physics by
means of the Smoothed Particles Hydrodynamics \citep[SPH,
][]{Lucy1977, Monaghan1992}, in a modern formulation based on the
review by \citet{Rosswog2009}, to which the reader is referred for
details. Anyway, some different features are present in our
implementation and we summarize them in the following, along with a
short review of the whole SPH algorithm.

To model fluid hydrodynamics by means of a discrete description  of
a continuous the fluid, in the SPH approach the properties of single
particles are \textit{smoothed} in real space through the kernel
function $W$, and thus \textit{weighted} by the contributions of
neighbouring particles. In this way, the physical properties of each
point in real space can be obtained by the summation over particles
of their individual, discrete properties. Note that, on the contrary
of what happens when softening the gravitational force (which is
gradually reduced to zero within the softening sphere), in this case
the smoothing sphere is the only ``active'' region, and only particles
inside this region contribute to the computation of the local
properties of the central particle.

Starting again from the interpolation rule described in Sect. \ref{soft}, but
replacing the softening length $\epsilon$ by a suitable
\textit{smoothing length} $h$, the integral interpolating  the
function $A(\vec{r})$ becomes
\begin{eqnarray}
A_i(\vec{r})=\int{A(\vec{r}')W(\vec{r}-\vec{r}',h)\vec{dr'}},
\label{sphinterp}
\end{eqnarray}

\noindent (the kernel function $W$ can be  the density kernel Eq.
\ref{spline}).

The relative discrete approximation becomes
\begin{eqnarray}
A_s(\vec{r}_i) = \sum_j{\frac{A_j}{\rho_j} m_j W(\vec{r}_i-\vec{r}_j,h_i)},
\label{sph1}
\end{eqnarray}

\noindent and the physical density of any SPH particle can be computed as

\begin{eqnarray}
\rho_i = \sum{m_j W(|\vec{r}|,h_i)}
\label{densitySPH}
\end{eqnarray}

\noindent (here and throughout this Section, it is implied that all
summations and computations are extended on SPH particles only).

A differentiable interpolating  function can be constructed from its
values at the interpolation points, i.e. the particles:
\begin{eqnarray}
\nabla A(\vec{r}) = \sum_j{\frac{A_j}{\rho_j} m_j \nabla W(\vec{r}-\vec{r}',h)}.
\label{grad1}
\end{eqnarray}

Anyway, better accuracy is found rewriting the formulae with the
density \textit{inside} operators and using the general rule
\begin{eqnarray}
\rho \nabla A = \nabla (\rho A) - A \nabla \rho
\label{gradients}
\end{eqnarray}

\noindent \citep[e.g.,][]{Monaghan1992}. Thus the divergence of
velocity is customarily estimated by means of
\begin{eqnarray}
\nabla \cdot \vec{v}_i = \sum_j m_j [(\vec{v}_j-\vec{v}_i) \cdot \nabla_i W_{ij}] / \rho_i,
\end{eqnarray}

\noindent where $\nabla_i W_{ij}$ denotes the gradient of
$W(\vec{r}-\vec{r}',h)$ with respect to the coordinates of a
particle $i$.


The dynamical evolution of a continuous fluid is governed by the
well known laws of conservation: the continuity equation which
ensures conservation of mass, the Euler equation which represents
the conservation of momentum, and the equation of energy
conservation (plus a suitable equation of state to close the
system). These equations must be written in discretized form to be
used with the SPH formalism, and can be obtained by means of the
Lagrangian analysis, following the same strategy used to obtain the
gravitational accelerations.

The continuity equation is generally replaced by Eq.
\ref{densitySPH} \footnote{Alternatively, it could be written as
\begin{eqnarray}
\frac{d\rho_i}{dt} = \sum_j m_j (\vec{v}_j-\vec{v}_i) \nabla_i W_{ij},
\label{continuitysph}
\end{eqnarray}

\noindent which is advantageous in case the simulated fluid has
boundaries or edges, but has the disadvantage of not conserving the
mass exactly.}. Like in the case of the  gravitational softening
length $\epsilon$, the smoothing length $h$ may in principle be a
constant parameter, but the efficiency and accuracy of the SPH
method is greatly improved adapting the resolution lengths to the
local density of particles. A self-consistent method is to adopt the
same algorithm described in Sect. \ref{soft}, i.e. obtaining $h$ from
\begin{eqnarray}
\frac{dh}{dt} = \frac{dh}{dn}\frac{dn}{dt},
\label{dhdt}
\end{eqnarray}

\noindent where of course $n$ is the local number density
\textit{of SPH particles only}, and relating $h$ to
$n$ by requiring that a fixed number
of kernel-weighted particles is contained within a smoothing sphere, i.e.
\begin{eqnarray}
h_i = \eta \left( \frac{1}{n_i} \right)^{1/3}.
\label{hn}
\end{eqnarray}

\noindent Note that in this case, while still obtaining the mass
density by  summing Eq. \ref{densitySPH}, one can  compute the
number density of particles at particle $i$'s location, i.e.:
\begin{eqnarray}
n_i = \sum_j W(|\vec{r}_{ij}|,h_i),
\label{nsum}
\end{eqnarray}

\noindent which is clearly formally equivalent to Eq.
\ref{rho0}\footnote{ Some authors \citep[e.g.][]{Hu2005, Ott2003}
proposed a different "number density" formulation of SPH in which
$\rho_i = m_i \times n_i$, and $n_i$ is obtained via Eq.
\ref{nsum}. While this can help in resolving sharp discontinuities
and taking into account the presence of multi-phase flows, it may as
well lead to potentially disastrous results if mixed unequal mass
particle are \textit{not} intended to model density discontinuities
but are instead used as mass tracers in a homogeneous field.}.


It is worth recalling here that in the first versions of SPH with
spatially-varying resolution, the spatial variation of the smoothing
length was generally not considered in the  the equations of motion,
and this resulted in secular errors in the conservation of entropy.
The inclusion of extra-correcting terms (the so-called
\textit{$\nabla h$ terms}) is therefore important to ensure the
conservation of \textit{both} energy and entropy. For example,
\citet{Serna1996} studied the pressure-driven expansion of a gaseous
sphere, finding that while the energy conservation is generally good
even without  the inclusion of the corrections (and sometimes get
slightly worse if these are included!), errors up to $\sim 5\%$ can
be found in the entropy conservation, while all this does not occur
 with the inclusion of the extra-terms. Although the effects of
entropy violation in SPH codes are not completely clear and need to
be analysed in much more detail, especially in simulations where
galaxies are formed in a cosmological framework, there are many
evidences that the problem must be taken into consideration.
\citet{Alimi2003} have analysed this issue in the case of the
collapse of isolated objects and have found that, if correcting
terms are neglected, the density peaks associated with central cores
or shock fronts are overestimated at a $\simeq 30\%$ level.

These $\nabla h$ terms were first introduced explicitly
\citep{Nelson1994}, with a double-summation  added to the canonical
SPH equations. Later, an implicit formulation was obtained
\citep{Monaghan2002, Springel2002}, starting from the Lagrangian
derivation of the equations of motion and self-consistently
obtaining correcting terms which accounts for the variation of $h$.
Obviously such terms are formally similar to those obtained for the
locally varying gravitational softening lengths (Sect. \ref{soft}).

Following \citet{Monaghan2002}, the Lagrangian for non-relativistic
fluid dynamics can be written as
\begin{eqnarray}
L=\int \rho \left (\frac{1}{2}v^2 - u \right) dV,
\end{eqnarray}

\noindent (the gravitational part is not included here since SPH
does not account for gravity), which in the SPH formalism becomes
\begin{eqnarray}
L_{SPH} = \sum_j m_j \left[ \frac{1}{2}v_j^2 - u_j \right].
\end{eqnarray}

As already made for Eq. \ref{EulerLagrange}, the equations of motion
of a particle can be obtained from the Euler-Lagrange equations,
giving
%
%
\begin{eqnarray}
\frac{d\vec{v}_i}{dt} = - \sum_j m_j \left( \frac{\partial
u}{\partial \rho} \right)\frac{\partial \rho_j}{\partial \vec{r}_i}.
\label{dvdt}
\end{eqnarray}

\noindent To obtain the term $\partial \rho_j/ \partial \vec{r}_i$,
one can note that using the density $\rho$ given by  Eq.
\ref{densitySPH} and  letting the smoothing length  vary according
to Eq. \ref{hn} one gets
\begin{eqnarray}
\frac{\partial \rho_j}{\partial \vec{r}_{i}} =
\frac{1}{\Omega_j} \sum_k m_k \nabla_i W_{ik}(h_i) \delta_{ij} -
m_i \nabla_j W_{ij}(h_j),
\end{eqnarray}

\noindent where $\delta_{ij}$ is the Kronecker delta function,
$\nabla_i W$ is the gradient of the kernel function $W$ taken with
respect to the coordinates of particle $i$ keeping $h$ constant, and
\begin{eqnarray}
\Omega_j = 1 - \frac{\partial h}{\partial \rho} \sum_k m_c
\frac{\partial W_{jk}(h_{j})}{\partial h_j}
\end{eqnarray}

\noindent is an extra-term that accounts for the dependence of the
kernel function on the smoothing length. Note that $\Omega$  is
formally identical to the $\Upsilon$ term introduced in Sect.
\ref{gravity}, replacing $\epsilon$ with $h$. This is obvious since
the underlying mathematics is exactly the same; both terms arise
when the motion equations  are derived from the  Lagrangian
formulation.

In case of particles with different mass, the  term $\partial
\rho_j/
\partial \vec{r}_i$ is
\begin{eqnarray}
\frac{\partial \rho_j}{\partial \vec{r}_i} = \sum_k m_k \nabla_j
W_{ij}(h_j)\left[ 1+ \frac{\zeta_j/m_k}{\Omega_j^*}\right]
(\delta_{ij}-\delta_{ik}), \label{drhodr}
\end{eqnarray}

\noindent where
\begin{eqnarray}
\zeta_i = \frac{\partial h_i}{\partial n_i}\sum_j m_j \frac{\partial
W_{ij}(h_i)}{\partial h_i}
\end{eqnarray}

\noindent and
\begin{eqnarray}
\Omega_i^* = 1 - \frac{\partial h_i}{\partial n_i} \sum_j
\frac{\partial W_{ij}(h_i)}{\partial h_i}.
\end{eqnarray}
(Price 2009, private communication).  These terms can be easily
computed for each particle while looping over neighbours to obtain
the density from Eq. \ref{densitySPH}.

The term $\partial u / \partial \rho$ in Eq. \ref{dvdt} can be
derived from the first law of thermodynamics, $dU = dQ - PdV$,
dropping the $dQ$ term since only adiabatic processes are
considered. Rewriting in terms of specific quantities, the volume
$V$ becomes volume ``per mass'', i.e. $1/\rho$, and $dV = d(1/\rho) =
-d\rho / \rho^2$. Thus,

\begin{eqnarray}
du = \frac{P}{\rho^2} d\rho,
\end{eqnarray}
\noindent  and, if no dissipation is present, the specific entropy
is constant, so

\begin{eqnarray}
\left(\frac{\partial u}{\partial t} \right)_s = \frac{P}{\rho^2}.
\label{dudrho}
\end{eqnarray}

Inserting Eqs. \ref{dudrho} and \ref{drhodr} into Eq. \ref{dvdt},
the equations of motion finally become
\begin{eqnarray}
\frac{d\vec{v}_i}{dt} & = & \sum_j m_j \times  \\
& & \left[ \frac{P_i}{\rho_i^2} \left( 1+\frac{\zeta_i/m_j}
{\Omega_i^*}\right) \nabla_i W_{ij}(h_i) + \right. \nonumber \\
& & \left. \frac{P_j}{\rho_j^2} \left( 1+\frac{\zeta_j/m_i}
{\Omega_j^*}\right) \nabla_i W_{ij}(h_j) \right], \nonumber
\label{motionvarmass}
\end{eqnarray}

The equation for the thermal energy equation is
\begin{eqnarray}
\frac{du_i}{dt} & = &  \\
& & \sum_j m_j \left[ \frac{P_i}{\rho_i^2} \left(
1+\frac{\zeta_i/m_j}{\Omega_i^*}\right) (\vec{v}_j-\vec{v}_i) \cdot
\nabla_i W_{ij}(h_i) \right]. \nonumber \label{thermalvarmass}
\end{eqnarray}

It can be shown that if all SPH particles have the same mass, Eqs.
\ref{motionvarmass} and \ref{thermalvarmass} reduce to the standard
Eqs. 2.10 and 2.23 in \citet{Monaghan2002} \footnote{As noted by
\citet{Schussler1981}, the gradient of the spline kernel Eq.
\ref{spline} can lead to unphysical clustering of particles. To
prevent this, \citet{Monaghan2000}  introduced a small artificial
repulsive pressure-like force between particles in  close pairs.
In \textsc{EvoL} a modified form of the kernel gradient is instead
adopted, as suggested by \citet{Thomas1992}:

\begin{eqnarray}
&& \frac{dW}{d\vec{r}} = -\frac{1}{4 \pi} \times
\begin{cases}
4 & \text{if $0 \leq u < 2/3$,} \\
3u(4-3u) &\text{if $2/3 \leq u < 1$,} \\
3(2-u)^2 &\text{if $1 \leq u < 2$,} \\
0 &\text{if $u \geq 2$,}
\end{cases}
\end{eqnarray}

\noindent with the usual meaning of the symbols.}.

%
%
%

Equations \ref{nsum} (density summation) and Eq. \ref{hn}  form a
non-linear system to be solved for  $h$ and $n$, adopting the same
method described in Sect. \ref{soft}.

A drawback  of this scheme can be encountered when a sink
of internal energy such as radiative cooling is included. In this
case, very dense clumps of cold particles can form and remain stable
for long periods of time. Since the contribution of neighbouring
particles to the density summation (Eq. \ref{nsum}) is weighted by
their distance from the active particle, if such clump is in the
outskirts of the smoothing region then each body will give a very
small contribution to the summation, and this will result in an
unacceptably long neighbour list. In this cases, the adoption of a
less accurate but faster method is appropriate. The easiest way to
select $h$ is to require that a constant number of neighbouring
particles is contained within a sphere of radius $\eta h$ (perhaps
allowing for small deviations). This type of procedure  was
generally adopted in the first SPH codes, and provided sufficiently
accurate results. If the scheme in question is adopted
(\textsc{EvoL} may choose between the two algorithms), the terms
$\nabla h$ are neglected, because the relation Eq. \ref{hn} is no longer
strictly valid. In the following, we will refer to this scheme as to
the \textit{standard SPH scheme}, and to previous one  based on the
$n-h$ (or $\rho-h$ relation)  as to the \textit{Lagrangian SPH
scheme}.

A test to check the ability of the different algorithms to capture
the correct description of an (ideally) homogeneous density field
has been performed, using particles of different masses mixed
together in the domain $0<x<1, 0<y<0.1, 0<z<0.1$, to obtain the
physical densities $\rho = 1$ if $x < 0.5$ and $\rho = 0.25$ if $x
\geq 0.5$. To this aim, a first set of particles were displaced on a
regular grid and were given the following masses:
\begin{displaymath}
m  =  0.9 \times 10^{-6} \mbox{    for } x  <  0.5; \\
m  =  2.25 \times 10^{-7} \mbox{    for } x  \geq  0.5.
\end{displaymath}
\noindent Then, a second set of particles were displaced on another
superposed regular grid, shifted along all three dimensions with
respect to the previous one by a factor $\Delta x = \Delta y =
\Delta z = 5 \times 10^{-3}$. These particles were then assigned the
following masses
\begin{displaymath}
m  =  0.1 \times 10^{-6} \mbox{    for } x  <  0.5; \\
m  =  0.25 \times 10^{-7} \mbox{    for } x  \geq  0.5.
\end{displaymath}

\begin{figure}[!htbp]
\centering
\includegraphics[width=8.5cm,height=6.5cm]{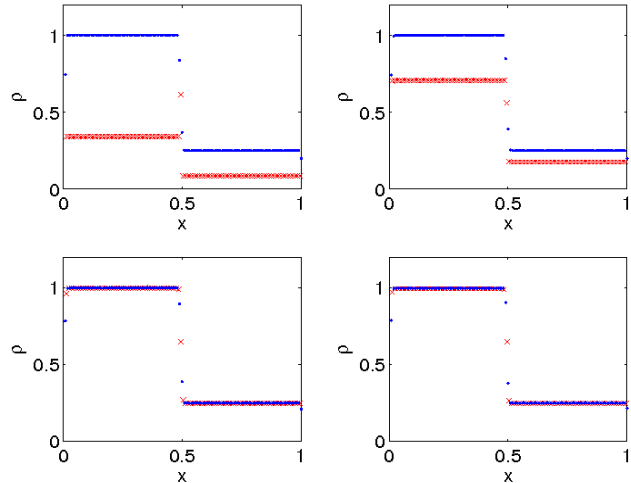}
\caption{Determination of the Initial density  adopting four different SPH
algorithms. Left to right, top to bottom: standard SPH, mass density
summation; Lagrangian SPH, mass density summation; standard SPH,
number density summation; Lagrangian SPH, number density
determination. Blue dots: high mass particles; red crosses: low mass
particles.} \label{densitydetermination}
\end{figure}

\noindent Then, four different schemes were used to compute the
density of particles:
\begin{itemize}
\item standard SPH, mass density summation
\item standard SPH, number density summation
\item Lagrangian SPH, mass density summation
\item Lagrangian SPH, number density summation
\end{itemize}

\noindent where the expressions ``mass density'' and ``number density''
refer to the different scheme adopted to compute the density of a
particle, i.e. to Eq. \ref{densitySPH} and \ref{nsum}, respectively.
Looking at Fig. \ref{densitydetermination}, it is clear that the
discrete nature of the regular displacement of particles gives
different results depending on the adopted algorithm. The mass
density formulation is not able to properly describe the uniform
density field: low-mass particles strongly underestimate their
density, on both sides of the density discontinuity. The situation
is much improved when a number density formulation is adopted.
Little differences can be noted between the determinations using the
constant neighbours scheme and the density-$h$ relation.

\subsection{On symmetrisation}

To ensure the conservation of energy and momenta, a proper
symmetrisation of the equations of motion is required. It is worth
noticing that while in the first formulations of SPH such
symmetrisation had to be imposed ``ad hoc'', in the Lagrangian
derivation it is naturally built-in without any further assumption.

Anyway, the symmetrisation is strictly necessary only when pairwise
forces act on couples of particles. Thus, it is not needed when
obtaining the density from Eq. \ref{nsum} and \ref{densitySPH}, or
when the internal energy of a particle is evolved according to Eq.
\ref{thermalvarmass}. Indeed, symmetrisation of the energy equation
may lead to unphysical negative temperatures. For example when cold
particles are pushed by a strong source of pressure: in this case,
the mechanical expansion leads to a loss of internal energy which,
if equally subdivided between the cold and hot phase, may exceed the
small energy budget of cold bodies \citep[see e.g.
][]{Springel2002}.

Since symmetrisation is not needed in the density summation
algorithm, the smoothing length of neighbouring particles is not
required to compute the density of a particle $i$. This allows an
exact calculation simply gathering position and masses of
neighbours, which are known at the beginning of the time-step. On
the other hand, after the density loop each particle has  its new
smoothing length $h$ and it is now possible to use these values in
the acceleration calculations, where symmetrisation is needed.
Anyway, while in the previous case only particles within $2h_i$ were
considered neighbours of the particle $i$, in this latter case the
symmetrisation scheme requires a particle $j$ to be considered
neighbour of particle $i$ if $r_{ij} < 2 \times \max(h_i, h_j)$
\footnote{It would also be possible to compute the interactions 
by only considering neighbours within $h_i$ (or $\epsilon_i$). 
To do so, each particle should ``give back'' to its 
neighbours its contribution to the acceleration. Anyway, this
approach is a more complicated with individual timesteps 
(where one has to compute inactive contributions to active 
particles), and in a parallel architecture. We therefore
decided to adopt the fully symmetric scheme in this release
of \textsc{EvoL}}.

Note that with this scheme \textit{two} routines of neighbour
searching are necessary: the first one in the density summation, and
the second in the force calculation. In practice, when the
evaluation of density is being performed, a \textit{gather} view of
SPH is adopted \citep[see][]{Hernquist1989}; thus, each particle $i$
searches its neighbours using only its own smoothing length $h_i$,
until the convergence process is completed. In this way, if a
particle finds too high or too low a number of neighbours (and
therefore must change its smoothing length to repeat the search) it
is not necessary to inform the other particles (including those
belonging to other processors) about the change that is being made.
This would in principle be necessary if a standard
\textit{gather/scatter} mode is adopted. Thus, each particle
evaluates its density independently, adjusting its smoothing length
on its own, and no further communication among processors is necessary,
after the initial gathering of particles positions.
Moreover, the search is initially performed using an effective
searching radius which is slightly larger than the correct value
$2h_i$. In this way, the neighbour list will include slightly more
particles than effectively needed. However, if at the end of the
loop, the smoothing length has to be increased by a small factor, it
is not be necessary to repeat the tree-walk to reconstruct the
neighbour list. Of course, if the adaptive softening length scheme
is adopted the tree-walk will be performed to upgrade $h$ and $\rho$
for SPH particles only, \textit{and} $\epsilon$ and $\rho_{tot}$ for
all particles.

Then, when forces (hydrodynamical forces and, if the adaptive
softening length scheme is adopted, gravitational forces too) are
evaluated, a new search for neighbours must be made, in which a
particle is considered neighbour to another one if the distance
between the two is lower than the maximum between the two searching
radii. During this second tree-walk, the gravitational force is
evaluated together with the construction of the particle's neighbour
list. Finally, correcting terms are computed for hydrodynamics
equations of motion ($\nabla h$ terms) and, if necessary, for
gravitational interactions (if softening lengths are adaptive),
summing the neighbouring particles' contributions.

\subsection{Smoothing and softening lengths for SPH particles} \label{187}

SPH particles under self-gravity have two parameters characterizing
their physical properties: the softening length $\epsilon$, which tunes
their gravitational interactions with nearby particles, and the
smoothing length $h$, used to smooth their hydrodynamical features.

In principle there is no need for these two quantities to be related
to one another, because they refer to two distinct actions, and have
somehow opposite functions as noted above. However, \citet{Bate1997}
claimed that a quasi-stable clump of gas could became  un-physically
stable against collapse if $\epsilon>h$, because in this case
pressure forces would dominate over the strongly softened gravity,
or on the other hand, if $\epsilon<h$, collapse on scales smaller
than the formal resolution of SPH may be induced, causing artificial
fragmentation. They recommend that gravitational and hydrodynamical
forces should be resolved at the same scale length, imposing
$\epsilon = h$ (requiring both of them to fixed, or $\epsilon$ to be
adaptive like $h$, with the introduction of the correcting terms
described above). In other studies \citep[e.g.][]{Thacker2000} the
softening length is fixed, but the smoothing length is allowed to
vary imposing a lower limit close to the value of the softening
length (this is a typical assumption in SPH simulations of galaxy
formation).

Anyway, \citet{Williams2004} have studied the effects of such
procedure, finding that in many hydrodynamical studies this can
cause a number of problems. The adoption of large values of $h$ can
result in over-smoothed density fields, i.e. a decrease in the
computed values of density $\rho$ and of density gradients $\nabla
\rho$, which in turn un-physically decreases the computed pressure
accelerations in the Euler equation. Problems with angular momentum
transfer may also arise. Therefore, they suggest to allow $h$ to
freely vary. This  eventually yields improved accuracy and also,
contrary to expectations, a significant saving in computational time
(because the number of neighbours is kept fixed instead of largely
increasing as it may happen in dense environments with fixed $h$).
Moreover, they pointed out that, in contrast with the claims by
\citet{Bate1997}, the smoothing length should not be kept equal to
the softening length, because in many physical situations (shocks
for example) hydrodynamical forces dominate over gravity  and likely
need to be properly resolved on size scales smaller than the
softening length.

Finally, a numerical problem arises when trying to keep $\epsilon=h$
in cosmological simulations. The presence of a Dark Matter component
makes it impossible to keep constant the number of both
hydrodynamical \textit{and} gravitational neighbours. For example, a
collapsed Dark Matter halo could retain few gaseous particles
(because of shocks and/or stellar feedback), and these will thus
have many gravitational neighbours but few hydrodynamic neighbours
with a unique softening/smoothing length. The opposite problem could
arise in the case of very dense clumps of cold gas that may form
behind shocks, without a corresponding concentration of Dark Matter.

Thus, in \textsc{EvoL} both $\epsilon$ and $h$ are let free to vary
independently from one another. Future tests may be done trying to
keep the two parameters linked, for example allowing for some
variations in both their ratio $\epsilon / h$ (imposing it to be
\textit{around}, but not exactly, 1) and in the number of
hydrodynamical and gravitational neighbours.

\subsection{Discontinuities}

\subsubsection{Artificial viscosity} \label{visco}

The SPH formalism in its straight form is not able to handle
discontinuities. By construction, SPH smooths out local
properties on a spatial scale of the order of the smoothing length
$h$. Anyway, strong shocks are of primary importance in a number of
astrophysical systems. Thus, to model a real discontinuity in the
density and/or pressure field \textit{ad hoc} dissipational,
pressure-like terms must be added to the perfect fluid equations, to
spread the discontinuities over the numerically resolvable length.
This is usually achieved by introducing  an \textit{artificial
viscosity}, a numerical artifact that is not meant to mimic physical
viscosity, but to reproduce on large scale the action of
smaller, unresolved scale physics, with a formulation consistent 
with the Navier-Stockes terms.

Among the many formulations that have been proposed, the most
commonly used is obtained adding a term $\Pi_{ij}$ to the Eqs.
\ref{motionvarmass} and \ref{thermalvarmass}, so that
\begin{eqnarray}
\frac{d\vec{v}_i}{dt} & = & \sum_j m_j \times  \\
& & \left[ \frac{P_i}{\rho_i^2}
\left( 1+\frac{\zeta_i/m_j}{\Omega_i^*}\right) \nabla_i W_{ij}(h_i) + \right. \nonumber \\
& & \left. \frac{P_j}{\rho_j^2}
\left( 1+\frac{\zeta_j/m_i}{\Omega_j^*}\right) \nabla_i W_{ij}(h_j)
+ \Pi_{ij} \bar{\nabla_i W_{ij}} \right], \nonumber
\label{motionvarmass1}
\end{eqnarray}

\noindent and
\begin{eqnarray}
\frac{du_i}{dt} & = & \\
& & \sum_j m_j \left[ \frac{P_i}
{\rho_i^2}\left( 1+\frac{\zeta_i/m_j}{\Omega_i^*}\right)
(\vec{v}_j-\vec{v}_i) \cdot \nabla_i W_{ij}(h_i) \right.  \nonumber \\
& & + \left. \frac{1}{2}\Pi_{ij} w_{ij} |\bar{\nabla_i W_{ij}}| \right]. \nonumber
\label{thermalvarmass1}
\end{eqnarray}

\noindent Here $w_{ij} = (\vec{v}_{ij} \cdot
\vec{r}_{ij})/|\vec{r}_{ij}|$,

\begin{eqnarray}
&& \Pi_{ij} =
\begin{cases}
\left( \frac{-\alpha\bar{c_{ij}}\mu_{ij}+\beta\mu_{ij}^2}
{\bar{\rho_{ij}}} \right) & \text{if  $\vec{v}_{ij}\cdot\vec{r}_{ij} < 0$,} \\
0 &\text{if $\vec{v}_{ij}\cdot\vec{r}_{ij} \geq 0$,}
\end{cases}
\label{mu_visc}
\end{eqnarray}

\noindent
\begin{eqnarray}
\mu_{ij} = \frac{h\vec{v}_{ij}\cdot\vec{r}_{ij}}{\vec{r_{ij}^2}+\eta^2},
\end{eqnarray}

\noindent and
\begin{eqnarray}
\bar{\nabla_i W_{ij}} & = & \\
& & \frac{1}{2} \left[ \nabla_i W_{ij}(h_i) \left(
1+\frac{\zeta_i/m_j}{\Omega_i^*} \right) \right. \\ \nonumber & &
\left. + \nabla_i W_{ij}(h_j) \left(
1+\frac{\zeta_j/m_i}{\Omega_j^*} \right) \right] \nonumber
\end{eqnarray}

\noindent \citep{Monaghan1988}. $\bar{c_{ij}}$ is the average sound
speed, which is computed for each particle  from a suitable equation
of state (EoS),  like the ideal monatomic gas EoS $P = (\gamma-1)
\rho u$, using $c_i = \sqrt{\gamma (\gamma -1) u_i}$ ($\gamma$ is
the adiabatic index); the parameter $\eta$ is introduced to avoid
numerical divergences, and it is usually taken to be $\eta = 0.1 h$.
$\alpha$ and $\beta$ are two free parameters, of order unity, which
account respectively for the shear and bulk viscosity in the
Navier-Stokes equation \citep{Watkins1996}, and for the von
Neuman-Richtmyer viscosity, which works in high Mach number
flows.


A more recent formulation is the so-called ``signal velocity''
viscosity \citep{Monaghan1997}, which works better for small pair
separations between particles, being a first-order expression of
$h/r$; it reads
\begin{eqnarray}
\Pi_{ij} = \frac{-\alpha v_{ij}^{sig}w_{ij}}{\bar{\rho_{ij}}},
\label{pvsig}
\end{eqnarray}

\noindent with the \textit{signal velocity} defined as
\begin{eqnarray}
v_{ij}^{sig}= 2 \bar{c_{ij}} - w_{ij}.
\label{vsig}
\end{eqnarray}

\noindent One or the other  of the two  formulations above can be
used in \textsc{EvoL}.

In general, artificial dissipation should be applied only where
really necessary. However, it has long been  recognized that
artificial viscosity un-physically boosts post-shock shear viscosity,
suppressing structures in the velocity field on scales well above
the nominal resolution limit, and damping the generation of
turbulence by fluid instabilities.

To avoid the overestimate of the shear viscosity,
\citet{Balsara1995} proposed to multiply the chosen expression of
$\Pi_{ij}$ by a function $\bar{f_{ij}}=(f_i+f_j)/2$; the proposed
expression for the ``limiter'' $f_i$ for a particle $i$ is
\begin{eqnarray}
f_i = \frac{|\nabla_i\cdot\vec{v}_i|}
{|\nabla_i\cdot\vec{v}_i|+|\nabla_i\times\vec{v}_i|+\eta c_i/h_i},
\end{eqnarray}

\noindent where $\eta \sim 10^{-4}$ is a factor introduced to avoid
numerical divergences. It can be seen that $f$ acts as a limiter
reducing the efficiency of viscous forces in presence of rotational
flows.

Another possible cure to the problem is as follows. The most
commonly used values for the parameters $\alpha$ and $\beta$ in Eq.
\ref{mu_visc} are $\alpha = 1$ and $\beta = 2$. Bulk viscosity is
primary produced by $\alpha$, and this sometimes over-smoothen the
post-shock turbulence. To cope with this, \citet{Morris1997} have
proposed a modification to Eq. \ref{mu_visc}, in which the parameter
$\alpha$ depends on the particle and changes with time according to:
\begin{eqnarray} \label{alphaswitch}
\frac{d\alpha_i}{dt} = - \frac{\alpha_i-\alpha_{min}}{\tau} + G_i,
\end{eqnarray}

\noindent where $\alpha_{min} = 0.01$ is the minimum allowed value,
$\tau = 0.1 h_i/v^{sig}$ is an $e$-folding time scale (with $v^{sig}
= \max_j[v_{ij}^{sig}]$), and $G_i$ is a source term which can be
parameterized as $G_i = 0.75 f_i \max[0,-|\nabla\cdot\vec{v}_i|]$.
This formulation embodies some different prescriptions \citep[e.g.
in][]{Rosswog2007,Price2008}. In Eq. \ref{pvsig}, one can then put
$\alpha = \frac{1}{2}(\alpha_i+\alpha_j)$. \citet{Dolag2005} have
shown that this scheme captures shocks as well as the original
formulation of SPH, but, in regions away from shocks, the numerical
viscosity is much smaller. In their high resolution cluster
simulations, this resulted in higher levels of turbulent gas motions
in the ICM, significantly affecting radial gas profiles and bulk
cluster properties. Interestingly, this tends to reduce the
differences between SPH and adaptive mesh refinement simulations of
cluster formation.

\subsubsection{Artificial thermal conductivity}

As \citet{Price2008} pointed out, an approach similar to that
described above should be used to smooth out discontinuities in
\textit{all} physical variables. In particular, an
\textit{artificial thermal conductivity} is necessary to resolve
discontinuities in thermal energy (even if only a few formulations
of SPH in literature take this into account).

The artificial thermal conductivity is represented by  the term
\begin{eqnarray}
\Pi^u = -\frac{\bar{\alpha_{ij}^u} v_{sig}^u (u_i-u_j)}{\bar{\rho_{ij}}}.
\end{eqnarray}

\noindent Here $\alpha^u$ is a conductivity coefficient,
$\frac{1}{2}(\alpha_i^u+\alpha_j^u)$ that varies with time according
to
\begin{eqnarray}
\frac{d\alpha_i^u}{dt} = - \frac{\alpha_i^u-\alpha^u_{min}}{\tau} + S_i^u,
\end{eqnarray}

\noindent where $\tau$ is the same time scale as above,
$\alpha^u_{min}$ is usually $0$, and the source term can be defined
as $S_i^u = h_i |\nabla^2 u_i|/\sqrt{u_i}$ \citep{Price2008}, in
which  the expression for the second derivative is taken from
\citet{Brookshaw1986}
\begin{eqnarray}
\nabla^2 u_i = \sum_j 2 m_j \frac{u_i - u_j}
{\rho_j} \frac{|\bar{\nabla_i W_{ij}}|}{r_{ij}}
\end{eqnarray}

\noindent which reduces particles noise;  $v_{ij}^{sig}$ can be
either the same quantity defined in Eq. \ref{vsig} or
\begin{eqnarray}
v_{ij}^{sig} = \sqrt{\frac{|P_i-P_j|}{\rho_{ij}}}
\end{eqnarray}

\noindent In passing, we note  that adopting the source term
suggested by \citet{Price2005}, $S_i^u = 0.1 h_i |\nabla^2 u_i|$,
would be less accurate in problems requiring an efficient thermal
conduction such as the standard shock tube test with un-smoothed
initial conditions, see Sect. \ref{test}.

The term $\Pi^u$ must finally be added to Eq. \ref{thermalvarmass1}, giving
\begin{eqnarray}
\frac{du_i}{dt} & = &\\ \nonumber
& & \sum_j m_j \left[ \frac{P_i}{\rho_i^2}\left( 1+\frac{\zeta_i/m_j}
{\Omega_i^*}\right) (\vec{v}_j-\vec{v}_i) \cdot \nabla_i W_{ij}(h_i) \right. \\ \nonumber
& & + \left. \left( \frac{1}{2}\Pi_{ij} + \Pi_{ij}^u \right) w_{ij} |\bar{\nabla_i W_{ij}}| \right].
\label{thermalvarmass2}
\end{eqnarray}

\noindent As in the case of artificial viscosity, it is worth noting
that this conductive term is not intended to reproduce a physical
dissipation; instead, it is a merely numerical artifact introduced
to smooth out unresolvable discontinuities.

\subsection{Entropy equation}

Instead of looking at the thermal energy, we may follow the
evolution  of a function of the total particle entropy, for instance
the \textit{entropic function}
\begin{eqnarray}
A(s) = \frac{P}{\rho^{\gamma}},
\end{eqnarray}

\noindent where $\gamma$ is the adiabatic index. This function is
expected to grow monotonically in presence of shocks,  to change
because of radiative losses or external heating, and to remain
constant for an adiabatic evolution of the gas. \citet{Springel2002}
suggest the following SPH expression for  the equation of entropy
conservation:
\begin{eqnarray}
\frac{d A_i}{dt} & = & - \frac{\gamma -1}
{\rho_i^{\gamma}} S(\rho_i,u_i) \\
& & + \frac{1}{2}\frac{\gamma -1}{\rho_i^{\gamma-1}}
\sum_j m_j \Pi_{ij} \vec{v}_{ij}\cdot \nabla_i W_{ij}, \nonumber
\label{SPHentr}
\end{eqnarray}

\noindent where $S$ is the \textit{source function} describing  the
energy losses or gains due to non adiabatic physics apart from
viscosity. If $S<<1$, the function $A(s)$ can only grow in time,
because of the shock viscosity. Eq. \ref{SPHentr} can be integrated
instead of Eq. \ref{thermalvarmass1} giving a more stable behaviour
in some particular cases. Note that internal energy and entropic
function of a particle can be related via
\begin{eqnarray}
u=\frac{A(s)}{\gamma-1} \rho^{\gamma-1}.
\end{eqnarray}

\noindent The entropic function can be used to detect situations of
shocks, since its value only depends on the strength of the viscous
dissipation.

In general, if the energy equation is used (integrated),  the
entropy is not exactly conserved, and viceversa. Moreover, if the
density is evolved according to the continuity Eq.
\ref{continuitysph} and the thermal energy according to  Eq.
\ref{thermalvarmass}, both energy and entropy are conserved; but in
this case the mass is not be conserved (to ensure this latter,
density must be computed with Eq. \ref{densitySPH}).
This problem should be cured with the inclusion of the $\nabla h$ terms
in the SPH method, as already discussed.

\subsection{X-SPH}

As an optional alternative, \textsc{EvoL} may adopt the smoothing of
the velocity field by replacing the normal equation of motion
$d\vec{r}_i/dt = \vec{v}_i$ with \begin{eqnarray}
\frac{d\vec{r}_i}{dt}=\vec{v}_i+\eta\sum_j m_j \left(
\frac{\vec{v_{ji}}}{\bar{\rho_{ji}}} \right) W_{ij}, \label{xsph}
\end{eqnarray}

\noindent \citep{Monaghan1992} with $\bar{\rho_{ji}} = (\rho_i +
\rho_j)/2$, $\vec{v_{ji}} = \vec{v}_j-\vec{v}_i$, and $\eta$
constant, $0 \leq \eta \leq 1$. In this variant of the SPH method,
known as \textit{X-SPH}, a particle moves with a velocity  closer to
the average velocity in the surroundings. This formulation can be
useful in simulations of weakly incompressible fluids, where it
keeps particles in order even in absence of viscosity, or to avoid
undesired inter-particle penetration.

\section{Description of the Code: Integration }\label{description_integration}

\subsection{Leapfrog integrator} \label{kdk}

Particles positions and velocities are advanced in time by means of a
standard \textit{leapfrog} integrator, in the so-called
\textit{Kick-Drift-Kick} (\textbf{KDK}) version,
where the \textbf{K} operator evolves
velocities and the \textbf{D} operator evolves positions:

\begin{eqnarray}
&& \mbox{\textbf{K$_{1/2}$}: } \vec{v_{n+1/2}} =
\vec{v_{n}}+\frac{1}{2}\vec{a(\vec{r_n})} \delta t \nonumber\\
&& \mbox{\textbf{D: }} \vec{r_{n+1}} =
\vec{r_n}+\vec{v_{n+1/2}} \delta t \nonumber\\
&& \mbox{\textbf{K$_{1/2}$}: } \vec{v_{n+1}} =
\vec{v_{n+1/2}}+\frac{1}{2} \vec{a(\vec{r_{n+1}})} \delta t. \nonumber
\end{eqnarray}

\noindent A predicted value of physical quantities at the end of each time-step 
is also predicted at the beginning the same step, to guarantee
synchronization in the calculation of the accelerations and other
quantities. In particular, in the computation of the viscous
acceleration on gas particles the synchronized predicted velocity 
$\vec{v_{n+1}} =\vec{v_{n}}+\vec{a(\vec{r_n})} \delta t $ is used.

Moreover, if the individual time-stepping  is adopted
(see below, Sect. \ref{inddt}), non-active particles use predicted
quantities to give their own contributions to forces and
interactions.

\subsection{Time-stepping} \label{tst}

A trade-off between accuracy and efficiency can be reached by a
suitable choice of the time-step. One would like  to obtain
realistic results in a reasonable amount of CPU time. To do so,
numerical stability must be ensured, together with a proper
description of all relevant physical phenomena, at the same time
reducing the computational cost keeping the  time-step as large as
possible.

To this aim, at the end of each step,  for each particle the optimal
time-step is given by the shortest among the following times:

\begin{eqnarray}
&& \delta t_{acc,i} = \eta_{acc} \sqrt{\frac{min(\epsilon_i,h_i)}{|\vec{a}_{i}|}}  \nonumber \\
&& \delta t_{vel,i} = \eta_{vel} \sqrt{\frac{min(\epsilon_i,h_i)}{|\vec{v}_{i}|}}  \nonumber \\
&& \delta t_{av,i} = \eta_{av} \sqrt{\frac{|\vec{v}_{i}|}{|\vec{a}_{i}|}} \nonumber
\end{eqnarray}

\noindent with obvious meaning of symbols; the parameters $\eta$'s
are of the order of $0.1$. The third of these ratios, under
particular situations, can assume extremely small values, and should
therefore be used with caution.

We can compute two more values, the first obtained from the well
known \textit{Courant condition} and the other constructed to avoid
large jumps in thermal energy:

\begin{eqnarray}
&& \delta t_{C,i} = \eta_C \frac{h_i}{h_i |\nabla \cdot
\vec{v}_{i}| + c_i +1.2(\alpha c_i + \beta \max_j |\mu_{ij}|)}  \nonumber \\
&& \delta t_{u,i} = \eta_u \left|\frac{u_i}{du_i/dt}\right| \nonumber
\end{eqnarray}

\noindent where $\alpha, \beta$ and $\mu$ are the quantities defined
in the artificial viscosity parametrization, see Sect. \ref{visco},
and $c$ is the sound speed.

If the simulation is run with a  \textit{global} time-stepping
scheme, all particles are advanced adopting the minimum of all
individual time-steps calculated in this way; synchronization is
clearly guaranteed by definition.

If desired, a minimum time-step can be imposed. In this case, however,
limiters must be adopted to avoid numerical divergences in accelerations
and/or internal energies. This of course leads to a poorer description of
the physical evolution of the system; anyway, in some cases a threshold
may be necessary to avoid very short time-steps due to violent high
energy phenomena.

\subsubsection{Individual time-stepping} \label{inddt}

Cosmological systems usually display a wide range of densities and
temperatures, so that an accurate description of very different
physical states is necessary within the same simulation. Adopting
the \textit{global} time-stepping scheme may cause a huge waste of
computational time, since gas particles in low-density environments
and collisionless particles generally require much longer time-steps
than the typical high-density gas particle.

In \textsc{EvoL} an \textit{individual} time-stepping
algorithm can be adopted, which makes a better use of the computational resources.
It is based on a powers-of-two scheme, as in \citet{Hernquist1989}.
All individual particles real time steps, $\delta t_{i,true}$, are
chosen to be a power-of-two subdivision of the largest allowed
time-step (which is fixed at the beginning of the simulation), i.e.
\begin{eqnarray}
\delta t_{i,true} = \delta t_{max} / 2^{n_i},
\end{eqnarray}

\noindent where $n_i$ is chosen as the minimum integer for which the
condition $\delta t_{i,true} \geq \delta t_i$ is satisfied.
Particles can move to a smaller time bin (i.e. a longer time step)
at the end of their own time step, but are allowed to do so  only if
that time bin is currently time-synchronized with their own time
bin, thus ensuring synchronization at the end of every largest time
step.

 A particle is then marked as ``active'' if $t_i + \delta
t_{i,true} = t_{sys}$, where the latter is the global system time,
updated as $t_{sys,new} = t_{sys,old} + \delta t_{sys}$, where
$\delta t_{sys} = min(\delta t_i)$. At each time-step, only active
particles re-compute their density and update their accelerations
(and velocities) via summation over the tree and neighboring
particles. Non-active particles keep their acceleration and velocity
fixed and only update their position using a prediction scheme at
the beginning of the step. Note that some other evolutionary
features (i.e. cooling or chemical evolution) are instead updated at
every time-step for \textit{all} particles.

%
%

Since in presence of a sudden release of very hot material (e.g.
during a Supernova explosion) one or a few particles may have to
drastically and abruptly reduce their time-step, the adoption of
individual time-stepping scheme would clearly lead to unphysical
results, since neighbouring non-active particles would not notice
any change until they become active again. This is indeed a key
issue, very poorly investigated up to now. To cope with this, a
special recipe has been added, forcing non-active particles to ``wake
up'' if they are being shocked by a neighbour active particle, or in
general if their time-step is too long (say more than 4-8 times
bigger) with respect to the minimum time-step among their
neighbours. In this way, when an active particle suddenly changes
its physical state, all of its neighbouring particles are soon
informed of the change and are able to react properly.
\citet{Saitoh2008} recently studied the problem, coming to similar
conclusions and recommending a similar prescription when modelling
systems in which sudden releases of energy are expected.

Note that when a non-active particle is waken up energy and momentum
are not perfectly conserved, since its evolution during the current
time-step is not completed. Anyway, this error is negligible if
compared to the errors introduced by ignoring this correction.

\subsection{Periodic boundary conditions} \label{periodism}

Periodic boundary conditions are implemented in \textsc{EvoL} using
the well known \citet{Ewald1921} method, i.e. adding to the
acceleration of each particle an extra-term due to the replicas of
particles and nodes, expanding in Fourier series the potential and
the density fluctuation, as described in \citet{Hernquist1991} and
in \citet{Springel2001}. If $\vec{x}_i$ is the coordinate at the
point of force-evaluation relative to a particle or a node $j$ of
mass $m_j$, the \textit{additional} acceleration that must be added
to account for the action of the infinite replicas of $j$ is given
by

\begin{eqnarray}
& & acc_{per}(\vec{x}) = m_j \left[ \frac{\vec{x}}
{|\vec{x}|^3} - \sum_n \frac{\vec{x}-\vec{n}L}{|\vec{x}-\vec{n}L|^3} \times \right.\\
& & \left( erfc(\omega|\vec{x}-\vec{n}L|) + \frac{2\omega|\vec{x}-\vec{n}L|}
{\sqrt{\pi}} \times e^{-\omega^2|\vec{x}-\vec{n}L|^2} \right) \nonumber\\
& & \left. - \frac{2}{L^2} \sum_{\vec{h}/=0} \frac{\vec{h}}
{|\vec{h}|^2} e^{ -\frac{\pi^2|\vec{h}|^2}{\omega^2 L^2} } sin
\left( \frac{2 \pi}{L} \vec{h} \cdot \vec{x} \right)  \right] \nonumber
\end{eqnarray}

\noindent where $\vec{n}$ and $\vec{h}$ are integer triplets, $L$ is
the periodic box size, and $\omega$ is an arbitrary number.
Convergence is achieved for $\omega = 2/L$, $|\vec{n}|<5$ and
$|\vec{h}|<5$ \citep{Springel2001}. The correcting terms
$acc_{per}/m_j$ are tabulated, and trilinear interpolation off the
grid is used to compute the correct accelerations. 
It must be pointed out that this interpolation significantly slows down the
tree algorithm, almost doubling the CPU time.

Periodism in the SPH scheme is obtained straightforwardly
finding neighbours across the spatial domain and taking the modulus 
of their distance, in order to obtain their nearest image with respect
to the active particle. In this way, no ghost particles need to be introduced.

\subsection{Parallelization}

\begin{figure*}[!thbp]
\centering
\includegraphics[width=14cm,height=8cm]{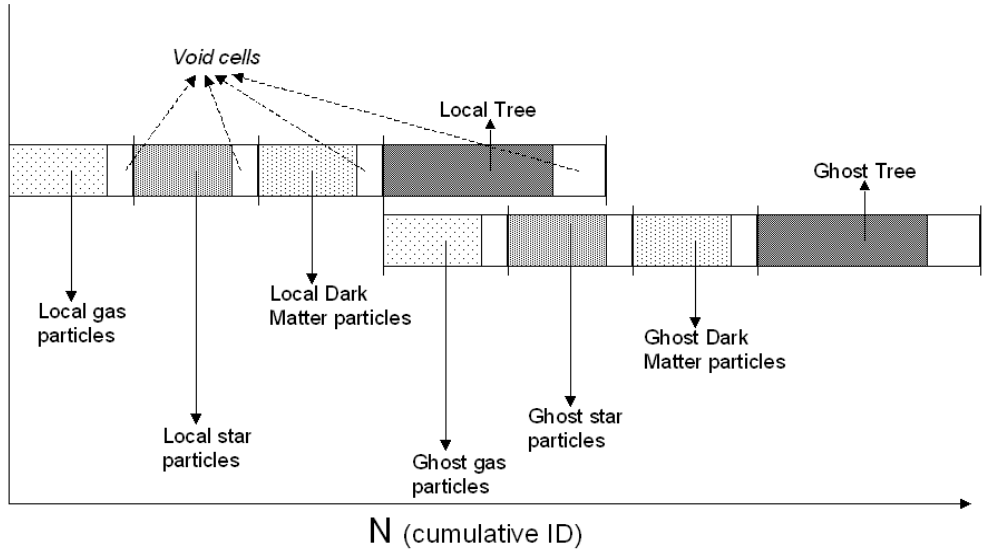}
\caption{Data structures in \textsc{EvoL}. Void cells are included
to allow for new particles to be created and/or imported from other
CPUs. Ghost structures are reallocated at each time-step.} \label{data}
\end{figure*}


\textsc{EvoL} uses the MPI communication protocol to run in parallel on multiple
CPUs. First of all, particles must be subdivided among the CPUs, so
that each processor has to deal only with a limited subset of the
total number of bodies. To this aim, at each time-step the spatial
domain is subdivided using the \textit{Orthogonal Recursive
Bisection} (ORB) scheme. In practice, each particle keeps track of
the time spent to compute its properties during the previous
time-step in which it has been active, storing it in a work-load
variable. At the next time-step the spatial domain is subdivided
trying to assign to each CPU the same total amount of work-load
(only considering active particles if the individual time-stepping
option is selected), re-assigning bodies near the borders among the
CPUs. To this aim, the domain is first cut along the $x$-axis, at
the coordinate $x_{cut}$ that better satisfies the equal work-load
condition among the two groups of CPUs formed by those with
geometrical centers $x_{centre} \leq x_{cut}$ and those
with $x_{centre}>x_{cut}$. In practice, each CPU exchanges
particles only if its borders are across $x_{cut}$. Then, the two
groups of CPUs are themselves subdivided into two new groups, this
time cutting along the $y$-axis (at two  different
coordinates, because they are independent from one another having
already exchanged particles along the $x$-axis). The process is
iterated as long as required (depending to the number of available CPUS)
cutting recursively along the three dimensions. It should be noted
that this scheme  allows to use only  a power-of-two number of
CPUs, whereas other procedures \citep[e.g. a Peano-Hilbert domain
decomposition, see e.g.][]{Springel2005} can more efficient
exploit  the available resources.

It may sometimes happen that a CPU wants to acquire more particles
than permitted by the dimensions of its arrays. In this case, data
structures are re-allocated using temporary arrays to store
extra-data, as described in the next Section.

Apart from the allocation of particles to the CPUs, the other task
for the parallel architecture is to spread the information about
particles belonging to different processors. This is achieved by
means of ``ghost'' structures, in which such properties are stored
when necessary. A first harvest among CPUs is made at each time step
\textit{before} computing SPH densities: at this point, positions,
masses and other useful physical features of nearby nodes and
particles, belonging to other processors, are needed. Each CPU
``talks''  only to another CPU per time, first importing \textit{all}
the data-structures belonging to it in temporary arrays, and then
saving within its ``ghost-tree'' structure only the data relative to
those nodes and particles which will be actually used. For example,
if a ``ghost node'' is sufficiently far away from the active CPU
geometrical position so that the gravitational opening criterion is
satisfied, there will be no need to open it to compute gravitational
forces, and no other data relative to the particles and sub-nodes
it contains will be stored in the ghost-tree.

A second communication among CPUs is necessary before computing
accelerations. Here, each particle needs to known the exact value of
the smoothing and softening lengths (which have been updated during
the density evaluation stage) as well as many other physical
values of neighbouring (but belonging to different CPUs) particles.
The communication scheme is exactly the same as before, involving
ghost structures. It was found that, instead of upgrading the
ghost-tree built before the density evaluation, a much faster
approach is to completely re-build it.

The dimensions of the ghost arrays are estimated at the beginning of
the simulation, and every $N$ time-steps (usually, $N=100$), by
means of a ``preliminary walk'' among all CPUs. In intermediate steps,
the dimensions of these arrays are modified on the basis of the
previous step evaluation and construction of ghost structures.

\subsection{Data structures} \label{dst}

All data structures are dynamically adapted to the size
of the problem under exploration. In practice, at the beginning of a
simulation,  all arrays are allocated so that their dimension is
equal to the number of particles per processor, plus some empty
margin to allow for an increase of the number of particles to be
handled by the processor (see Fig. \ref{data}).

Anyway, whenever a set of arrays within a processor becomes full of
data and has no more room for new particles, the whole data
structure of that processor is re-scaled increasing its dimensions,
to allow for new particles to be included. Of course, also the
opposite case (i.e. if the number of particles decreases, leaving
too many void places) is considered.

To this aim, the data in excess are first stored in temporary
\textit{T} arrays. Then, the whole ``old'' data structure \textit{O}
is copied onto new temporary \textit{N} arrays, of the new required
dimensions. Finally, \textit{T} arrays are copied into the
\textit{N} structure; this substitutes the \textit{O} structure,
which is then deleted. Clearly, this procedure is quite
memory-consuming. Anyway, it ideally has to be carried out only a
few times per simulation, at least in standard cases. Moreover, it
is performed \textit{after} the ``ghost'' scheme described above has
been completed, and ghost-structures have been deallocated, leaving
room for other memory-expensive routines.

\section{Tests of the code} \label{test}

An extended set of hydrodynamical tests have been performed to check
the performance of \textsc{EvoL} under different demanding
conditions:

\begin{itemize}
\item Rarefaction and expansion problem \citep[][ 1D]{Einfeldt1991};
\item Shock tube problem \citep[][ both in 1D and in 3D]{Sod1978};
\item Interacting blast waves \citep[][ 1D]{Woodward1984};
\item Strong shock collapse \citep[][ 2D]{Noh1987};
\item Kelvin-Helmholtz instability problem (2D);
\item Gresho vorticity problem \citep[][ 2D]{Gresho1990};
\item Point-like explosion and blast wave (Sedov-Taylor problem, 3D);
\item Adiabatic collapse \citep[][ 3D]{Evrard1988};
\item Isothermal collapse \citep[][ 3D]{Boss1979};
\item Collision of two politropic spheres (3D);
\item Evolution of a two-component fluid (3D).
\end{itemize}

Except where explicitly pointed out, the tests have been run on a
single CPU, with the global time-stepping algorithm, and adopting
the following parameters for the SPH scheme: $\eta = 1.2$,
$\alpha_{max}=2$, $\alpha_{min}=0.01$, $\beta = 2\alpha$,
$\gamma=5/3$, and equation of state $P = (\gamma -1) \rho u$. Where
gravitation is present, the opening angle for the tree code has been
set $\theta = 0.8$. The tests in one and two dimensions have been
run adopting the kernel formulations shown in Appendix A. The exact
analytical solutions have been obtained using the software
facilities freely available at the website
\textit{http://cococubed.asu.edu/code\_pages/codes.shtml}.

\subsection{Rarefaction wave}

We begin our analysis  with a test designed to check the
the code in extreme, potentially dangerous
low-density regions, where some iterative Riemann solvers can return
negative values for pressure/density. In the Einfeldt rarefaction
test \citep{Einfeldt1991} the two halves of the computational domain
$0<x<1$ move in opposite directions, and thereby create a region of
very low density near the initial velocity discontinuity at $x=0.5$.
The initial conditions of this test are

\begin{eqnarray}
\begin{cases}
\rho = 1.0, \mbox{    } P  =  0.4, \mbox{    } v_x = -2.0, & \text{for $x  < 0.5$,} \\
\rho = 1.0, \mbox{    } P  =  0.4, \mbox{    } v  =  2.0,  & \text{for $x \geq  0.5$.}
\end{cases}
\end{eqnarray}

\begin{figure}[!htbp]
\centering
\includegraphics[width=8.5cm,height=6.5cm]{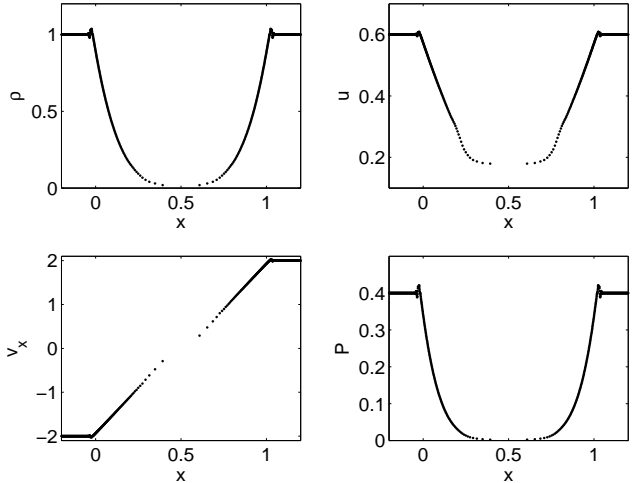}
\caption{Left to right, top to bottom: density, internal energy,
x-velocity, and pressure profiles at t=0.2 for the Einfeldt test.
All quantities are in code units. }
\label{einfeldt}
\end{figure}

The results at $t=0.2$ for a 1-D, 1000-particle calculation are
shown in Fig. \ref{einfeldt}. Clearly, the low-density region is
well described by the code.

\subsection{Shock Tube}

The classic Riemann problem to check the shock capturing capability
is the Sod's shock tube test \citep{Sod1978}. We run many different
cases changing numerical parameters, to check the response to
different settings. All tests have been started from
\textit{un-smoothed} initial conditions, leaving the code to smear
out  the initial discontinuities.

\subsubsection{1-D tests}

The first group of tests have been performed in one dimension.  1,000
equal mass particles have been displaced on the $x$ axis, in the
domain $0 \leq x \leq 1$, and changing inter-particle spacing, to
obtain the following setting:

\begin{eqnarray}
\begin{cases}
\rho = 1.0, \mbox{    } u  =  1.5, \mbox{    } v = 0, & \text{for $x  < 0.5$,} \\
\rho = 0.25, \mbox{    } u  =  1.077, \mbox{    } v  =  0,  & \text{for $x \geq  0.5$.}
\end{cases}
\end{eqnarray}

\noindent ($m=6.25 \times 10^{-4}$). In this way, particles
belonging to the left region of the tube initially have $P=1$, whereas
particles on the right side of the discontinuity have $P=0.1795$, as
in the classic \citet{Monaghan1983} test. We performed four runs:

\begin{itemize}
\item T1 - standard viscosity (with $\alpha=1$ and $\beta=2$) and
no artificial conduction;
\item T2 - standard viscosity plus artificial conduction;
\item T3 - viscosity with variable $\alpha$ plus artificial conduction;
\item T4 - the same as T2, but with the same density discontinuity obtained
by using particles with different masses on a regular lattice instead
of  equal mass particles shrinking their spacing. To do so,
particles in the left part of the tube have the mass
$m = 10^{-3}$, whereas those in the right part have $m = 2.5 \times 10^{-4}$.
\end{itemize}

\begin{figure}[!htbp]
\centering
\includegraphics[width=8.5cm,height=6.5cm]{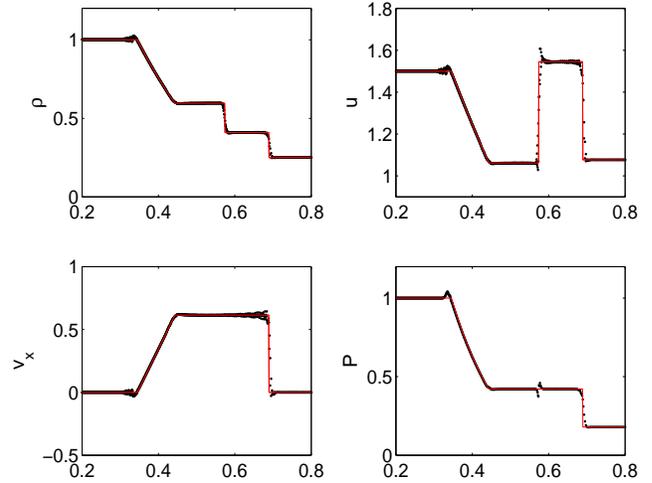}
\caption{Left to right, top to bottom: density, internal energy,
x-velocity and pressure profiles at t=0.12 for test T1 see text for
details. Red solid line: exact solution. Alla quantities are in code units.} \label{tube1}
\end{figure}

\begin{figure}[!htbp]
\centering
\includegraphics[width=8.5cm,height=6.5cm]{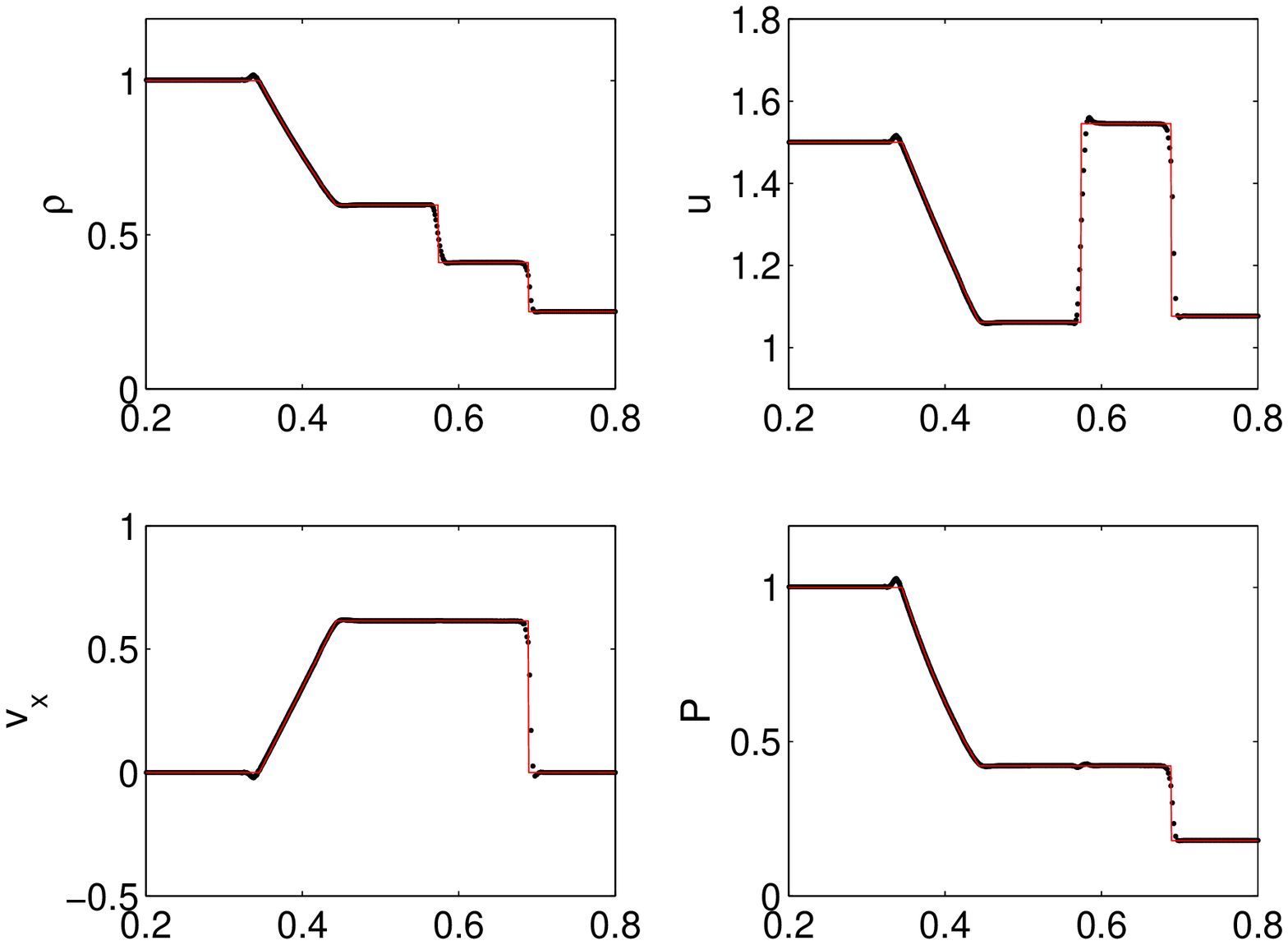}
\caption{the same as in Fig. \ref{tube1}, but for the test T2 (see text for details).} \label{tube2}
\end{figure}

\begin{figure}[!htbp]
\centering
\includegraphics[width=8.5cm,height=6.5cm]{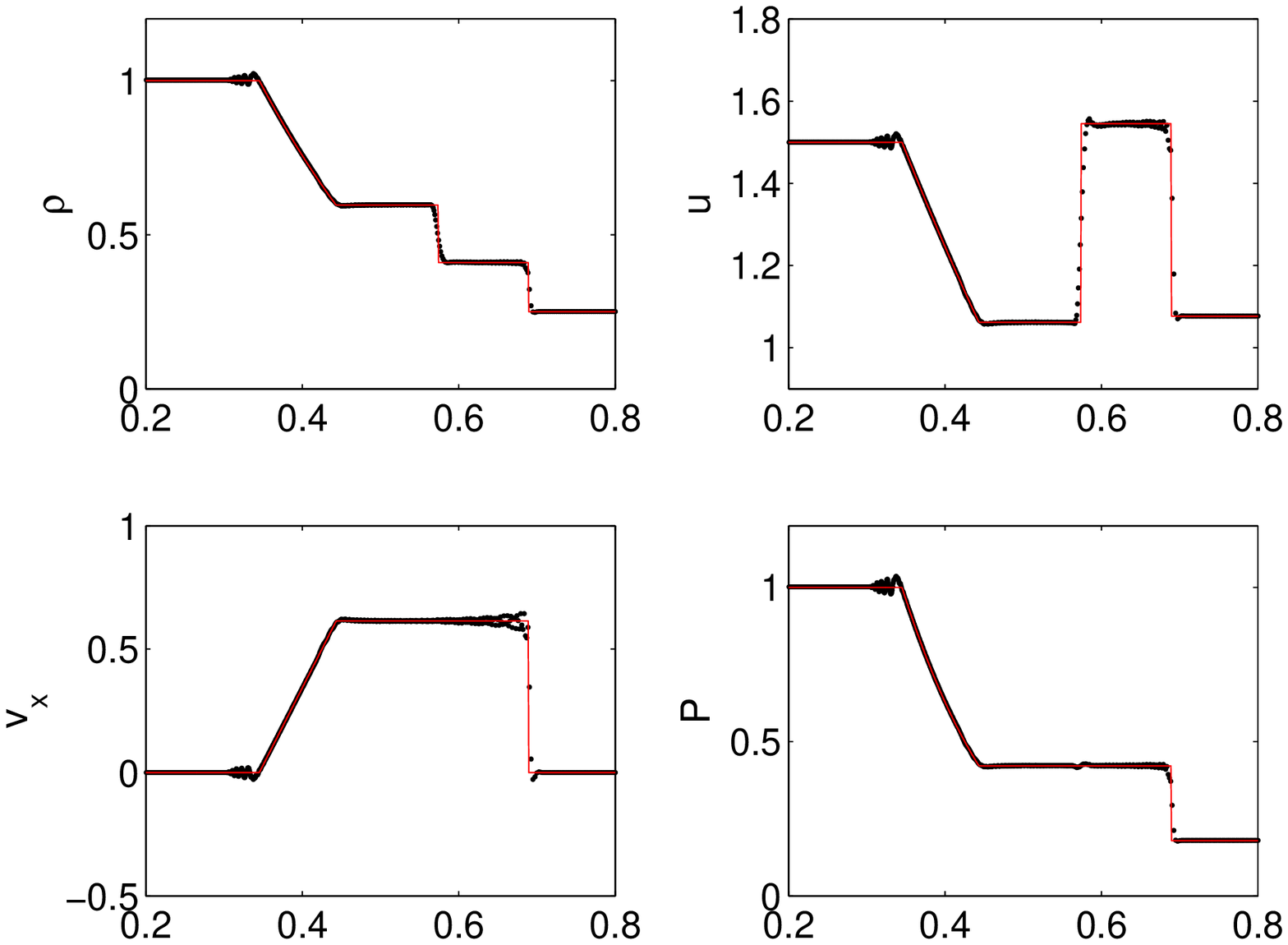}
\caption{The same as in Fig. \ref{tube1}, but for the test T3 (see text for details).} \label{tube3}
\end{figure}

\begin{figure}[!htbp]
\centering
\includegraphics[width=8.5cm,height=6.5cm]{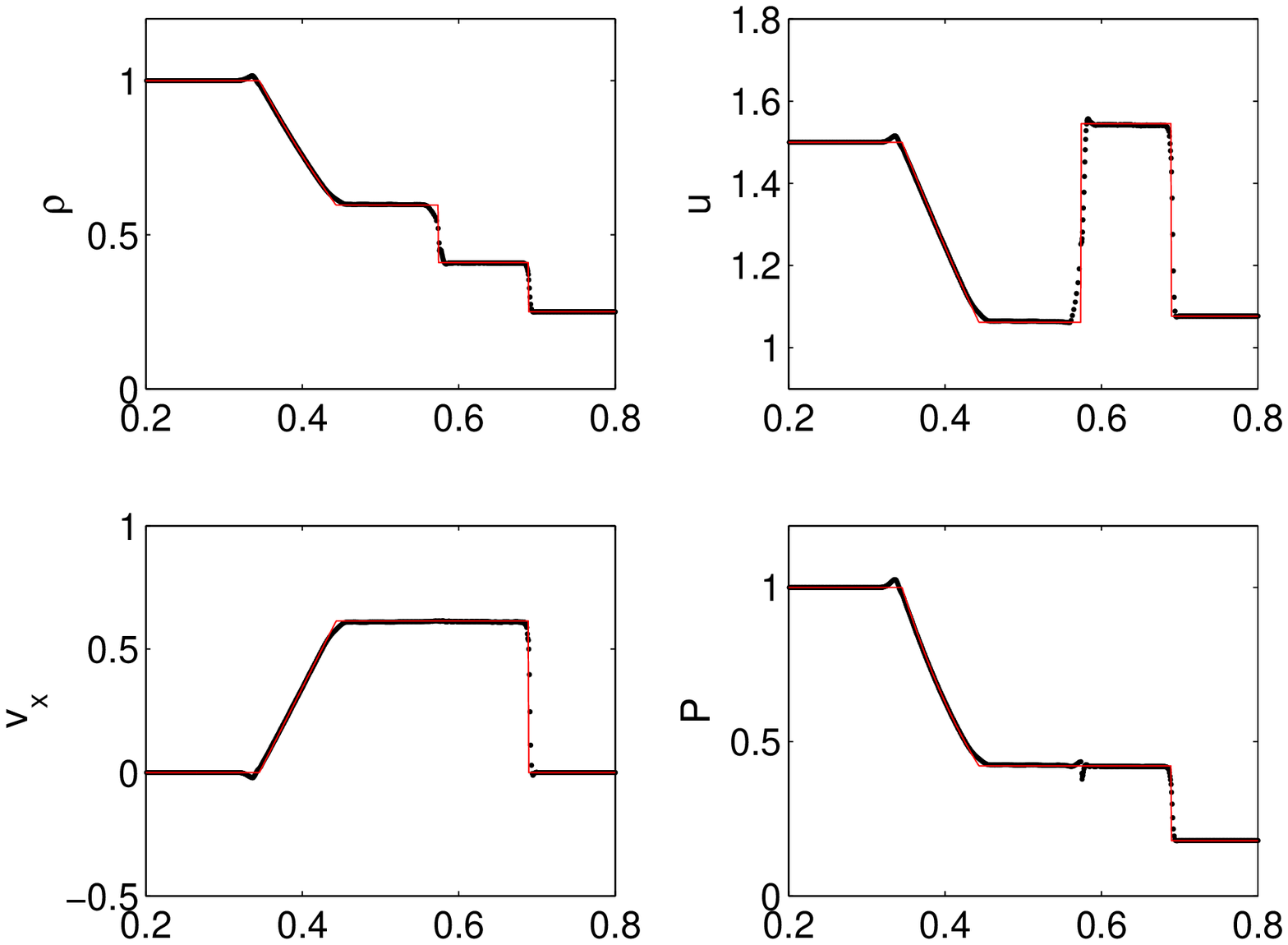}
\caption{The same as in Fig. \ref{tube1}, but for the test T4 (see text for details).} \label{tube4}
\end{figure}

\begin{figure}[!htbp]
\centering
\includegraphics[width=8.5cm,height=6.5cm]{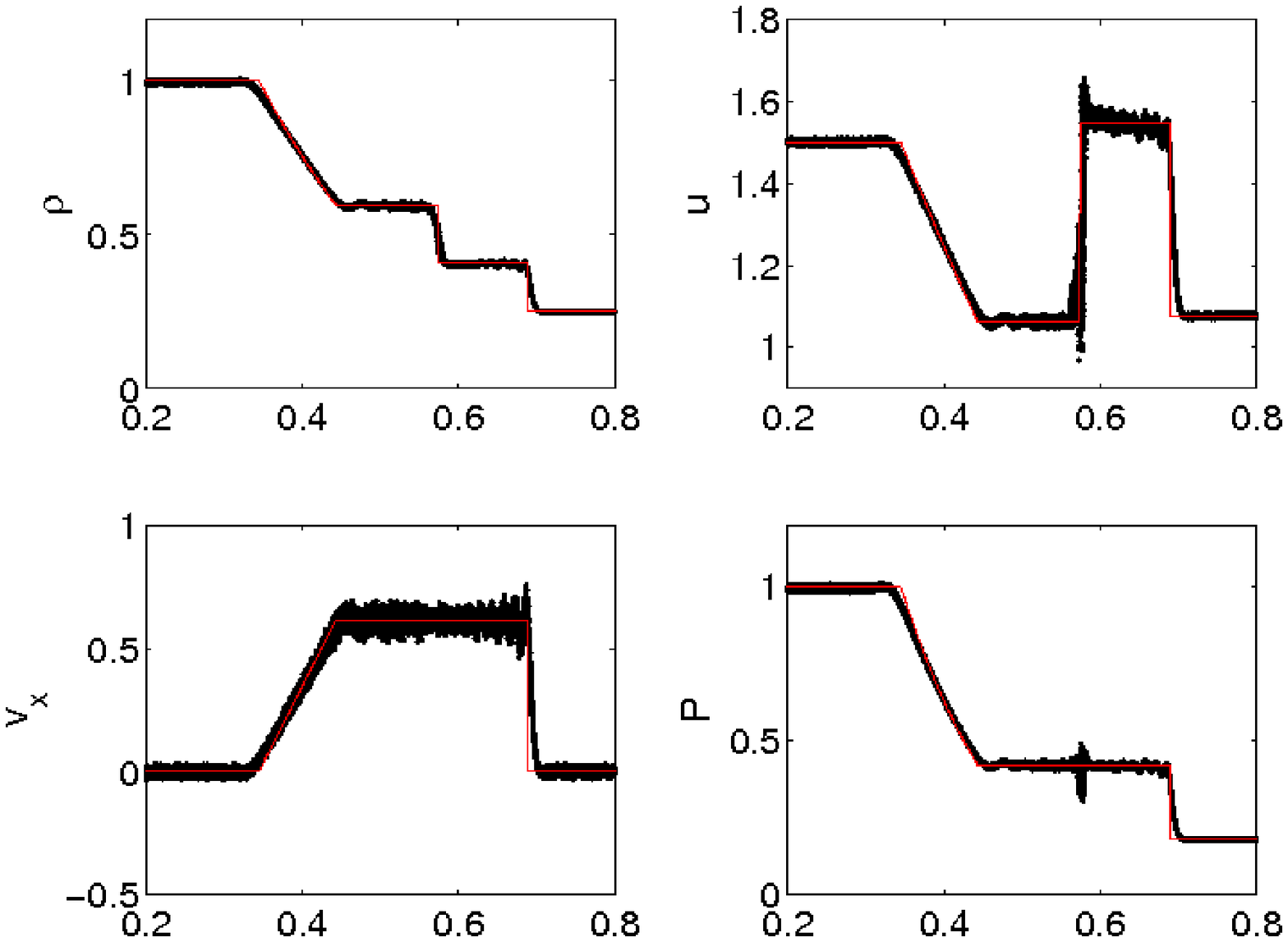}
\caption{The same as in  Fig. \ref{tube1}, but for the test T5 (see text for details).} \label{tube31}
\end{figure}

\begin{figure}[!htbp]
\centering
\includegraphics[width=8.5cm,height=6.5cm]{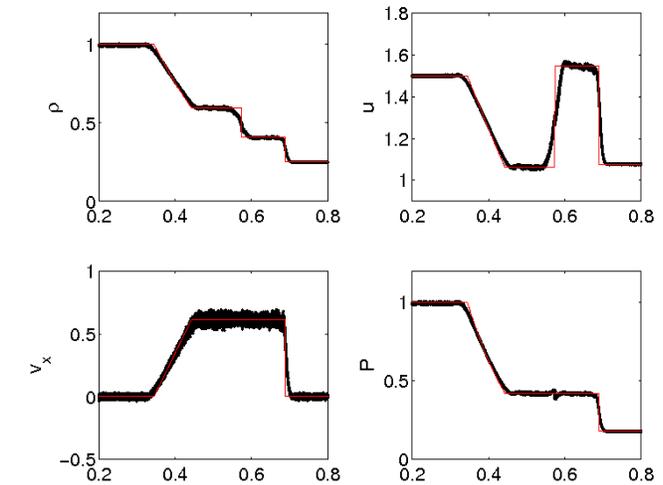}
\caption{The same as in  Fig. \ref{tube1}, but for the test T6 (see text for details).} \label{tube32}
\end{figure}

Figs. \ref{tube1} through  \ref{tube4} show the density, internal energy,
$x$-velocity and pressure $x$-profiles at $t=0.12$ for the four
cases (in code units).

The overall behaviour is good in all cases. We note the  typical \textit{blip}
in the pressure profile and the wall-like overshoot in  thermal energy
at the contact discontinuity in the T1 run. The
introduction of the artificial thermal conductivity (T2) clearly
cures the problem at the expense of a slightly shallower thermal
energy profile. Introducing the variable $\alpha$-viscosity (T3),
the reduced viscosity causes post-shock oscillations in the density
(and energy), and makes the velocity field noisier and
turbulent, as expected.

Little differences can be seen in the T4 run, except for a slightly
smoother density determination at the central contact discontinuity,
which also makes the \textit{blip} problem worse. Thus,
the best results are  obtained with the T2 configuration.

\subsubsection{3-D tests}

As pointed out by \citet{Steinmetz1993}, performing shock tube
calculations with 1-D codes, with particles initially displaced on a
regular grid, makes the effects of numerical diffusivity essentially
negligible, but in realistic cosmological 3-D simulations the
situation is clearly more intrigued. To investigate this issue, we
switch to a 3-D description of   the shock tube problem.
The initial conditions are set using a relaxed
\textit{glass} distribution of $\sim 60,000$ particles, in a box of
dimensions $[0, 1]$ along the $x$ axis and (periodic) $[0, 0.1]$
along the $y$ and $z$ axis. Particles are assigned a mass
$m_{right}=1.71 \times 10^{-7}$ and $m_{left}=2.5 \times m_{right}$
on the two sides on the discontinuity at $x=0.5$, with all other parameters
as in the 1-D case. A first run (T5) was
performed adopting the T1 configuration (standard viscosity, no artificial
conduction), while a second one (T6) was
performed switching on the artificial conduction term.

%

Figs. \ref{tube31} and \ref{tube32} show the state of the system at
$t=0.12$ in the two cases (note that \textit{all
particles} and not mean values are plotted). As in the 1-D case, the overall
agreement with the theoretical expectation is  quite good.
However,  the inclusion of the conductive term
on one hand yields sharper profiles and reduces the pressure
\textit{blip}, on the other hand gives a smoother description of
contact discontinuities (see the density  profile at $x=0.55$).
We argue that part of the worse results may be due to the use of particles with
different mass as a  similar result has been found with the  corresponding 1-D
test (see T4 above). A limit on the  conduction efficiency could give
better results, but and \textit{ad hoc} adjustment should be applied to each case.

\begin{figure*}[!thpb]
\centering
\includegraphics[width=12cm,height=7cm]{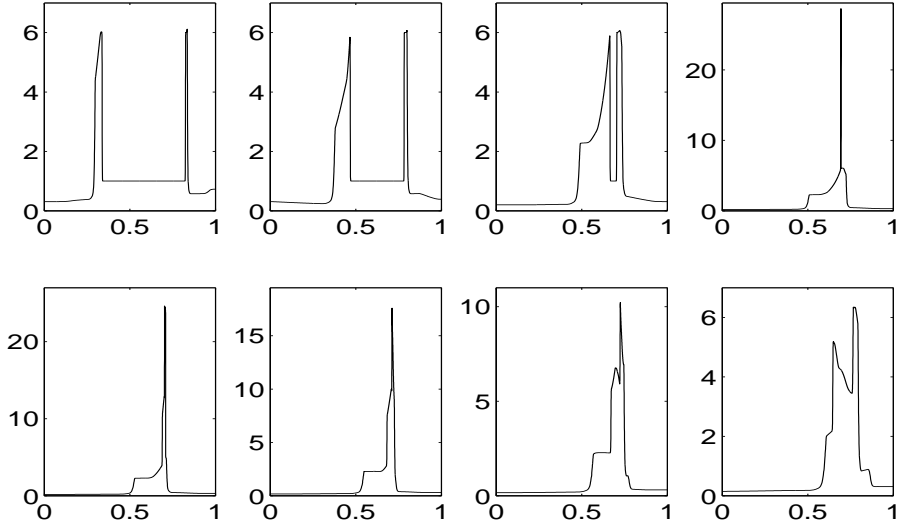}
\caption{Left to right, top to bottom: density profiles at
$t=0.010$, $t=0.016$, $t=0.026$, $t=0.028$, $t=0.030$, $t=0.032$,
$t=0.034$, $t=0.038$ for the Woodward high-resolution test (see text for
details). All quantities are in code units.} \label{woodward1}
\end{figure*}

\subsection{Interacting blast waves}

\citet{Woodward1984} proposed a severe test in 1-D, i.e.  two strong blast
waves develop out of a gas of
density $\rho = 1$ initially at rest in the domain $0<x<1$, with
pressure $P = 1000$ for $x < 0.1$, $P = 100$ for $x > 0.9$ and $P =
0.01$ in between (the adiabatic index is $\gamma = 1.4$). The boundary
conditions are reflective on both sides; this can be achieved introducing
ghost particles created on-the-fly. Evolving in time, the two blast
waves give rise to multiple interactions of strong shocks,
rarefaction and reflections.

We performed two runs: one at  high resolution with 1000 equally spaced particles,
and one at low resolution with only 200 particles. In both cases, the code
included the artificial conduction, and viscosity with constant $\alpha =1$.

Fig. \ref{woodward1} shows the density $x$-profiles for the high
resolution case, at $t=0.010$, 0.016, 0.026, 0.028, 0.030, 0.032,
0.034, 0.038. The results can be compared with those of
\citet{Woodward1984}, \citet{Steinmetz1993} and
\citet{Springel2009}. The complex evolution of the shock and
rarefaction waves is reproduced with good accuracy and with sharp
profiles, in nice agreement with the theoretical expectations. The
very narrow density peak at $t=0.028$, which should have $\rho
\approx 30$, is well reproduced and only slightly underestimated.

\begin{figure}[!hbtp]
\centering
\includegraphics[width=6.5cm,height=6.5cm]{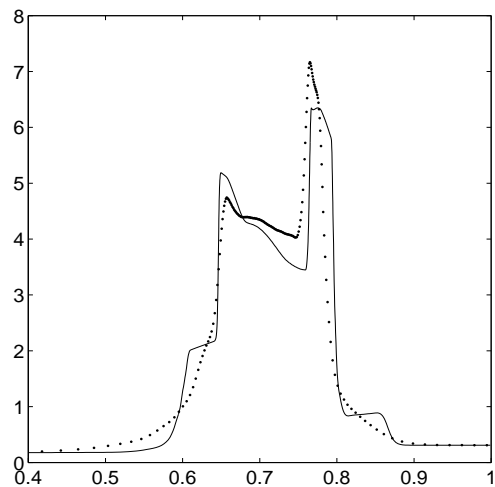}
\caption{Comparison between high-resolution (solid line) and
low-resolution (dots) density profiles at $t=0.038$ for the Woodward
test (see text for details). The quantities are in code units.} \label{woodward3}
\end{figure}

Fig. \ref{woodward3} shows the comparison between the density profiles
of the high and low resolution cases, at $t=0.038$. The
low-resolution one shows a substantial smoothing of the contact
discontinuities; nevertheless, the overall behaviour is still
reproduced, shock fronts are located at the correct positions, and
the gross features of the density profile are roughly reproduced.

\subsection{Kelvin-Helmoltz instability}

The  Kelvin-Helmoltz (KH) instability problem is a very demanding test.
Recently, \citet{Price2008} replied to the claim by \citet{Agertz2007} that
the SPH algorithm is not able to correctly model contact
discontinuities like the Kelvin-Helmoltz (KH) problem. He showed how
the inclusion of the conductive term is sufficient to correctly
reproduce the instability. We try to reproduce his results running
a similar test, using the 2-D version of \textsc{EvoL}.

We set up a regular lattice of particles in the periodic domain
$0<x<1$, $0<y<1$, using $256^2$ particles. The initial conditions are
similar to those adopted by \citet{Price2008}. The masses of
particles are assigned in such a way that $\rho = 2$ for $|y - 0.5|
< 0.25$, and $\rho = 1$ elsewhere \citep[note that in ][\, this
density difference was instead obtained changing the disposition of
equal mass particles]{Price2008}. The regions are brought to
pressure equilibrium with $P = 2.5$. In this case, although the
initial gradients in density and thermal energy are not smoothed,
the thermal energy is assigned to the particles \textit{after} the
first density calculation, to ensure a continuous initial pressure
field.

The initial velocity of the fluid is set as follows:

\begin{eqnarray}
&& v_x =
\begin{cases}
0.5 & \text{if $|y-0.5| < 0.25$,} \\
-0.5  & \text{if $|y-0.5| \geq 0.25$,} \\
\end{cases}
\end{eqnarray}

\noindent and

\begin{eqnarray}
v_y & = & w_0 \sin(4 \pi x) \times \\
&& \left\{ \exp \left[ -\frac{(y-0.25)^2}{2\sigma^2} \right] +
\exp \left[ -\frac{(y-0.75)^2}{2\sigma^2} \right] \right\}, \nonumber
\end{eqnarray}

\noindent where we use $\sigma = 0.05/\sqrt{2}$, and $w_0 = 0.1$
or $w_0 = 0.25$ for two different runs (K1 and K2, respectively).
In this way, a small perturbation in the velocity field is forced
near the contact discontinuity, triggering the instability.

\begin{figure*}[!htbp]
\centering
\includegraphics[width=12cm,height=10cm]{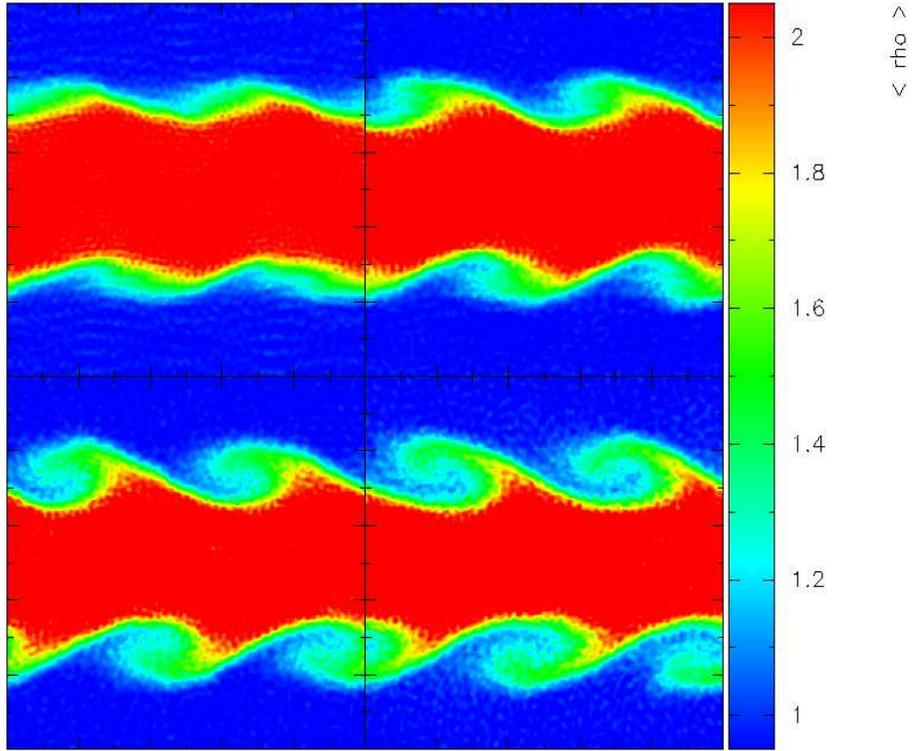}
\caption{Kelvin-Helmoltz instability according to test K1. Left to right,
top to bottom: density field at $t=0.768$, $t=1.158$,
$t=1.549$ and $t=1.940$. All quantities are in code units.} \label{kh1}
\end{figure*}

\begin{figure*}[!htbp]
\centering
\includegraphics[width=12cm,height=10cm]{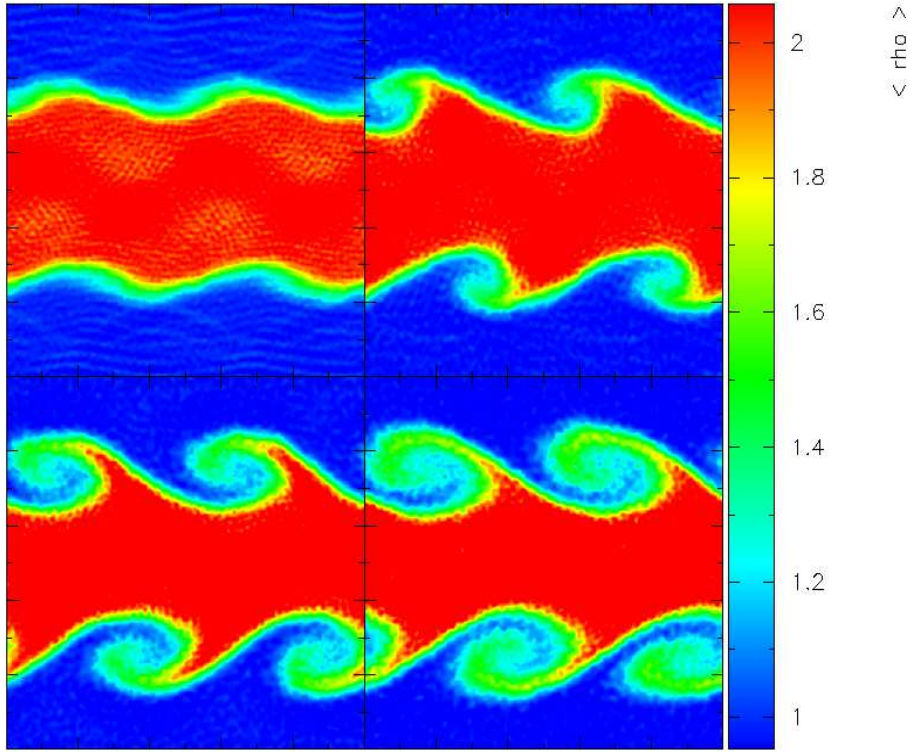}
\caption{Kelvin-Helmoltz instability according to test K2. Left to right,
top to bottom: density field at $t=0.377$, $t=0.767$, $t=1.158$
and $t=1.549$. All quantities are in code units.} \label{kh2}
\end{figure*}

\begin{figure*}[!htbp]
\centering
\includegraphics[width=12cm,height=5cm]{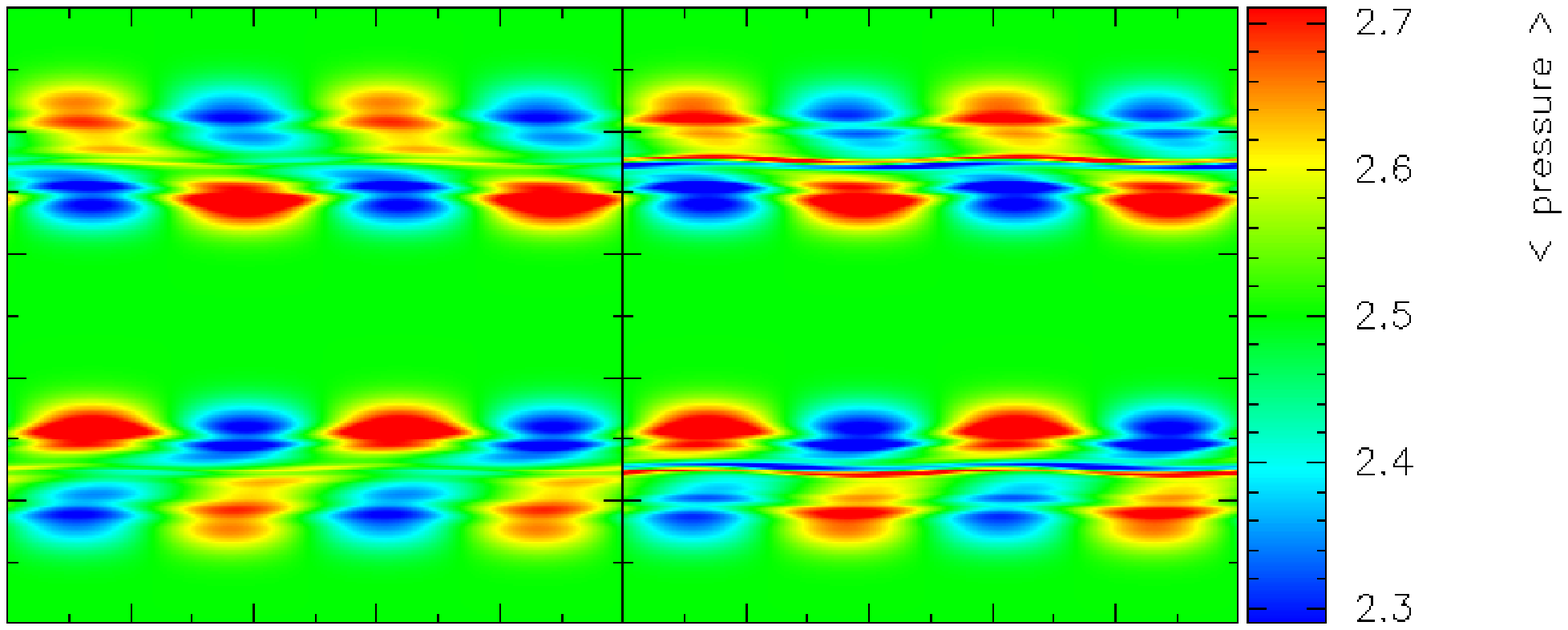}
\includegraphics[width=12cm,height=5cm]{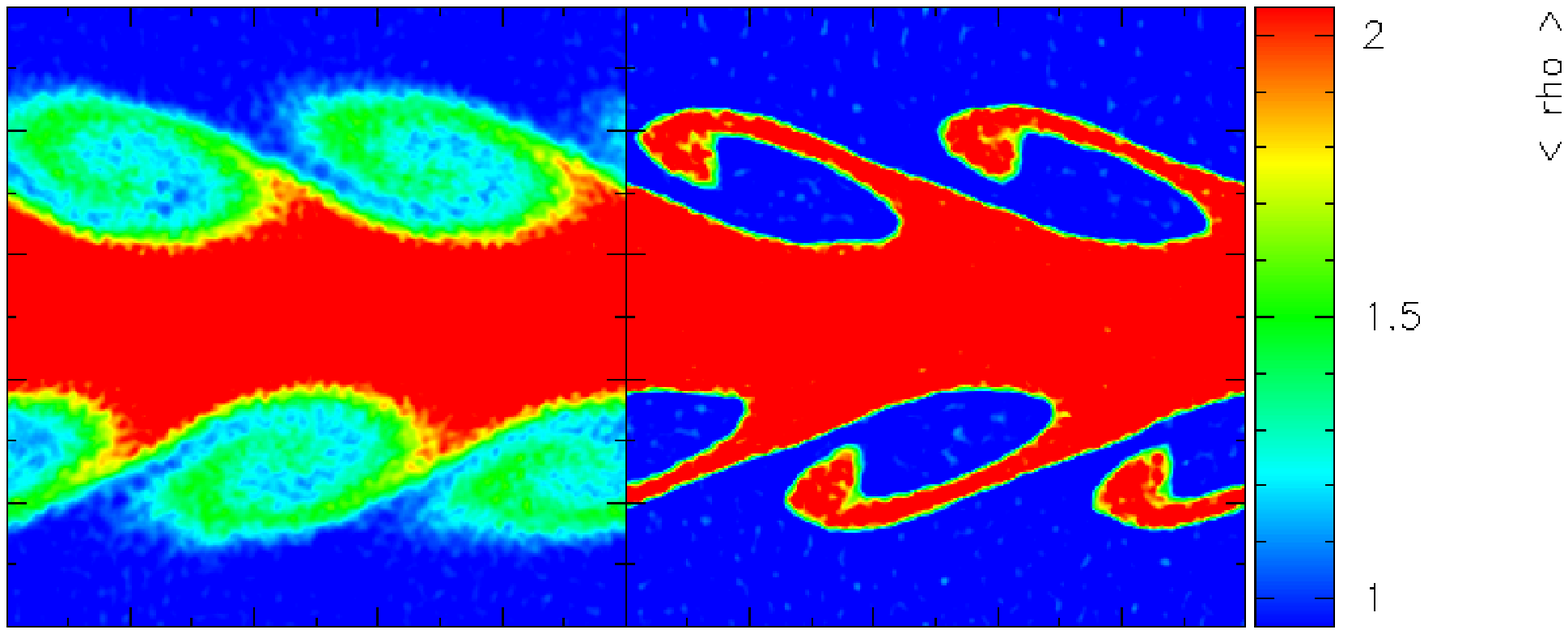}
\caption{Kelvin-Helmoltz instability according to test K2, with (left)
and without (right) artificial thermal conductivity.
Top panels: initial pressure field. Bottom panels:
density fields at $t=1.79$. All quantities are in code units.} \label{kh4}
\end{figure*}

Figs. \ref{kh1} and \ref{kh2} show the temporal evolution of the
instability in the two cases. Clearly, our code is able to reproduce
the expected trend of whirpools and mixing between the fluids; the
evolution is faster for a stronger initial perturbation. For
comparison, we also run a test with the K2 initial conditions, but
switching off artificial conduction; Fig. \ref{kh4} shows the
comparison between the final density fields (at $t=1.79$), and
between the initial pressure fields. In the latter, the
\textit{blip} that inhibits  mixing is clearly visible at the
discontinuity layer: the resulting spurious surface tension keeps
the two fluids separated, a problem efficiently cured by the
artificial conduction term.

\subsection{Strong shock collapse}

The strong shock collapse \citep{Noh1987} is another very demanding
test, and is generally considered a serious challenge for AMR
numerical codes. We run the test in 2-D, displacing $\sim 125,000$
particles on a regular grid and retailing those with $r \leq 2$ to
model a gas disk, with initially uniform density $\rho = 1$, and
negligible internal energy; we then add a uniform radial velocity
towards the origin, $v_{in} = -1$. A spherical shock wave with
formally infinite Mach number develops from the centre and travels
outwards, leaving a constant density region inside the shock (in
2-D, $\rho_{int}=16$). The test was run on 4 parallel CPUs.

\begin{figure}[!htbp]
\centering
\includegraphics[width=8.5cm,height=6.5cm]{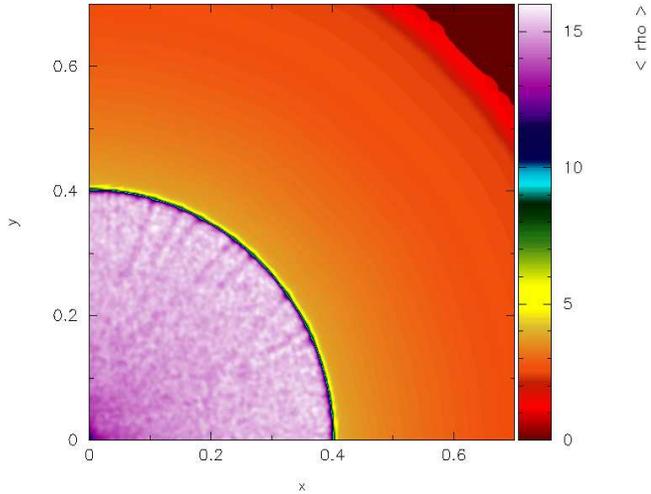}
\caption{The Noh's problem: density field in the first quadrant
at $t=1.2$. All quantities are in code units. See text for details.} \label{noh1}
\end{figure}

\begin{figure}[!htbp]
\centering
\includegraphics[width=7cm,height=6cm]{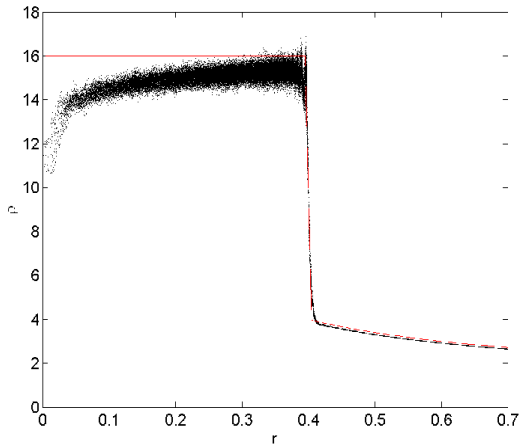}
\caption{The Noh's problem: density vs. radius relation for all particles at $t=1.2$.
Solid line: theoretical expectation. All quantities are in code units.
See text for details.} \label{noh2}
\end{figure}

In Fig. \ref{noh1}, the density in the $x>0$, $y>0$ region is
shown at $t = 1.2$,             
while in Fig. \ref{noh2} the density-radius relation at the same
time is plotted for all particles. There is good agreement with the
theoretical expectation, although a strong scatter is present in
particles density within the shocked region, where moreover the
density is slightly underestimated \citep[anyway, this is a common
feature in this kind of test: see e.g. ][]{Springel2009}.

Apart from these minor flaws, the code is able to handle the very
strong shock without any particular difficulty.

\subsection{Gresho vortex problem}

The test for the conservation of angular momentum and vorticity, the
so-called Gresho triangular vortex problem \citep{Gresho1990}, in
its 2-D version \citep{Liska2003}, follows the evolution of a vortex
initially described by an azimuthal velocity profile

\begin{eqnarray}
&& v_{\phi}(r) =
\begin{cases}
5r & \text{if $0 \leq r < 0.2$,} \\
2-5r & \text{if $0.2 \leq u < 0.4$,} \\
0 & \text{if $r \geq 0.4$,}
\end{cases}
\end{eqnarray}

\noindent in a gas of constant density $\rho = 1$.
The centrifugal force is balanced by the pressure gradient given
by an appropriate pressure profile,

\begin{eqnarray}
&& P(r) = \\
&& \begin{cases}
5 + 12.5 r^2 & \text{if $0 \leq r < 0.2$,} \\
9 + 12.5 r^2 - 20r + 4 \ln(5r) & \text{if $0.2 \leq u < 0.4$,} \\
3 + 4 \ln(2) & \text{if $r \geq 0.4$,}
\end{cases} \nonumber
\end{eqnarray}

\noindent so that the vortex should remain independent of time.

As noted by \citet{Springel2009}, this test seems to be a serious
obstacle for SPH codes. Due to shear forces, the angular momentum
tends to be transferred from the inner to the outer regions of the
vortex, eventually causing the rotational motion to spread within
the $r>0.4$ regions; these finally dissipate their energy because of
the interactions with the counter-rotating regions of the periodic
replicas. With the \textsc{Gadget3} code, the vortex did not survive
up to  $t=3$. AMR codes generally give a better description,
although under some conditions (for example, the addition of a
global bulk velocity, which should leave the system unchanged due to
its galileian invariance) they also show some flaws.

\begin{figure}[!htbp]
\centering
\includegraphics[width=8cm,height=7cm]{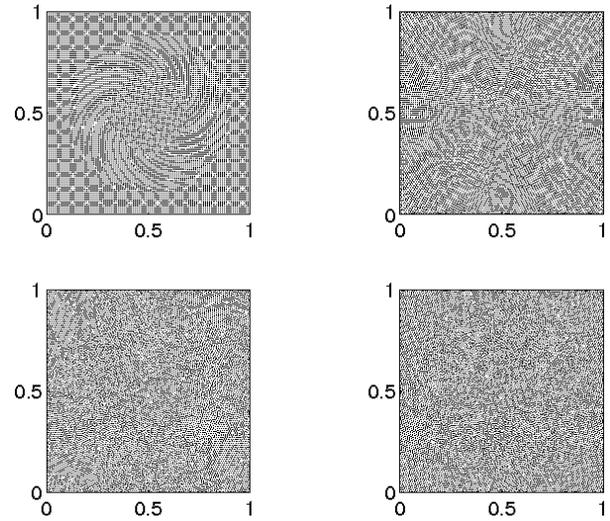}
\caption{Positions of all particles at $t=0.02$ (top) and $t=2.7$
(bottom) for the two different initial configurations for the Gresho
test: grid (left) and rings (right). All quantities are in code units.
See text for details.}
\label{gresho1}
\end{figure}

We tried to run the test starting from two different initial
settings of particles, since as pointed out by \citet{Monaghan2006}
the initial particle setting may have some influence on the
subsequent evolution of the system. Thus, in the first case we set
up a regular cartesian lattice of $\sim 10,000$ particles (in the
periodic domain $-0.5<x<0.5, -0.5<y<0.5$), and the second one  put
the same number of particles on concentric rings, allowing for a
more ordered description of the rotational features of the gas (see
Fig. \ref{gresho1}).

\begin{figure}[!htbp]
\centering
\includegraphics[width=8cm,height=6cm]{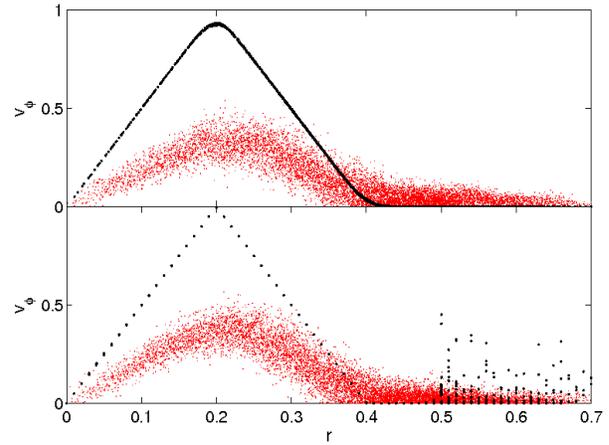}
\caption{Initial (black dots) and final (red small points)
azimuthal velocity profiles for the two Gresho tests: grid (top) and
rings (bottom). All quantities are in code units. See the text for details.} \label{gresho2}
\end{figure}

\begin{figure}[!htbp]
\centering
\includegraphics[width=8.5cm,height=4cm]{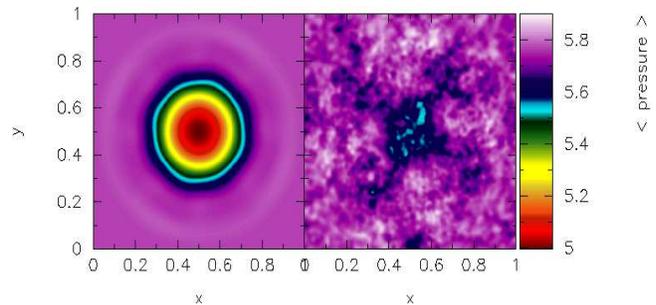}
\caption{Initial ($t=0.02$, left) and final ($t=2.7$, right) pressure fields
in the grid Gresho test. Note the increased pressure in the
outskirts (white) and in the central regions of the vortex. All quantities are in code units.
See the text for details.} \label{greshopress}
\end{figure}

The results are shown in Fig. \ref{gresho2}. We find a residual
vorticity still present at $t \sim 3$, although strongly degraded
and reduced in magnitude. As expected, a substantial amount of  the
angular momentum has been passed to the outer regions of the volume
under consideration. This is confirmed by comparing the initial to
the final pressure fields (Fig. \ref{greshopress}): at $t=2.7$ the
pressure exterted by the outskirts is higher, due to the increased
temperature of the regions confining with the periodic replicas.
This leads to a compression of the internal region, which in turn
must increase its pressure as well.

However, we could not find in literature other direct comparisons
for this test with an SPH code. It must be pointed out that the
resolution in these tests is quite low, and a larger number of
particles may help improving the results.

\subsection{Point explosion}

The presence of the artificial conduction becomes crucial in the
point explosion experiment, a classic test to check the ability of a
code to follow the evolution of strong shocks in three dimensions.
In this test, a spherically symmetric Sedov-Taylor blast-wave
develops and must be modelled consistently with the known analytical
solution.

We run the test in 3-D. Our initial setting of the SPH particles is
a uniform grid, in a box of size $[-0.5, 0.5]$ along each dimension,
with uniform density $\rho = 1$ and negligible initial internal
energy. The grid is made of $31^3$ particles. At $t=0$ an additional
amount $E=1$ of thermal energy is given to the central particle; the
system is subsequently let free to evolve self-consistently (without
considering self-gravity). It must be noted that, again, the central
additional thermal energy is \textit{not} smoothed, so to obtain
extreme conditions.

The following test runs are made (using the Lagrangian SPH scheme):
\begin{itemize}
\item S1 - constant viscosity with $\alpha=1$ viscosity, no artificial conduction;
\item S2 - variable viscosity with $\alpha$ plus artificial conduction;
\item S3 - as S2, but without the $\nabla h$ correcting terms;
\item S4 - as S2, with the individual time-stepping scheme;
\item S5 - as S2, with higher resolution ($51^3$ particles).
\end{itemize}

\begin{figure}[!htbp]
\centering
\includegraphics[width=6cm,height=6cm]{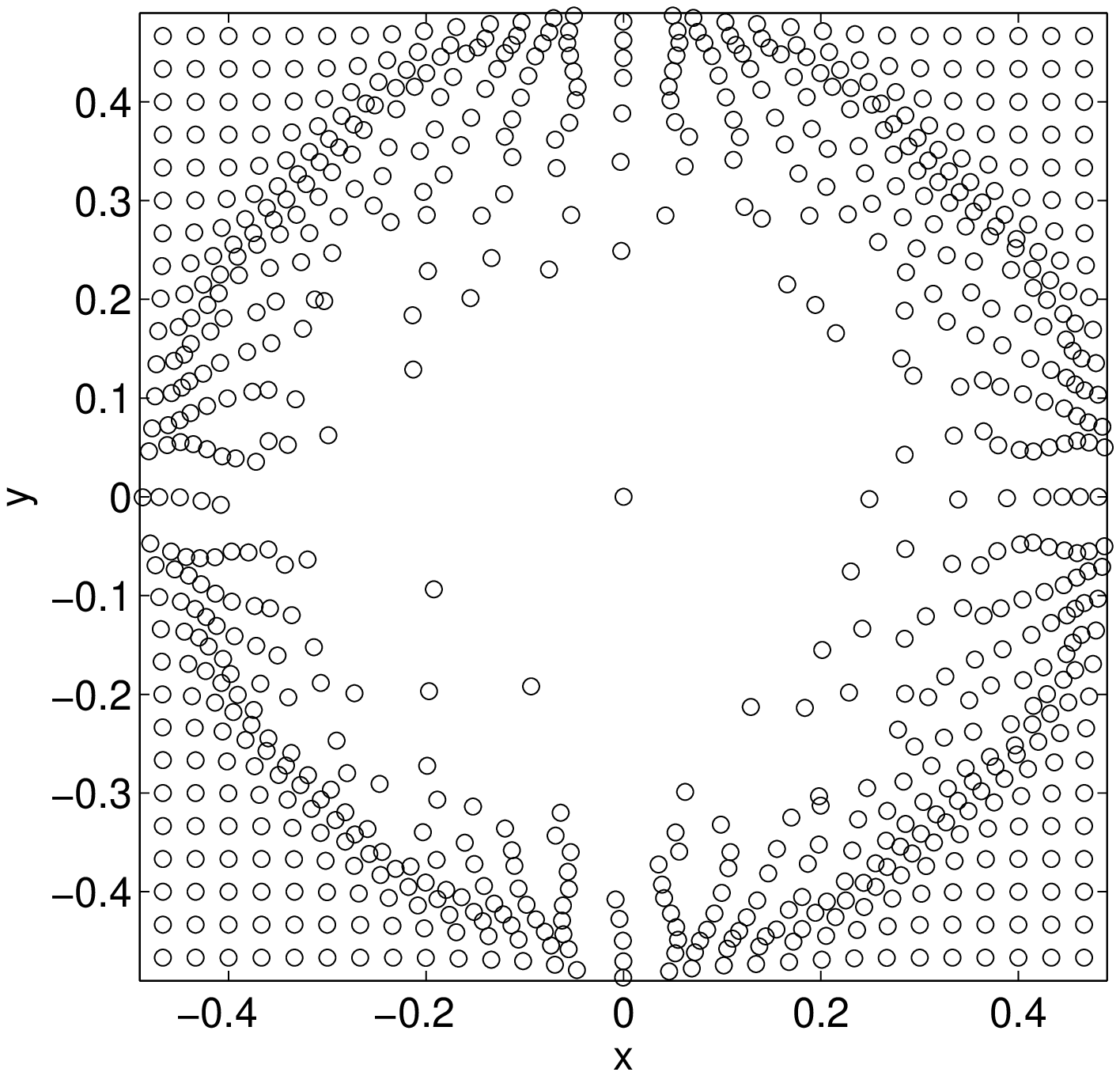}
\includegraphics[width=8cm,height=6cm]{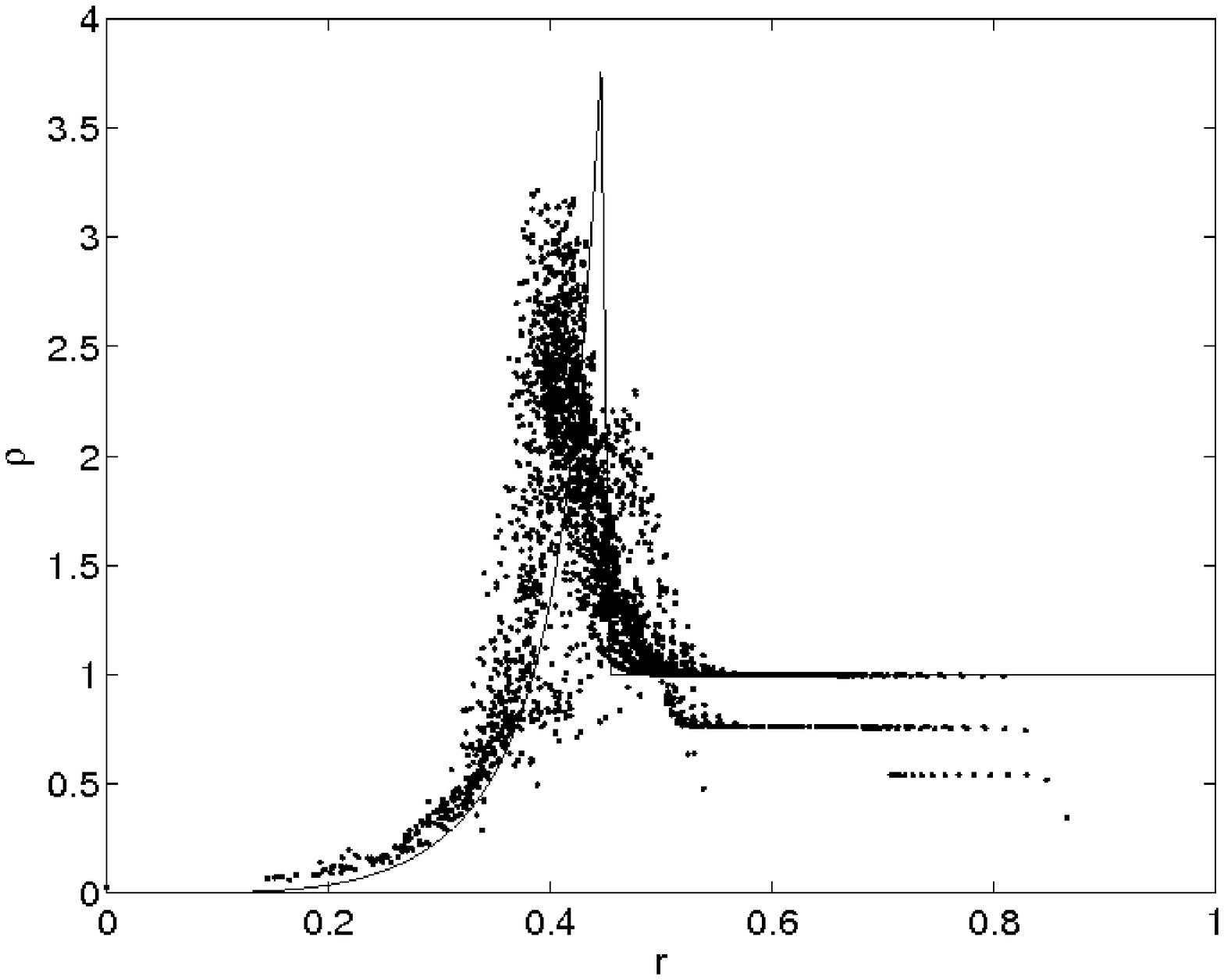}
\caption{Top: spatial distribution  of particles in a slice taken at z=0 and t=0.09 for
the Sedov blast wave of test S1. Bottom: Radial density profile for the same test.
Points: SPH particles; solid line: analytical prediction. All quantities are in code units.}  \label{sedov1}
\end{figure}

\begin{figure}[!htbp]
\centering
\includegraphics[width=6cm,height=6cm]{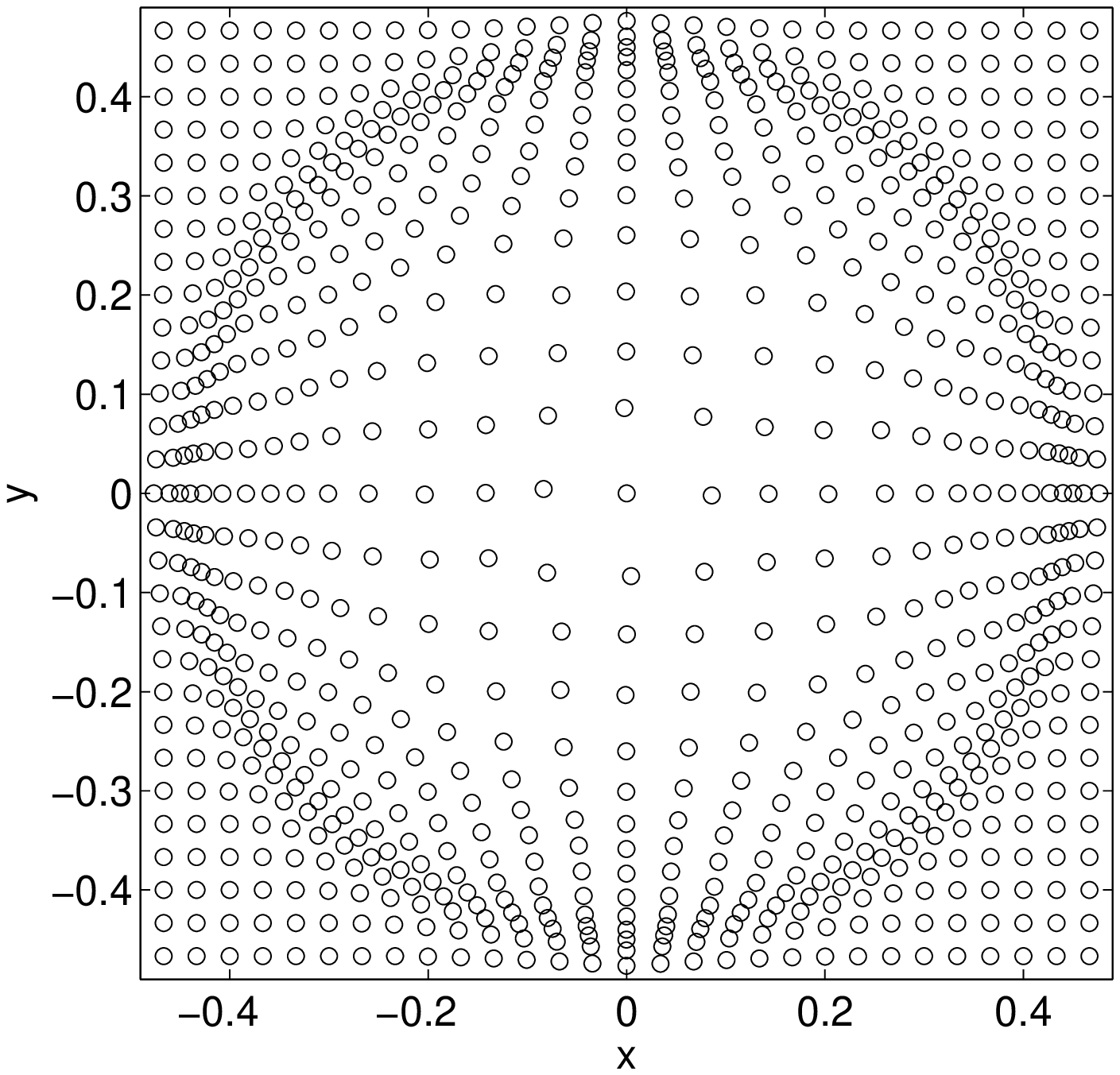}
\includegraphics[width=8cm,height=6cm]{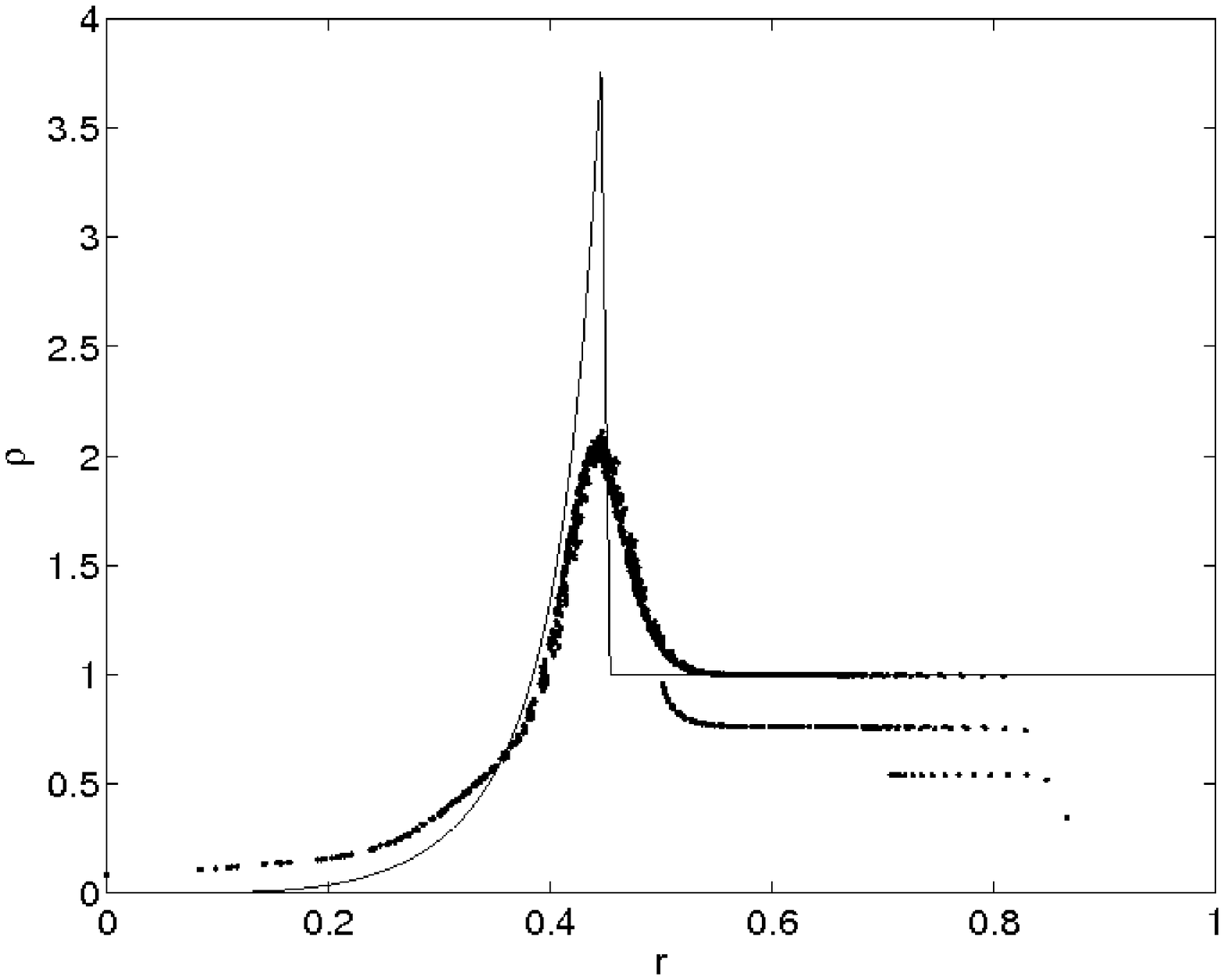}
\caption{The same as in Fig. \ref{sedov1}, but for the S2 test.} \label{sedov2}
\end{figure}

\begin{figure}[!htbp]
\centering
\includegraphics[width=7cm,height=5cm]{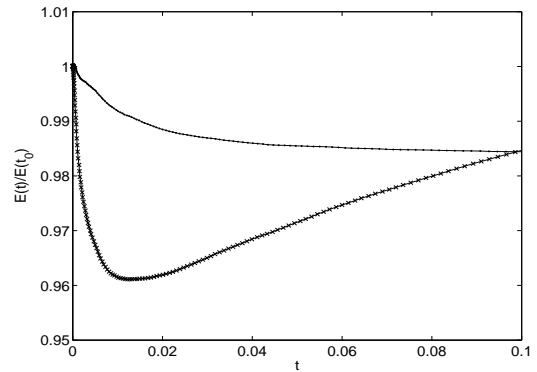}
\caption{Energy conservations for Sedov test with artificial
conductivity and variable viscosity. Points: with $\nabla h$ terms;
crosses: without $\nabla h$ terms. All quantities are in code units.} \label{sedovene}
\end{figure}

\begin{figure}[!htbp]
\centering
\includegraphics[width=6cm,height=6cm]{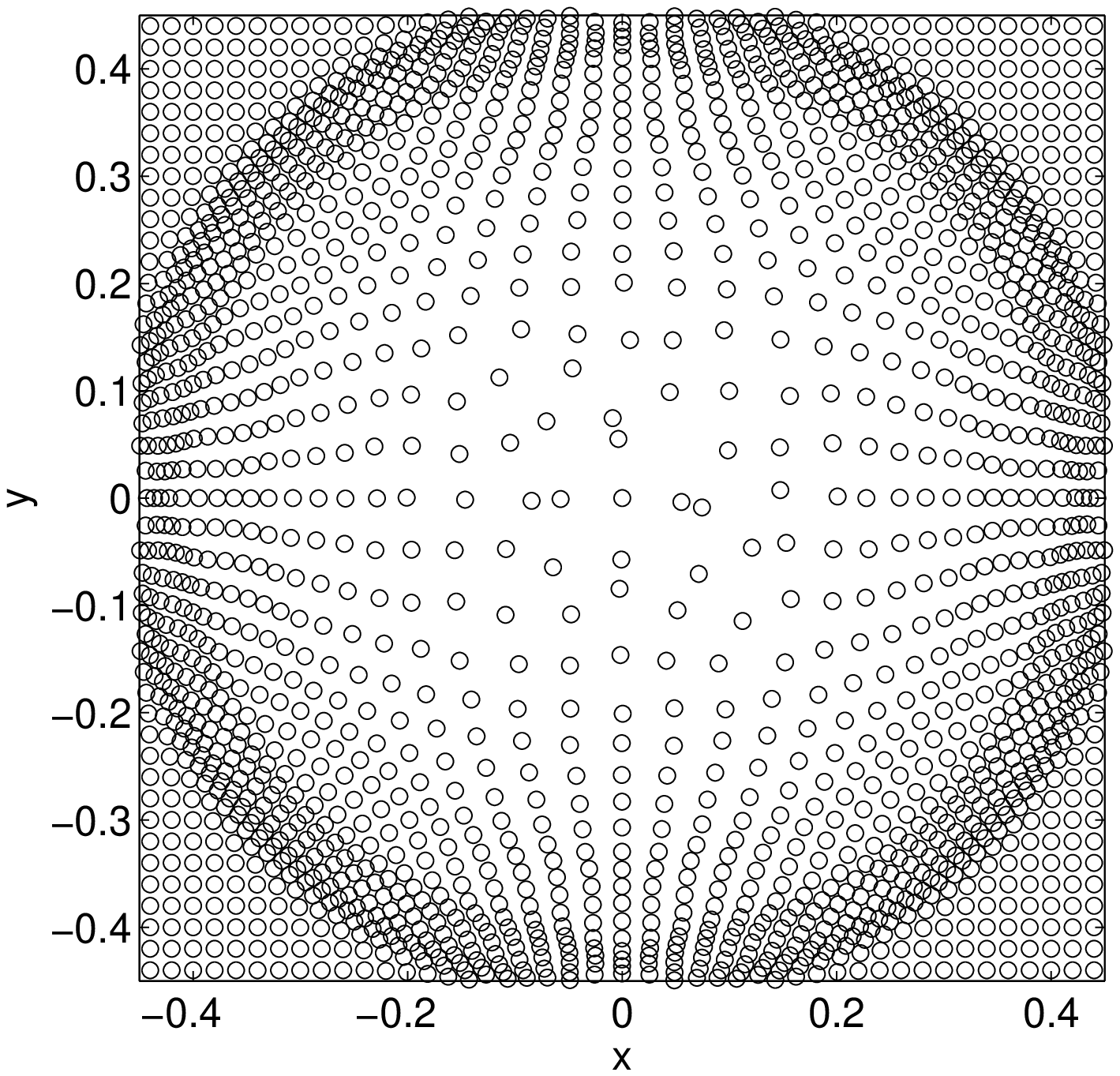}
\includegraphics[width=8cm,height=6cm]{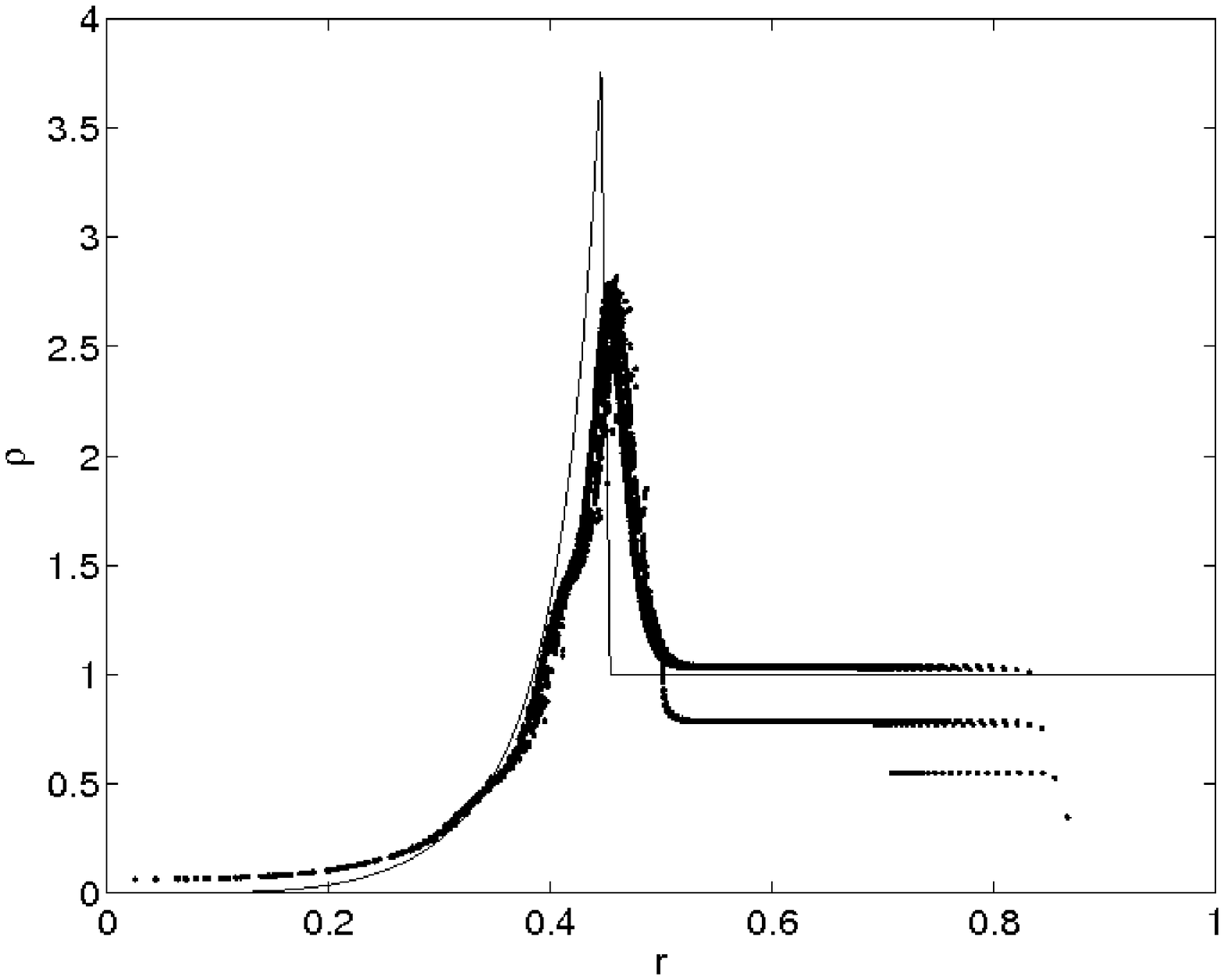}
\caption{The same as in Fig. \ref{sedov1}, but for the test S5.} \label{sedov5}
\end{figure}

Fig. \ref{sedov1} shows the $xy$ positions  of particles belonging
to the $z=0$ plane at $t=0.09$ for the S1 test, and the
corresponding radial density profile, compared to the analytical
expectation. A strong noise (which develops soon after the initial
explosion) is evident in the particle positions, due to the huge
unresolved discontinuity in thermal energy. The description of the
shock evolution  is quite poor, although its gross features are
still captured.

Fig. \ref{sedov2} presents the same data for the S2 run, in which
both the viscosity  with variable $\alpha$ and artificial conduction
are switched on. The increased precision in the description of the
problem is immediately evident: the conductive  term acts smoothing
out the thermal energy budget of the central particle, strongly
reducing the noise and allowing a symmetric and ordered dynamical
evolution. The radial density profile (in which single particles,
and not averages, are plotted) shows that the shock is well
described. The density peak is a factor of $\sim 2$ lower than the
analytical expectation, due to the intrinsic smoothing nature of the
SPH technique.

Another test (S3) is carried dropping the $\nabla h$ correcting
terms (see Sect. \ref{description_SPH}). While the differences in particle
positions and density profiles are almost negligible, the comparison
of the total energy conservations $E(t)/E(t_0)$ in the two cases
(Fig. \ref{sedovene}) shows that the correction helps in containing
losses due to wrong determinations of the gradient.

A fourth test (S4) is also  run to check the behaviour of the
individual time-stepping algorithm; the results are virtually
indiscernible from the those  of the previous case (not shown here).
This is  noteworthy as all particles must be ``waken up'' (using the
method described in Sect. \ref{inddt}), being still and cold at the
beginning of the simulation.

Finally, a test (S5) with increased numerical resolution is
performed (Fig. \ref{sedov5}). As expected, in this case, while the
particles positions again accurately describe the evolution of the
system, the shock boundary is sharper, and closer to the peak value
of the analytical solution.

\subsection{Self-gravitating collapse}

The collapse of a gaseous self-gravitating sphere is another
standard SPH 3-D test \citep{Evrard1988}. The combined action of
hydrodynamics and self-gravity leads the system, a sphere of gas
initially far from equilibrium, with negligible internal energy and
density profile $\rho \propto r^{-1}$, to a collapse in which most
of its kinetic energy is converted into heat. An outward moving
shock develops, a slow expansion follows and, at late times, a
core-halo structure develops with nearly isothermal inner regions
and the outer regions cooling adiabatically. To set the initial
conditions, 10,000 particles of mass $1.25 \times 10^{-2} M_{\odot}$
are placed in a sphere of radius $5\times10^{-6}$ Mpc, in such a way that
their initial density radial profile scales as  $1/r$. Another
sphere with  40,000 particles and consequently smaller particles'
masses (higher resolution) is also prepared. Two cases are examined:
namely the collisionless collapse and the adiabatic collapse.

%
%

\begin{figure}[!htbp]
\centering
\includegraphics[width=8.5cm,height=10cm]{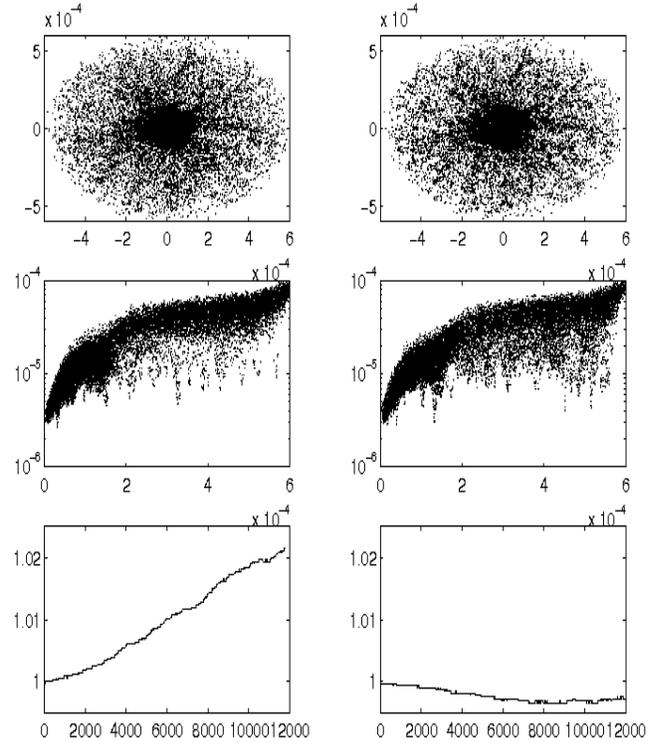}
\caption{Collisionless collapse with adaptive softening lengths.
Top: spatial distribution of particles projected onto the $xy$ plane
after 12,000 steps; middle: softening lengths (code units) as a
function of the radial coordinate; bottom: temporal conservation of
the total energies of the systems $E(t)/E(t_0)$, for the
un-corrected (left) and corrected (right) tests.} \label{adeps}
\end{figure}

\subsubsection{Collisionless collapse}

To check the performance of the adaptive softening length algorithm,
tests are run switching off hydrodynamical interactions to  mimic
the collisionless collapse. A first case  drops the hydrodynamical
interactions, includes the adaptive softening algorithm but neglects
the correcting terms presented in Section \ref{soft}.  It is meant
to provide the reference case. The second one includes also the
effects of these correcting terms.

Fig. \ref{adeps} shows, in the upper panels, the spatial
distribution of particles projected onto the $xy$ plane after during
the collapse of the sphere, for the un-corrected (left) and
corrected (right) cases: no evident difference is present. The same
holds for (middle panels) the value of the softening lengths of
particles as a function of the radial coordinate. However, the
temporal conservation of the total energies of the systems
$E(t)/E(t_0)$ in the two cases are quite different (bottom panels of
Fig. \ref{adeps}). If the correcting terms are not included a
secular error in the energy conservation develops, reaching $2\%$,
while the error is $\sim 10$ times smaller (albeit not completely
eliminated) with the introduction of the extra-terms.


\subsubsection{Adiabatic collapse}


The standard hydrodynamical tests \citep{Evrard1988} are calculated with the following settings:

\begin{itemize}
\item E1 - constant softening lengths ($\epsilon_i = 0.1 R \times (m_i / M)^{0.2}$,
with $R$ and $M$ total radius and mass of the sphere at the
beginning of the simulation); Lagrangian SPH scheme, with constant
viscosity ($\alpha=1$)  and no artificial conduction;
\item E2 - as E1, except for using adaptive softening lengths;
\item E3 - adaptive softening lengths, viscosity with variable $\alpha$, and
artificial conduction;
\item E4 - as E3, higher resolution (40,000 particles).
\end{itemize}

\begin{figure}[!htbp]
\centering
\includegraphics[width=8.5cm,height=6.5cm]{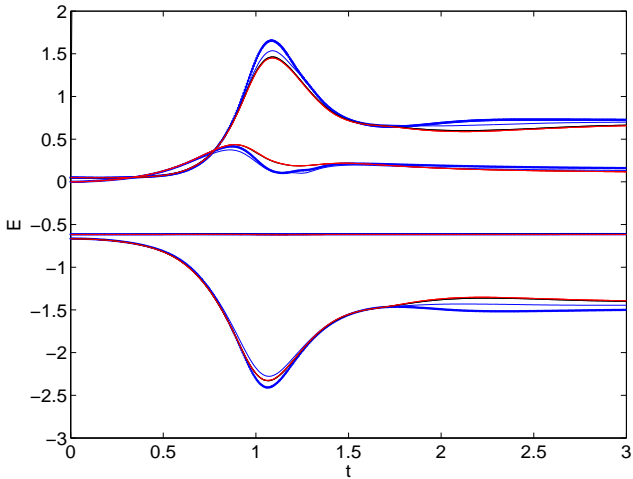}
\caption{Energy trends in the cold collapse test. Light lines: red: test E1; black:
test E2; blue: test E3. Heavy dotted blue line: test E4 (see text for details).
Top to bottom: thermal, kinetic, total and potential energies. All quantities are in code units.} \label{evrard}
\end{figure}

\begin{figure}[!htbp]
\centering
\includegraphics[width=8.5cm,height=6.5cm]{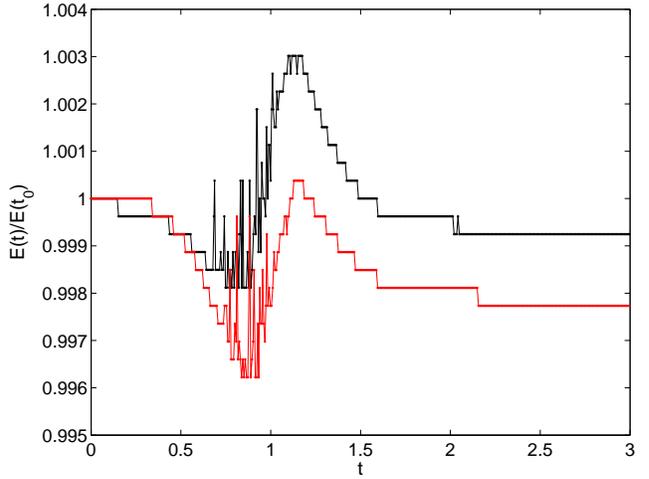}
\caption{Magnification of the energy conservations $E(t)/E(t_0)$ in the cold collapse
tests. Black: testE1; red: testE2. See text for details.}
\label{evrardene}
\end{figure}

\begin{figure}[!htbp]
\centering
\includegraphics[width=7.5cm,height=6.5cm]{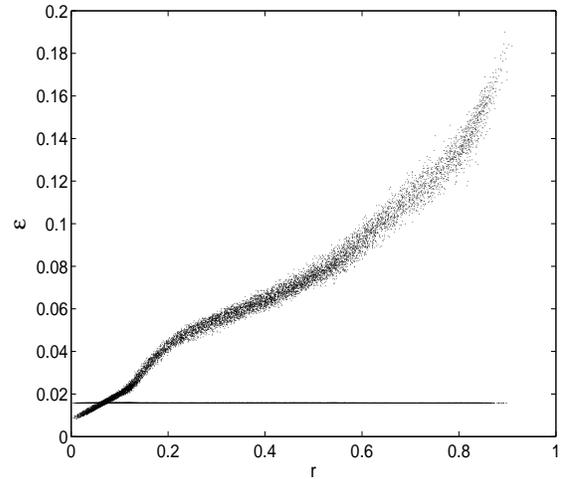}
\caption{Softening lengths in the cold collapse  test T2 at $t=0.8$. The solid line is
the value of the constant softening length in test T1. All quantities are in code units.}
\label{evrardeps}
\end{figure}

The global energy trends are shown in Fig. \ref{evrard}, while Fig.
\ref{evrardene} shows a magnification of the energy conservation
$E(t)/E(t_0)$ for the E1 and E2 runs.
The energy conservation is always good, reaching a maximum error of
$\sim 0.4\%$ at the moment of maximum compression. For comparison,
in the same test the finite differences method by \citet{Evrard1988}
ensures conservation at $\simeq 1\%$, while \citet{Harfst2006} finds
a maximum error of $\sim 3\%$ with a standard SPH implementation.
Almost no difference can be noted between the variable and the
constant $\epsilon$ runs, despite a variation of more than two
orders of magnitude in the softening length values in the E2 run
(see Fig. \ref{evrardeps}).

\begin{figure}[!htbp]
\centering
\includegraphics[width=8.5cm,height=6.5cm]{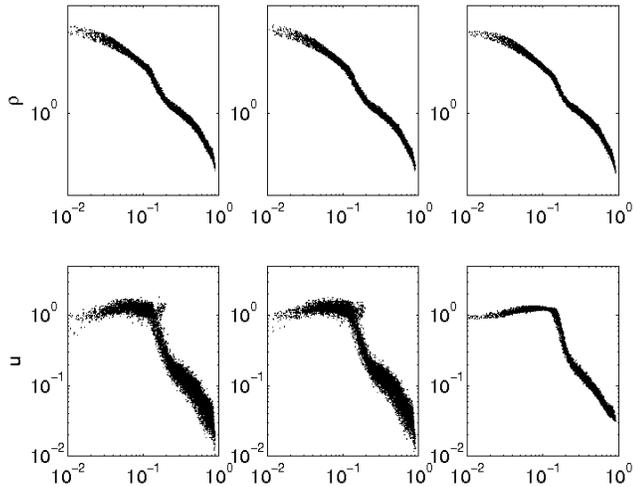}
\caption{Density (top) and internal energy (bottom) profiles at
$t=0.8$ in the cold collapse E1 (left), E2 (center) and E3 (right)
tests. All particles are plotted.} \label{evrardsnap}
\end{figure}

\begin{figure}[!htbp]
\centering
\includegraphics[width=8.5cm,height=6.5cm]{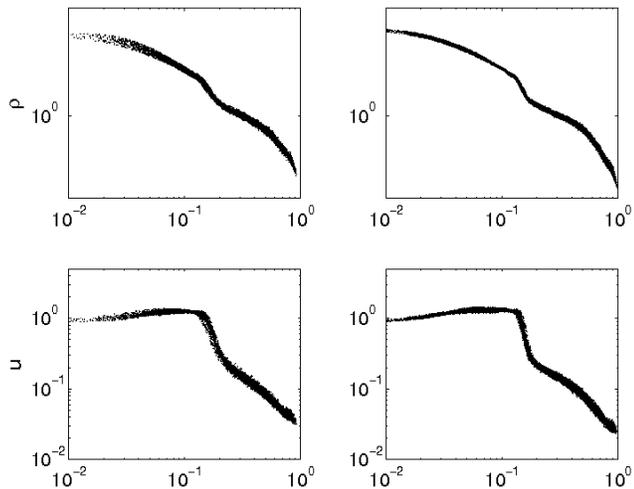}
\caption{Comparison between cold collapse E3 (low-res, left) and E4
(hi-res, right) test: density (up) and internal energy (bottom)
profiles at $t=0.8$.} \label{evrardsnap2}
\end{figure}

The density and internal energy profiles at $t=0.8$, for the E1 to
E3 runs, are shown in Fig. \ref{evrardsnap}; the comparison between
the E3 and the hi-res E4 runs is shown in Fig. \ref{evrardsnap2}
(the shock is moving outward leaving an inner, heated core). The
discontinuity in density is smoothed out by the kernel smoothing,
which in the current implementation requires a high number of
neighbors ($\sim 60$), but the effects of the inclusion of the
artificial conductive term are evident. We didn't find any particular
problem with the inclusion of the conductive term in a self-gravitating
system.


We also run a X-SPH test, adopting $\eta = 0.25$. While the energy
trends are essentially identical to the ones the previous tests (not
shown), the energy conservation is slightly worse, reaching an error
of $\sim 2.5\%$. Indeed, \citet{Monaghan2002} pointed out that a
more complex treatment of the equations of motion should be adopted
to achieve perfect energy conservation.

\subsection{Isothermal collapse} \label{bosstest}

Introduced by \citet{Boss1979} and later revisited by
\citet{Burkert1993}, the test on the isothermal collapse  couples
hydrodynamics and gravity, also adding rotation. It develops in a
complex game of collapse and fragmentation. We run the test in
the version proposed by \citet{Bate1997}.

A homogeneous sphere of cold gas with initial radius
$R=5\times10^{16}$ cm and mass $M = 1 M_{\odot}$ is
modelled retailing $\sim 78,000$ particles out of a uniform lattice.
The sphere is put in solid body rotation around the $z$ axis, with
angular velocity $\Omega=7.2\times 10^{-13}$ rad s$^{-1}$, and the
otherwise flat density field is perturbed by $10 \%$ in such a way
that

\begin{eqnarray}
\rho(\phi) = \rho_0 [1 + 0.1 cos(2 \phi)], \nonumber
\end{eqnarray}

\noindent where $\phi$ is the azimuthal angle about the rotation
axis and $\rho_0 = 3.82 \times 10^{-18}$ g cm$^{-3}$ (we achieve the
density perturbation by varying the masses of particles instead of
their positions). The gas is described by an isothermal equation of
state, $P = c_s^2 \rho$, with initial sound speed $c_s = 1.66 \times
10^4$ cm s$^{-1}$. It is assumed to be optically thin so that no
mechanical heating can occur, and the adiabatic index is $\gamma =
1.4$.

The test is very challenging and has been used to check the
behaviour of both AMR and SPH codes. In particular, \citet{Bate1997}
used it to study the effects of  wrong calibrations of the smoothing
and softening lengths in SPH (see Sect. \ref{187}). We run the test
in four different cases:
\begin{itemize}
\item B1 - Lagrangian SPH with constant softening lengths
($\epsilon \simeq 5.26 \times 10^{14}$ cm, an intentionally large
value);
\item B2 - Lagrangian SPH with adaptive softening lengths (no limits
imposed on $\epsilon$ and $h$);
\item B3 - as B2, but imposing a minimum smoothing
\textit{and} softening length $h_{min}=\epsilon_{min}=10^{14}$ cm,
to avoid artificial clumping \citep[as in ][]{Bate1997};
\item B4 - as B2, with the X-SPH method (adopting $\eta = 0.25$).
\end{itemize}

\begin{figure}[!htbp]
\centering
\includegraphics[width=8.5cm,height=6.5cm]{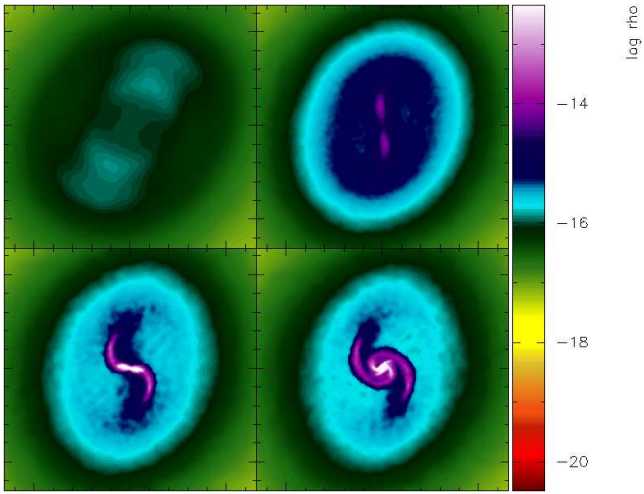}
\caption{Evolution of the density field in the central region for
the test B1. Plotted is the region within $1.54 \times 10^{16}$ cm
from the origin, on the $xy$ plane. Left to right, top to bottom:
$t=1.0$, $1.15$, $1.23$, $1.26$, in units of free fall time, $t_{ff}
=  1.0774 \times 10^{12}$ s.} \label{boss1}
\end{figure}

\begin{figure}[!htbp]
\centering
\includegraphics[width=8.5cm,height=6.5cm]{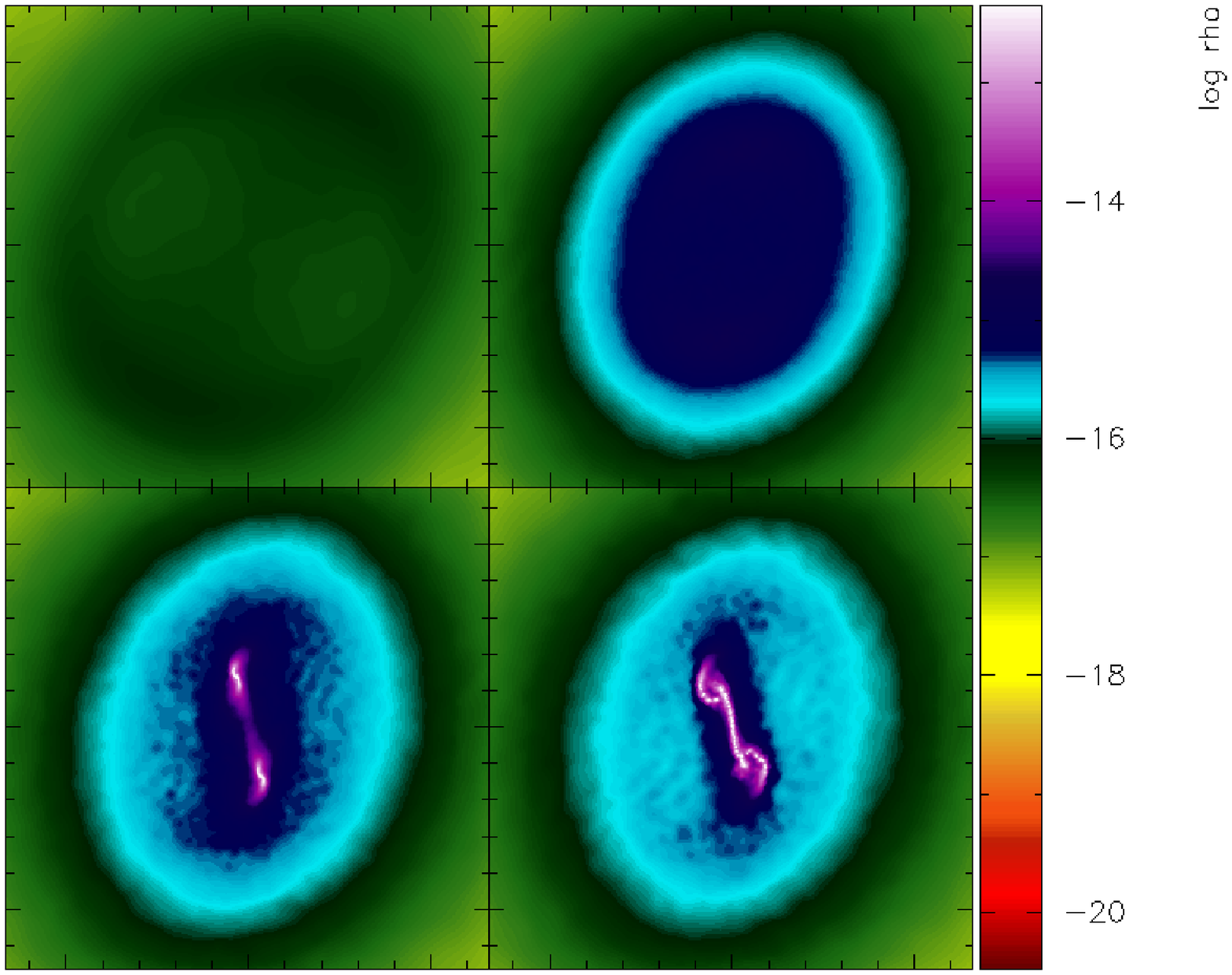}
\caption{The same as in Fig. \ref{boss1}, but for test B2.}
\label{boss2}
\end{figure}

\begin{figure}[!htbp]
\centering
\includegraphics[width=8.5cm,height=6.5cm]{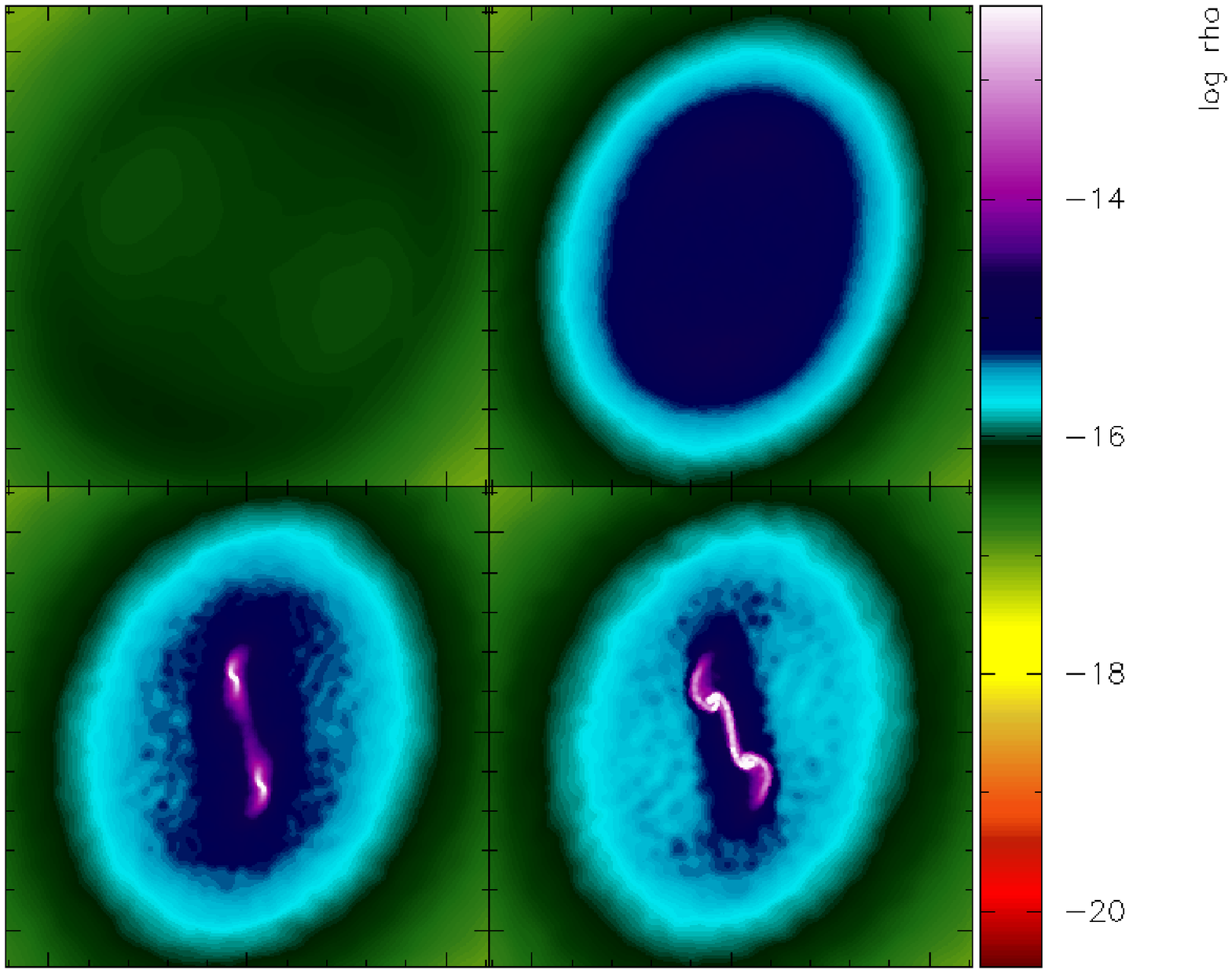}
\caption{The same as in Fig. \ref{boss1}, but for test B3.}
\label{boss3}
\end{figure}

\begin{figure}[!htbp]
\centering
\includegraphics[width=8.5cm,height=6.5cm]{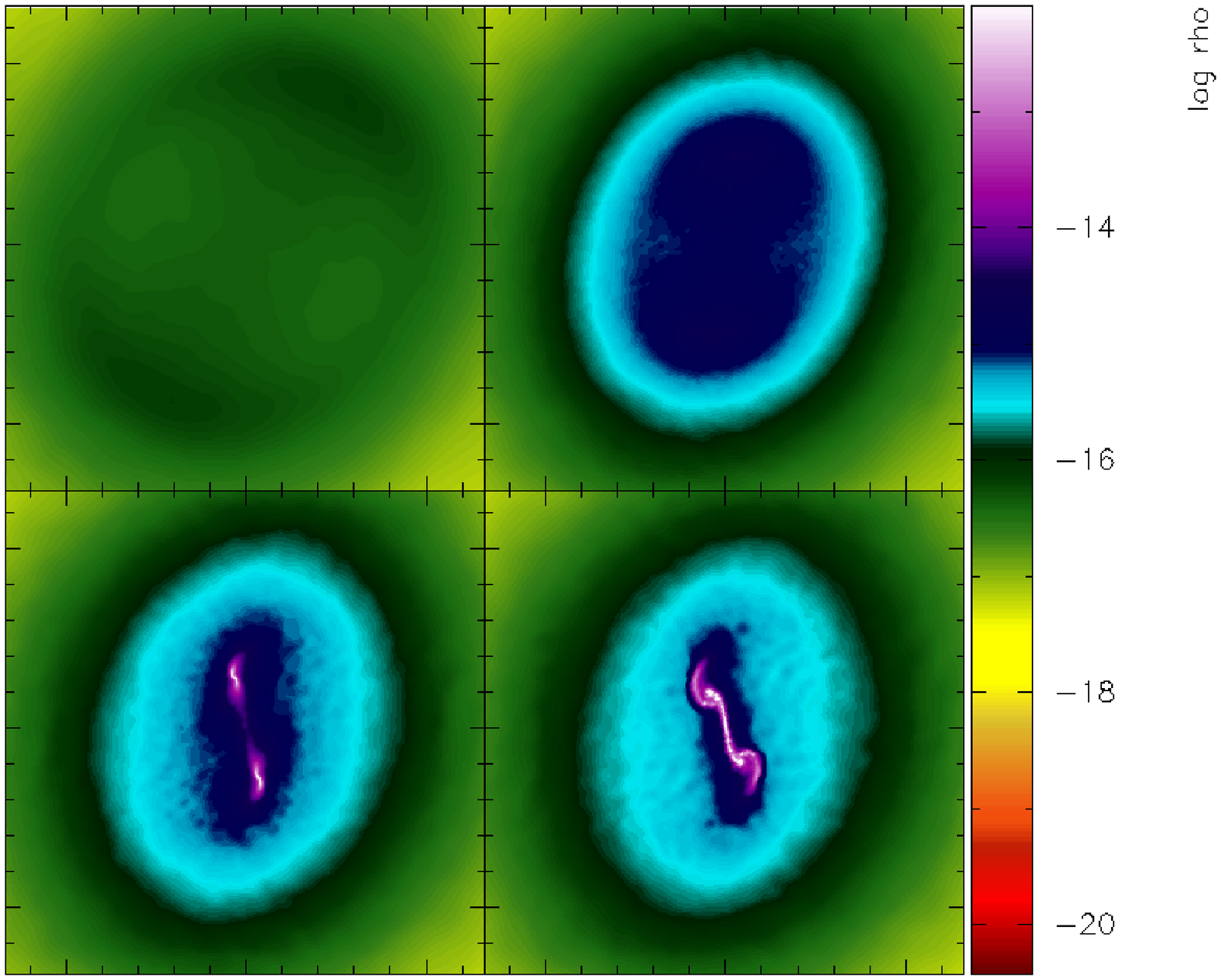}
\caption{The same as in Fig. \ref{boss1}, but for test B4.}
\label{boss4}
\end{figure}

\begin{figure}[!htbp]
\centering
\includegraphics[width=8.5cm,height=6.5cm]{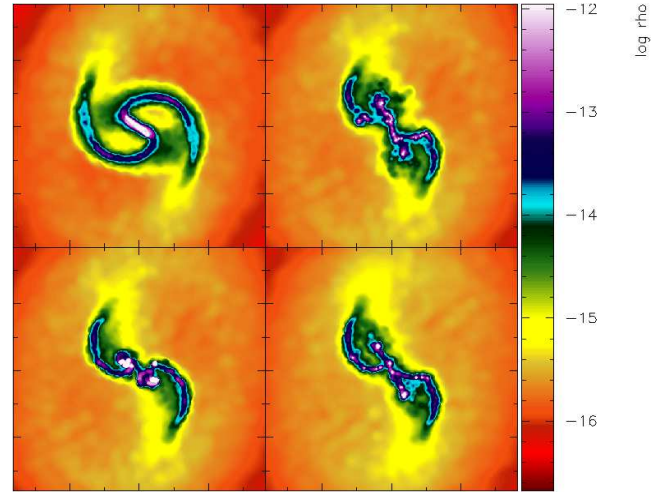}
\caption{Final density field at $t=1.29$ (in units of free fall
time) in the central region. Left to right, top to bottom: tests B1,
B2, B3, B4. See text for details.} \label{bossf}
\end{figure}

Figs. \ref{boss1} to \ref{boss4} plot the temporal evolution of the
density field in the $z=0$ plane; shown are snapshots at $t=1.0$,
$1.15$, $1.23$, and $1.26$, in units of free fall time, $t_{ff} =
1.0774 \times 10^{12}$ s. Fig. \ref{bossf} plots the final density
field at $t = 1.29$ for the four tests. These plots can be compared
to those by \citet{Bate1997} and \citet{Springel2005}.

The large value of $\epsilon$ assumed in the B1 test clearly
demonstrates how a wrong choice of the softening length can result
in catastrophic errors. The evolution of the system is clearly
different from the grid solution used by \citet{Bate1997} as a proxy
of the exact solution of the problem. Because of the excessive
softening, clumps may be  stable against collapse and a rotating
structure forms, resembling a barred spiral galaxy. More realistic
choices for $\epsilon$ may give better results, but the opposite
problem (artificial clumping of unresolved structures) may arise in
case of too small a softening length. Indeed, the case B2, in which
the adaptive algorithm is adopted (note that $\epsilon=h$), gives
much better results, but  disorderly clumped sub-structures may
form.

Imposing a minimum $\epsilon$ (=$h$, case B3) cures the problem (note
however that the evolution of the system seems to be somewhat faster than 
the grid solution). Anyway, this prescription cannot be generalised. A more
physically sounded solution may be the introduction of a
\textit{pressure floor} in unresolved regions, i.e. where the Jeans
mass of the gas is not resolved by the SPH technique. This issue
will be discussed in the forthcoming companion paper (Merlin et al.
2009, in preparation).

Finally, the X-SPH test B4 gives results that are essentially
undiscernible from those of case B2, apart from a slightly less
disordered density field in the regions surrounding the central
collapsed structure.

\begin{figure}[!htbp]
\centering
\includegraphics[width=7cm,height=6.5cm]{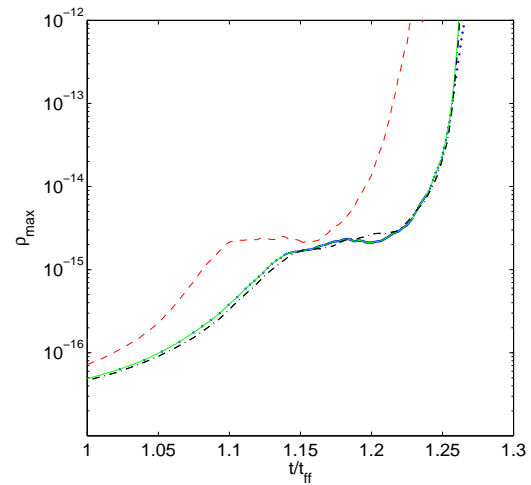}
\caption{Maximum density in the isothermal collapse tests. B1: red
dashed line; B2: green solid line; B3: blue thick dots; B4: black
dot-dashed line. See text for details.} \label{bossrho}
\end{figure}

Fig. \ref{bossrho} plots the temporal evolution of the maximum value
of the density. This plot can be compared with Fig. 5a in
\citet{Bate1997} and Fig. 12 in \citet{Springel2005}. The results
for all tests are very similar and agree quite well with their SPH
tests of comparable resolution (with the plateau at $\rho \simeq 2
\times 10^{-15}$ g cm$^{-3}$ reached at $t = 1.14$), except for the
case B1  which clearly evolves faster.


\begin{figure}[!htbp]
\centering
\includegraphics[width=8.5cm,height=6.5cm]{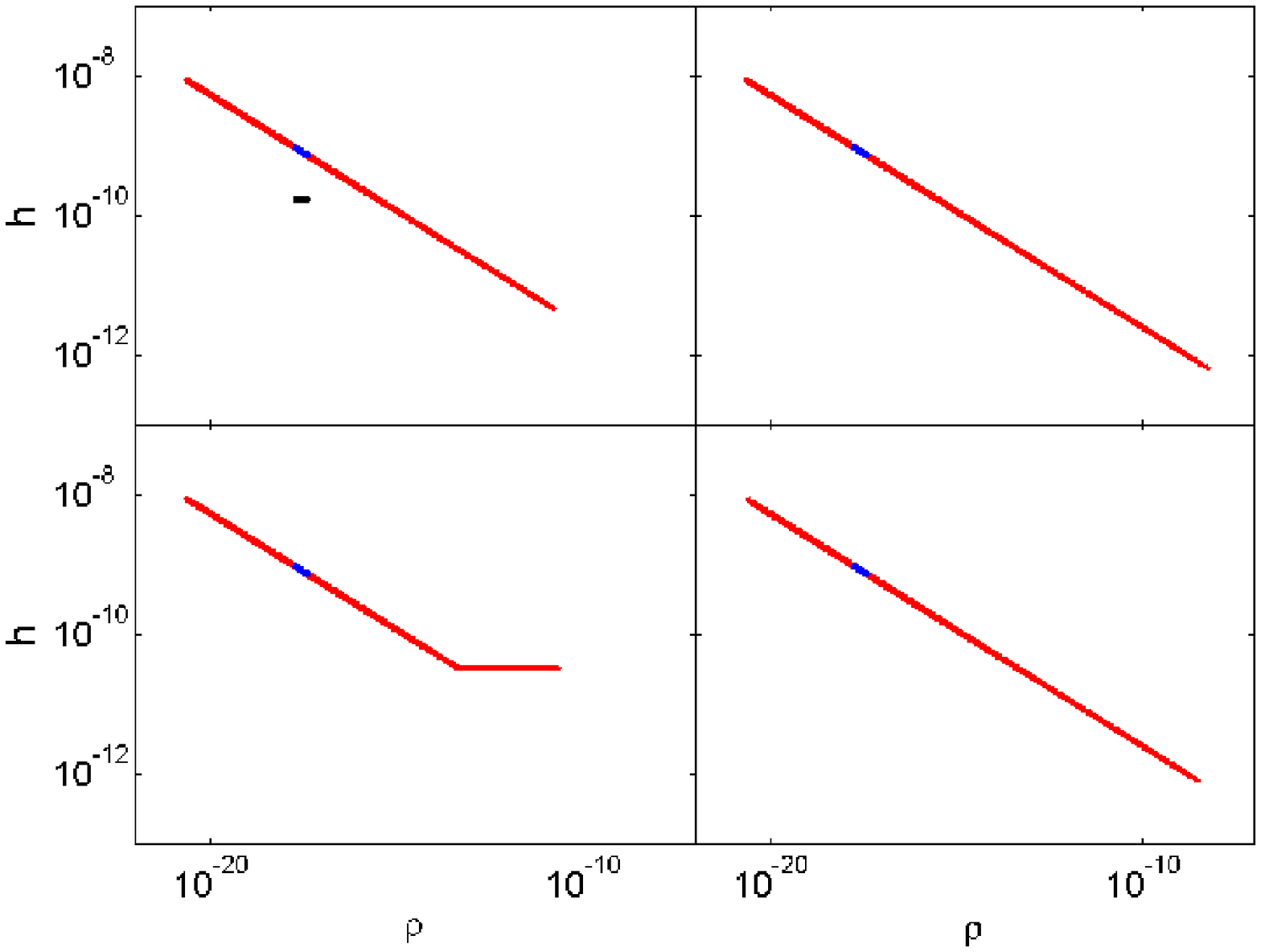}
\caption{Initial (blue) and final (red) smoothing length $h$ of all
particles in the four isothermal collapse tests (left to right, top
to bottom: B1, B2, B3, B4). The softening length $\epsilon$ is equal
to $h$ in the B2, B3 and B4 runs; the black dots in the top left
panel show its constant value in the test B1. The abrupt truncation
in the bottom left panel is due to the imposed $h_{min}$
($=\epsilon_{min}$) in the test B3.} \label{bosseps}
\end{figure}

For reference, Fig. \ref{bosseps} shows the initial and final values
of both $\epsilon$ and $h$ (which are equal in the cases B2, B3 and
B4) as a function of $\rho$ in the four runs.

\subsubsection{Parallel run}

\begin{figure}[!htbp]
\centering
\includegraphics[width=8.5cm,height=3.8cm]{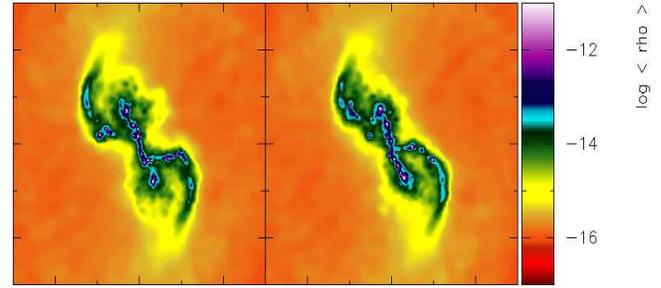}
\caption{Comparison between the final density fields for
a single CPU (left) and a 16 parallel CPUs (right) runs of
the B2 test. See the text for details.}
\label{boss16p1}
\end{figure}

\begin{figure}[!htbp]
\centering
\includegraphics[width=8.5cm,height=4cm]{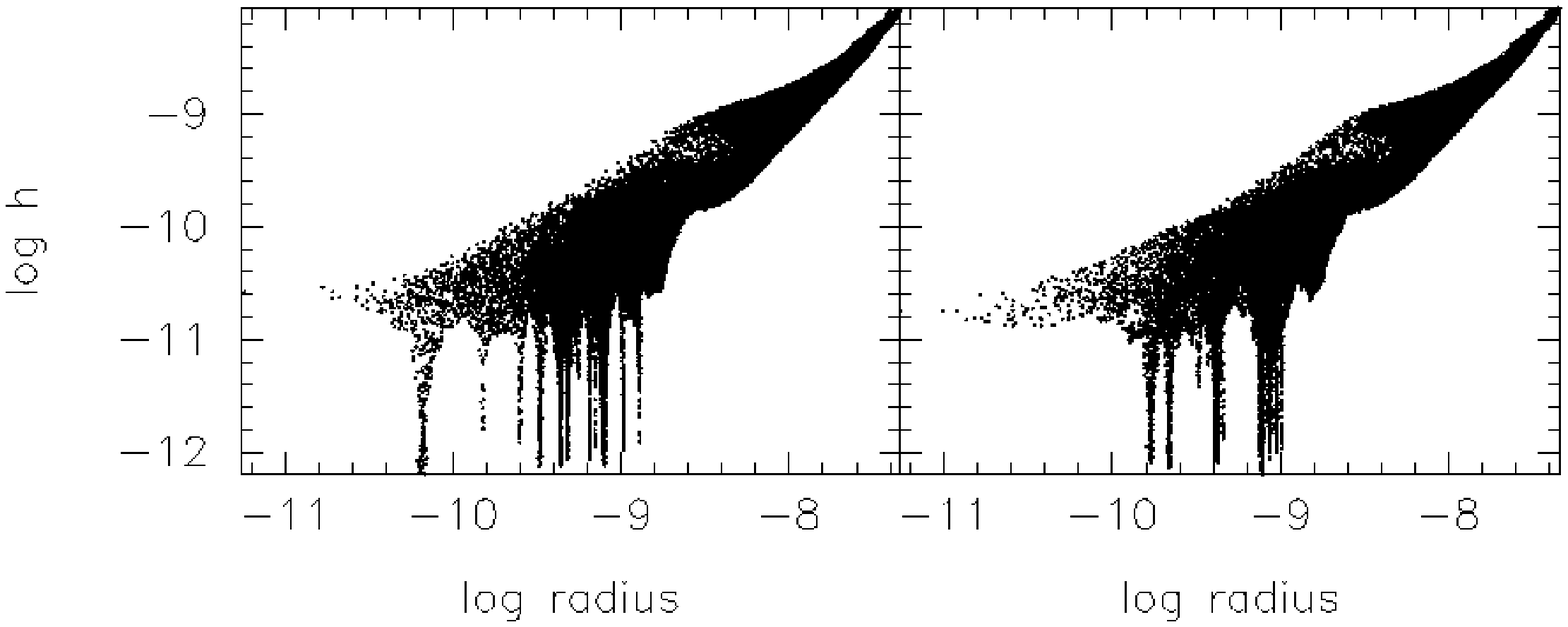}
\caption{Comparison between the final $h$-radius relation,
for a single CPU (left) and a 16 parallel CPUs (right) runs
of the B2 test. See the text for details.}
\label{boss16p2}\end{figure}

Finally, we re-computed the case B2 test  using 16 parallel CPUs.
Fig. \ref{boss16p1} shows the comparison between the final density
field for this run and for the original one with single CPU.
Although the agreement is good, and no sign of a multi-processor
subdivision of the domain is present, some differences in local
morphological features are present; these are more evident looking
at Fig. \ref{boss16p2}, where the comparison is made for the
smoothing lengths values. While the gross features and the minimum
and maximum values are very similar, differences in the local
clumping of particles are clearly present.

We ascribe this flaw to the parallel Oct-Tree scheme. The global
tree constructed by $N>1$ CPUs is slightly different from the one
built by a single processor on the same spatial domain: the regions
in which two or more different CPUs interact are described by a
different cell architecture depending on $N$ (note that the reason
for this is that the cells have to be always cubic; indeed, the
problem should not occur if a binary tree, in which cells have
adaptive shapes, was used). This causes a slightly different
approximation of the gravitational force, due to the opening cell
criterion, giving in turn a slightly different dynamical evolution.

\begin{figure}[!htbp]
\centering
\includegraphics[width=8.5cm,height=3.8cm]{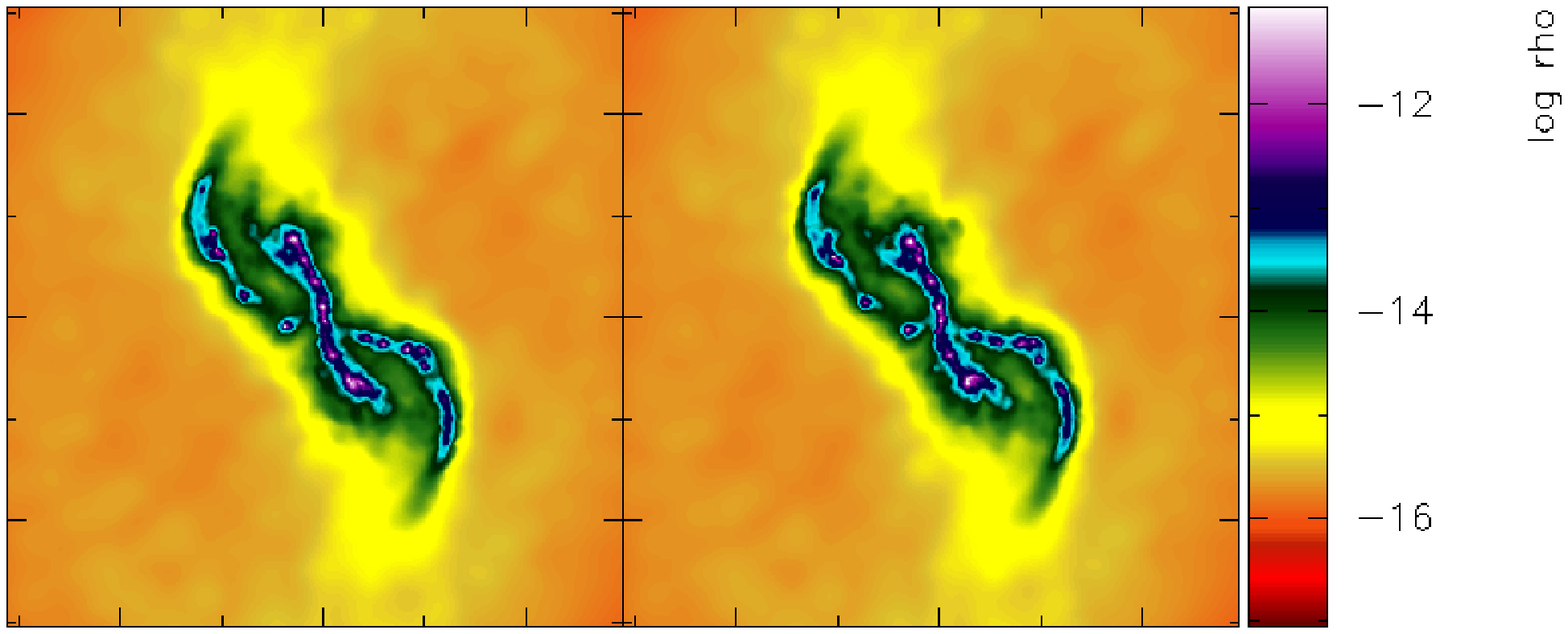}
\caption{Comparison between the final density fields for a
single CPU (left) and a 16 parallel CPUs (right) runs of
the B2 test, in a run with tree opening angle $\theta = 0.1$.
See the text for details.}
\label{boss16p3}
\end{figure}

\begin{figure}[!htbp]
\centering
\includegraphics[width=8.5cm,height=4cm]{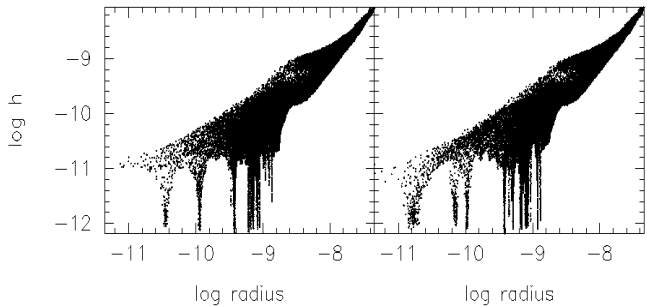}
\caption{Comparison between the final $h$-radius relation,
for a single CPU (left) and a 16 parallel CPUs (right) runs of
the B2 test, in a run with tree opening angle $\theta = 0.1$.
See the text for details.}
\label{boss16p4}\end{figure}

To prove the validity of this speculation, we run again the same
tests reducing the opening angle $\theta$ to $0.1$ instead of the
standard 0.8. In this way, the approximation given by the adoption
of the tree structure should become of negligible importance (at
the expenses of a much longer computation time). Indeed, looking
at Figs. \ref{boss16p3} and \ref{boss16p4}, we can notice an
improvement in that the density fields are more similar, although
still not exactly identical. An ideal run with $\theta = 0$
(i.e., exact particle-particle interaction) should give almost
identical distributions of particles.

We must point out that this flaw may be of non-negligible importance
when dealing with phenomena that directly depend on the local density
field, such as the star formation and feedback processes. The same
simulation run on a different number of CPUs may give slightly
different results because of this issue.



\subsection{Collision of two gas spheres}

The collision between two gaseous Plummer spheres is very crucial
test for the energy conservation. The tests are run in 3-D, with
adaptive softening lengths, variable $\alpha$ viscosity, and thermal
conduction.

Each of the two spheres (of unit mass, unit radius, and negligible
initial internal energy) are set up randomly selecting 10,000
particles in a way that their initial density radial profile
resembles the Plummer profile, $\rho \propto (1+r)^{-5/2}$. Their
centres are  on the $x$ axis.

In a first run (C1) we follow a head-on collision, assigning a
relative velocity of $1.5$ in the $x$ direction to the spheres. In a
second run (C2) we added a shear component, giving relative
velocities $|\Delta v_x|=0.15$ and $|\Delta v_y|=0.075$.

According to \citet{Hernquist1993} the violent head-on collision of
two polytropic stars does not conserve the energy within  $\sim 10
\%$. Modern formulations of SPH \citep[see e.g.][]{Rosswog2007} can
handle the problem with much more accuracy, reducing the error below
$0.1 \%$.

\begin{figure*}[!htbp]
\centering
\includegraphics[width=10cm,height=5cm]{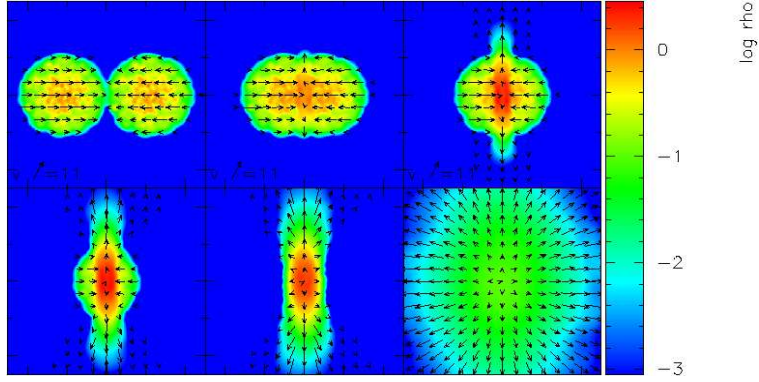}
\caption{Evolution of the density field in the $z=0$ plane for the
collision  test C1. Left to right, top to bottom: $t=0.05$, $0.15$,
$0.2$, $0.25$, $0.3$, $0.5$. Superposed is the velocity field.}
\label{coll1}
\end{figure*}

\begin{figure}[!htbp]
\centering
\includegraphics[width=8.5cm,height=6.5cm]{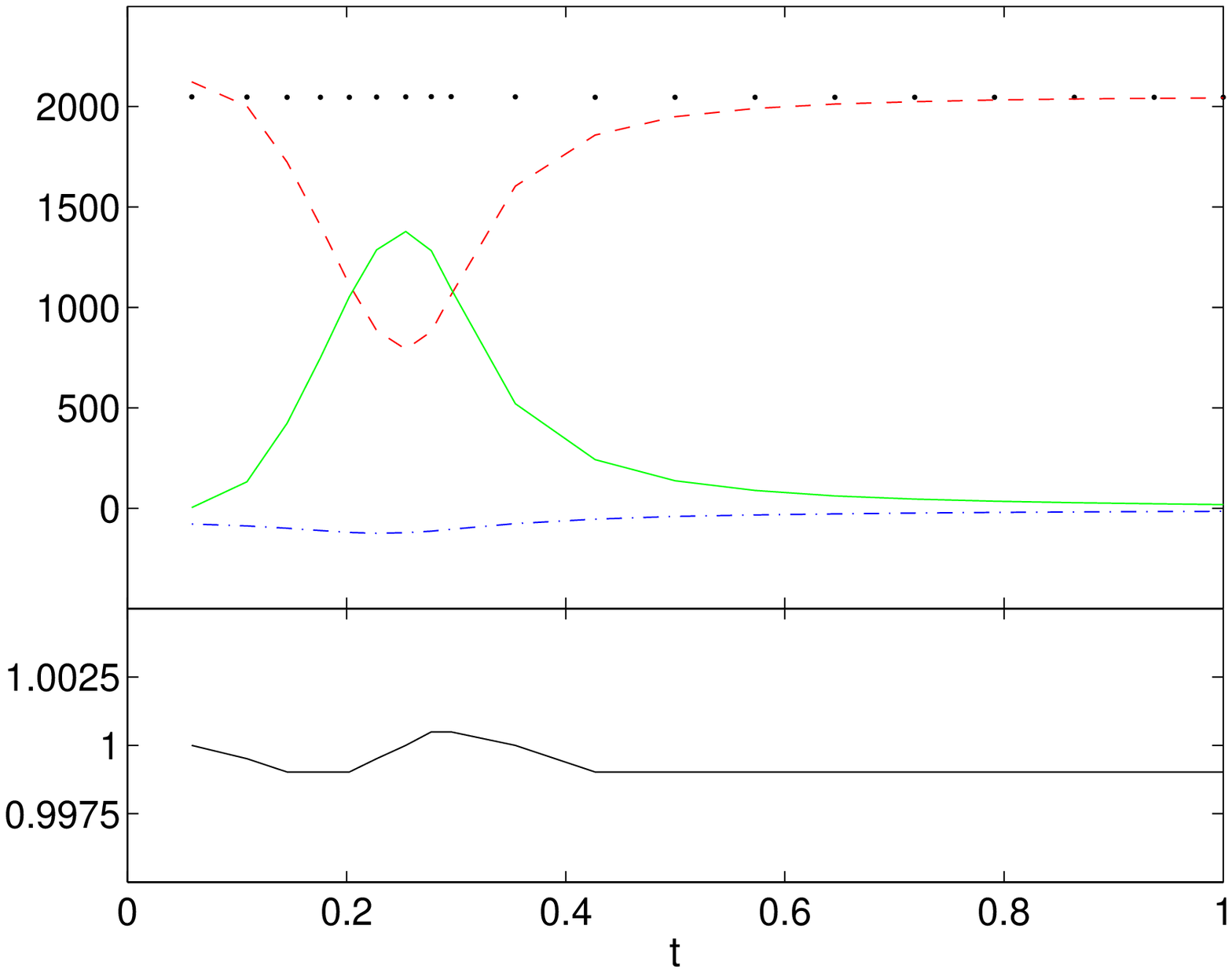}
\caption{Top: energy trends in the collision  test C1; black dots:
total energy, blue dot-dashed line: potential, red dashed line:
kinetic, green solid line: thermal. Bottom: magnification of the
total energy conservation $E(t)/E(t_0)$.} \label{coll1ene}
\end{figure}

\begin{figure*}[!htbp]
\centering
\includegraphics[width=10cm,height=8cm]{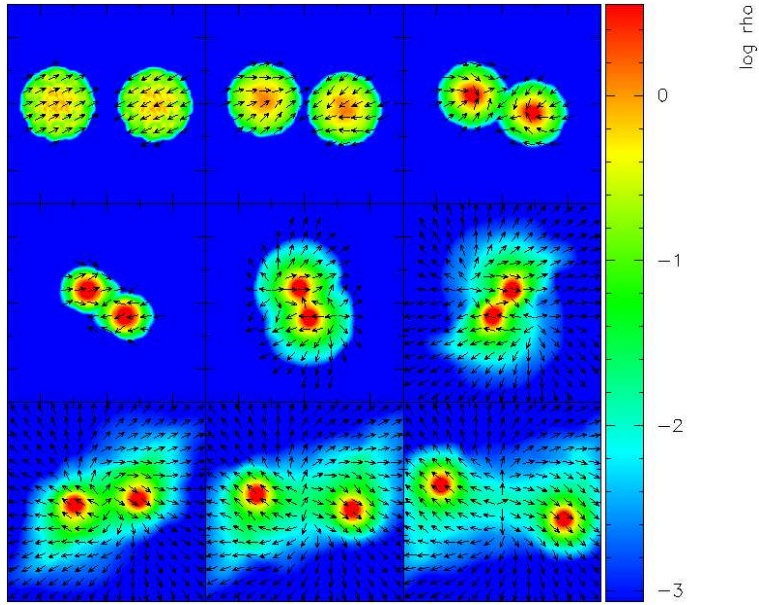}
\caption{Evolution of the density field in the $z=0$ plane for the
collision test C2. Left to right, top to bottom: $t=0.03$, $0.38$,
$0.77$, $1.16$, $1.55$, $1.94$, $2.72$, $3.5$, $4.29$. Superposed is
the velocity field.} \label{coll7}
\end{figure*}
\begin{figure}[!htbp]
\centering
\includegraphics[width=8.5cm,height=6.5cm]{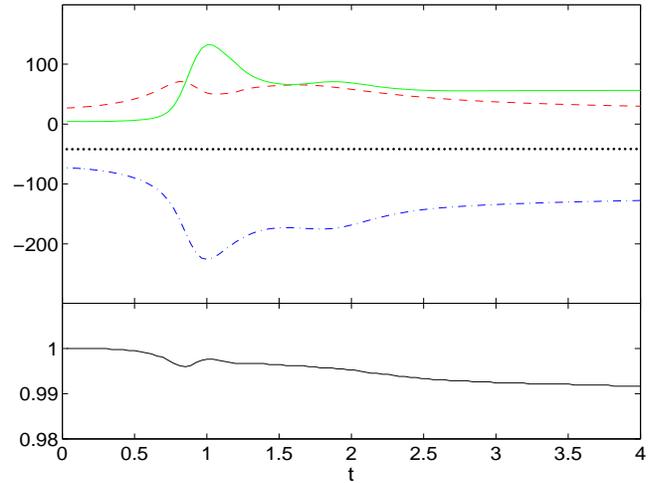}
\caption{Top: energy trends in the collision test C2; black dots:
total energy, blue dot-dashed line: potential, red dashed line:
kinetic, green solid line: thermal. Bottom: magnification of the
total energy conservation $E(t)/E(t_0)$.} \label{coll7ene}
\end{figure}

Figs. \ref{coll1} and \ref{coll7} plot the density and velocity
fields at different times, on a slice taken at $z=0$. In our tests,
the energy conservation is good (see Figs. \ref{coll1ene} and
\ref{coll7ene}), with a maximum error of $\sim 0.2 \%$ in the
head-on collision, and $\sim 1 \%$ error when the shear component is
added.

\subsection{Two-components evolution}

As a final test, we studied the evolution of a two-components fluid
in order to investigate the behaviour of the adaptive smoothing
and softening length algorithm in presence of more than one material.
We set up two spheres of the same radius, $R=10^3$ pc, and mass,
$M=4\times10^4 M_{\odot}$, using for each sphere the particle distribution
adopted adiabatic collapse test by Evrard. One of the two systems
was considered as made of collisionless bodies, with negligible pressure,
thus mimicking the behaviour of a Dark Matter halo; the other one
was instead considered as made of gas, with an initial temperature
of 1000 K. The system was the let free to evolve under the action
of self-gravity and hydrodynamical interactions, in four different
runs with the following settings:
\begin{itemize}
\item TC1 - constant softening lengths, $\epsilon_i = 0.1 R \times (m_i / M)^{0.2}$; null initial velocities
\item TC2 - adaptive softening lengths, null initial velocities
\item TC3 - as TC2, with an initial solid body rotation $\Omega=2.5\times10^{-17}$ rad s$^{-1}$, for both components
\item TC4 - as TC3, but with counter-rotating initial tangential velocities (same magnitude).
\end{itemize}

While the collisionless component soon collapses under self-gravity,
later reaching relaxation, the gas undergoes a phase of expansion
because of its internal pressure, but in the central region,
where it also interacts with collapsing Dark Matter, reaching very
high densities.

The free expansion of a system of particles is by itself a demanding
test, and a loss of total energy is a common flaw for Lagrangian codes.
Furthermore, the rotation imposed in the last two runs makes the
problem more complicated. In our tests, a further complication was
given by the dynamical interaction within the central regions of the
system, where the hydrodynamical pressure of the gas struggled against
self-gravity and the (stronger) gravitational action of the collisionless
component. Moreover, when the adaptive $\epsilon$ scheme was used,
gas particles had to cope with a large variation of their interaction
sphere, due to the presence of Dark Matter particles in the central
regions, and to their absence in the outskirts, where in addition
gas particles are far away from one another.


\begin{figure}[!htbp]
\centering
\includegraphics[width=8.5cm,height=6.5cm]{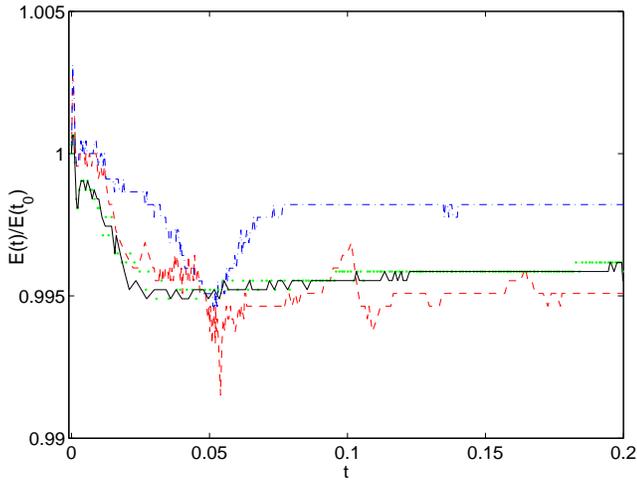}
\caption{Two components test, energy conservation $E(t)/E(t_0)$
for the four runs: dashed red line, TC1; blue dot-dashed line,
TC2; green dots, TC3; black solid line, TC4.}
\label{twocompene}
\end{figure}
\begin{figure}[!htbp]
\centering
\includegraphics[width=8.5cm,height=6cm]{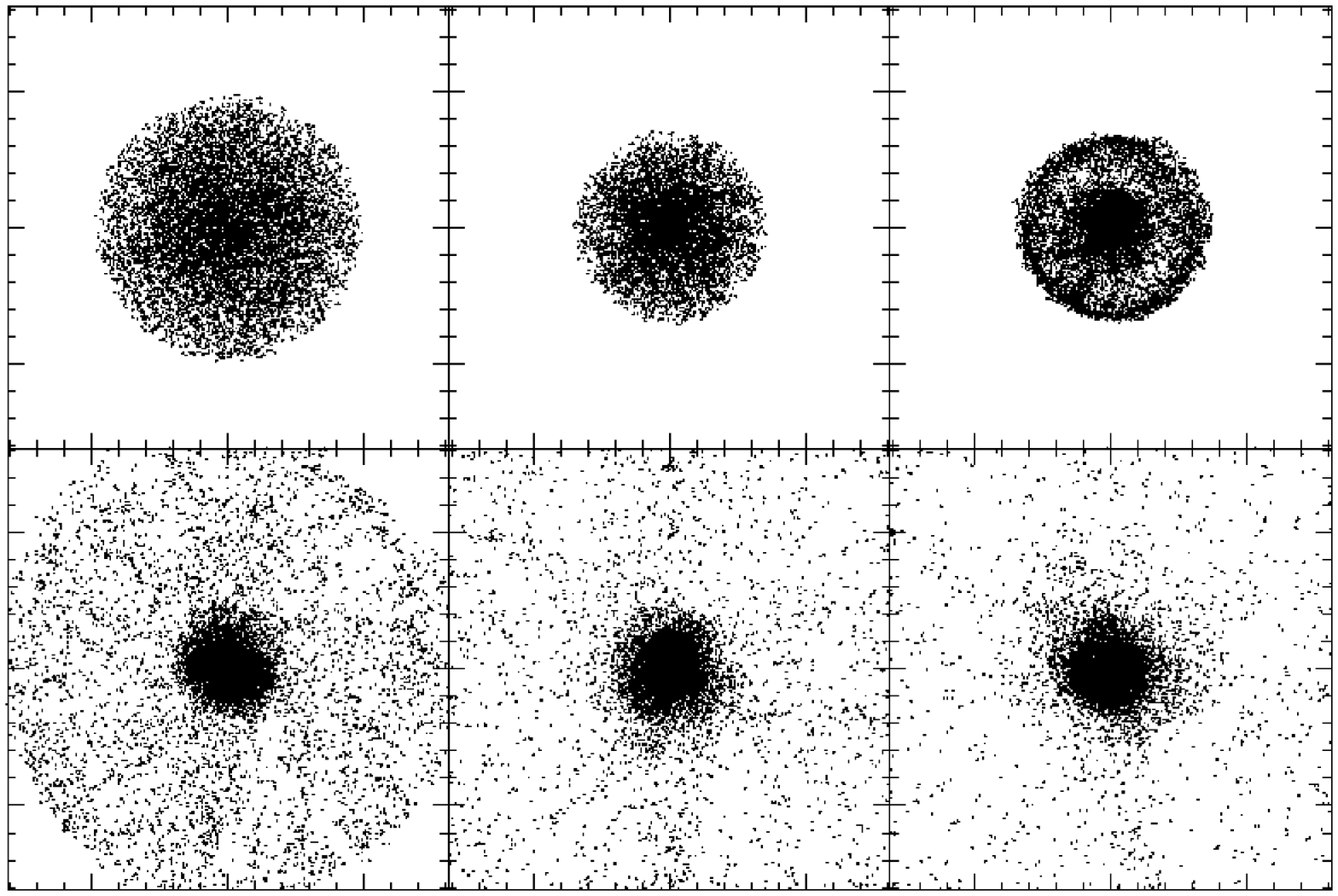}
\includegraphics[width=8.5cm,height=6cm]{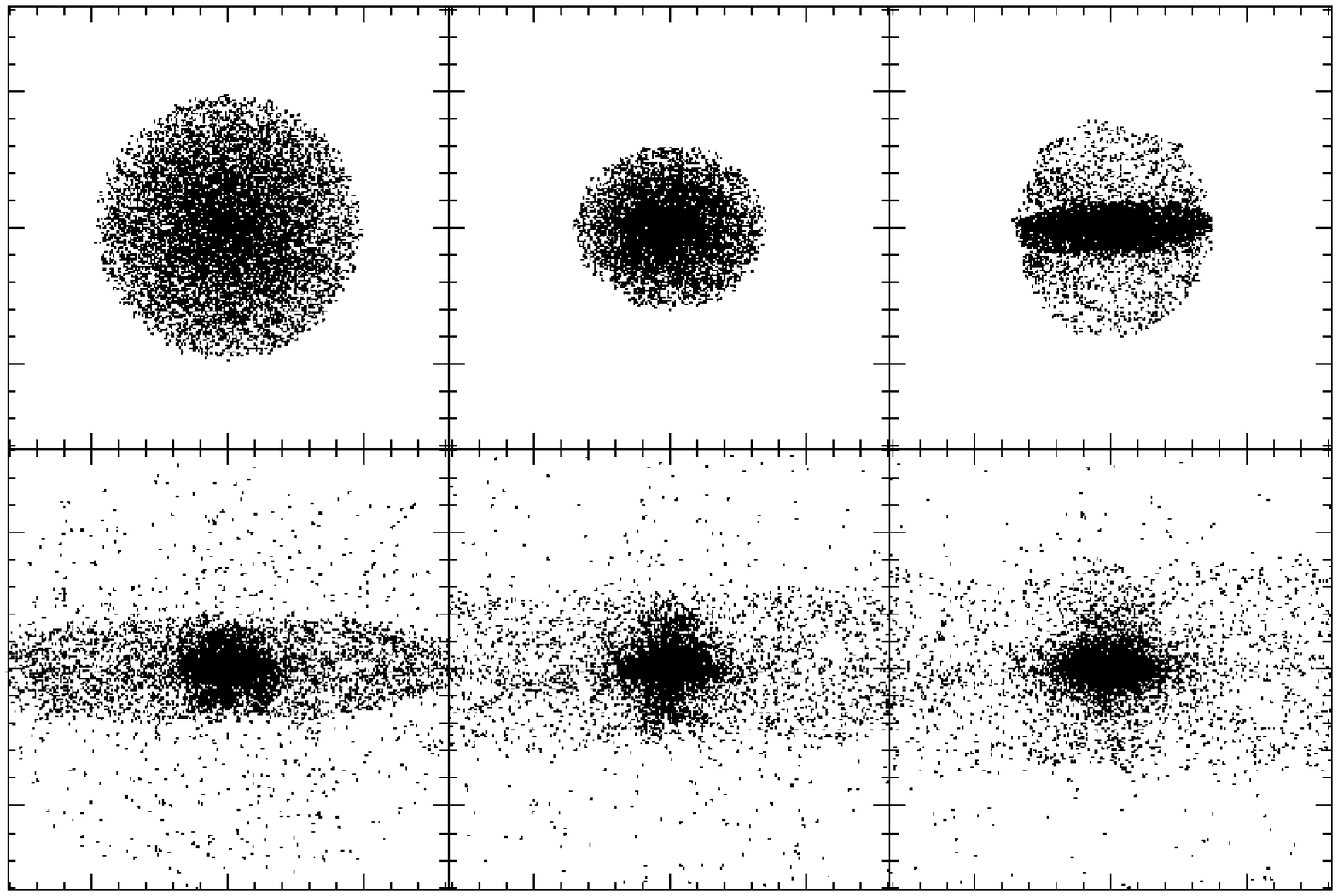}
\caption{Two components test, case TC3. Dark Matter particles
projected onto the $xy$ (top panels) and $xz$ (bottom panels)
planes: left to right, top to bottom, $t=0.5$, 1.0, 1.5, 2.0, 2.5 Gyr.}
\label{tc1d}
\end{figure}
\begin{figure}[!htbp]
\centering
\includegraphics[width=8.5cm,height=5.2cm]{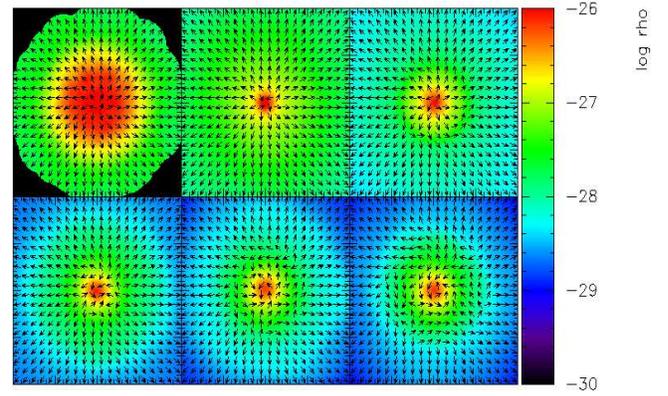}
\caption{Two components test, case TC3. Gas density and velocity
fields in a slice taken at $z=0$: left to right, top to bottom,
$t=0.5$, 1.0, 1.5, 2.0, 2.5 Gyr.}
\label{tc1g}
\end{figure}

\begin{figure}[!htbp]
\centering
\includegraphics[width=8.5cm,height=4cm]{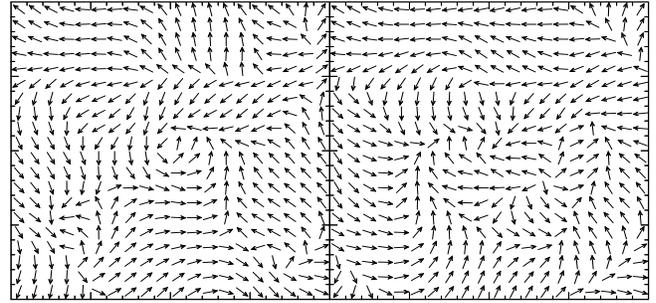}
\caption{Two components test: comparison between the central velocity fields in cases
TC3 (left) and TC4(right), in a slice taken at $z=0$, at $t=2.5$ Gyr.}
\label{tc12g}
\end{figure}

\begin{figure}[!htbp]
\centering
\includegraphics[width=8.5cm,height=8.5cm]{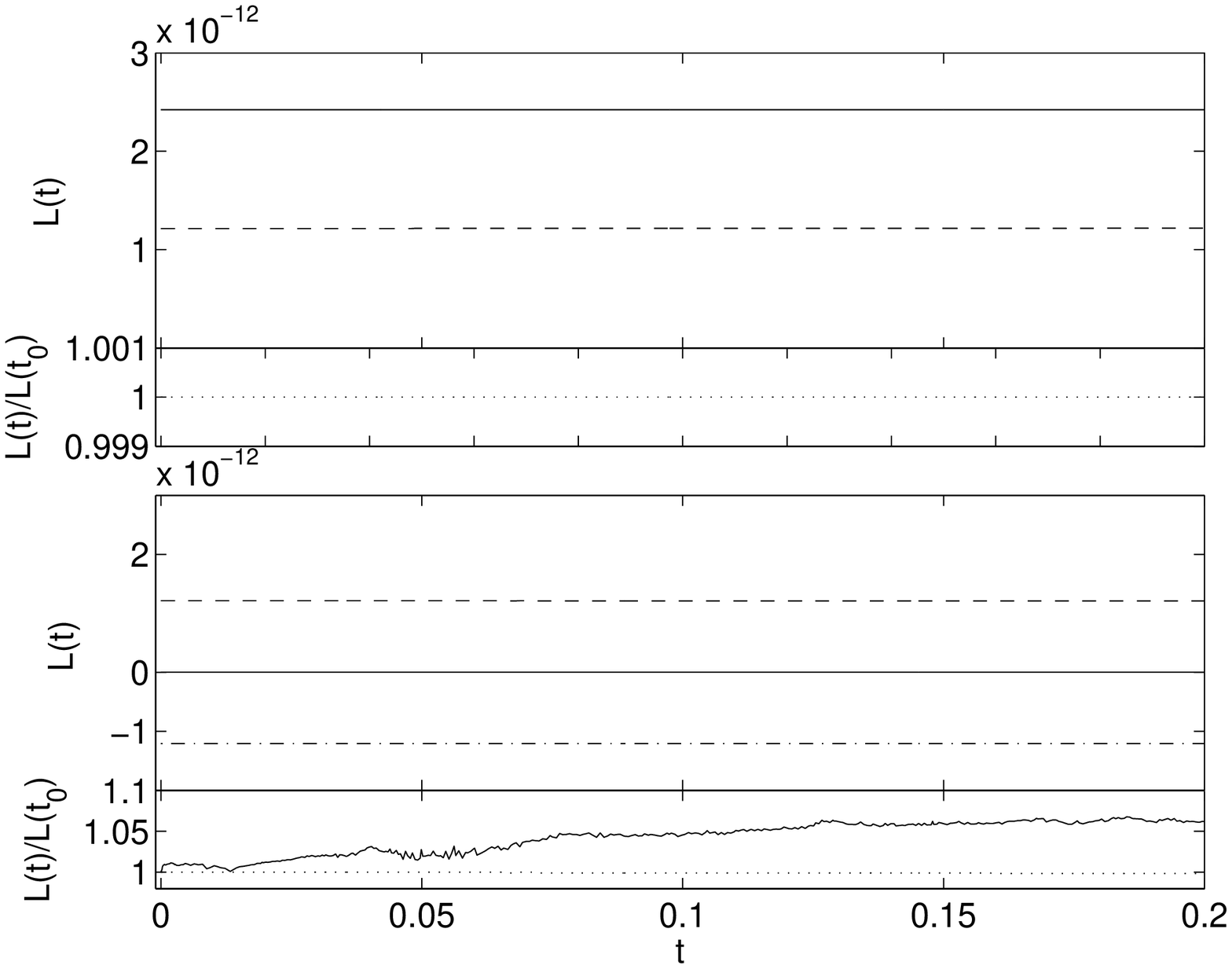}
\caption{Two components test, angular momenta conservations.
Top panels: test TC3, bottom panels: test TC4. Shown are the
total (solid lines), gas (dashed lines) and Dark Matter
(dash-dotted lines) angular momenta as a function of time
$L(t)$, and conservations as a function of time $L(t)/L(t_0)$
(same meaning of the symbols). Note that in the TC3 test the
gas and Dark Matter lines are superposed because they have
identical angular momentum.}
\label{tcangm}
\end{figure}


Fig. \ref{twocompene} plots the energy trends for all the four
tests. The conservation is very good in all cases, with a
maximum error well below $1 \%$; quite surprisingly, the worst
performance is given by the constant $\epsilon$ scheme, while
the corresponding test with adaptive softening performs better
staying  below a $\sim 0.5 \%$ error.

Fig. \ref{tc1d} shows the evolution of the Dark Matter component
in the test TC3, projected on the $xy$ and on the $xz$ planes
(the case TC4 run gives almost undistinguishable plots). Clear is
the early formation of a disk due to the initial rotational velocity,
and a subsequent expansion which leaves a dense core with a loosely
disky shape. On the other hand, the gas component expands with
almost radial motion (Fig. \ref{tc1g}), but in the central regions
the interaction with the collisionless component plays an important
role. Indeed, looking at Fig. \ref{tc12g} where a magnification of
the velocity field in the central region at late times is shown for
the tests TC3 and TC4, the difference in the two cases is clear:
in the counter-rotation run the gas in the innermost zone has
inverted its sense of motion and co-rotates with the Dark Matter,
creating a complex pattern of velocities.

The conservation of angular momenta in the tests TC3 and TC4
is shown in Fig. \ref{tcangm}. We find a maximum error of $\sim 6\%$
in the relative error for the total momentum in the counter-rotation
test; in this case the absolute violation is anyway small because
of the very small total initial momentum. In all the other cases
the conservation is always very satisfactory.

\section{Conclusions}\label{Conclusions}

We have presented the basic features of \textsc{EvoL}, the new
release of the Padova parallel N-body code for cosmological
simulations of galaxy formation and evolution. In this first paper,
the standard gravitational and hydrodynamical algorithms have been
extensively reviewed and discussed, as well as the parallel
architecture and the data structures of the code. \textsc{EvoL}
includes some interesting options such as  adaptive  softening
lengths with self-consistent extra-terms to avoid large errors in
energy conservation, $\nabla h$ terms in the SPH formalism, variable
$\alpha$ viscosity formulation, and artificial thermal conduction
terms to smooth out pressure at contact discontinuities.

We have also performed and presented an extended series of standard
hydrodynamical tests to check the ability of the code to handle
potentially difficult situations. The results are encouraging. Almost all
tests have given results in nice agreement with theoretical
expectations and previous calculations with similar codes, sometimes
even showing better performance. In particular,
we showed how the inclusion of an artificial thermal conduction term
as suggested by \citet{Price2008} significantly improves
the modeling of demanding problems such as the Kelvin-Helmoltz instability
or the Sedov-Taylor point-like explosion. 
Furthermore, the adoption of a variable softening lentghs algorithm allows for 
a higher degree of adaptivity without resulting in appreciable losses in terms
of precision and conservation of energy.
While some typical flaws of SPH codes are still present (e.g., problems
in the conservation of vorticity because of spurious shear viscosity),
these new features clearly improve the method and positively aid in curing
well-known drawbacks of the SPH algorithm.

It must be however pointed out
that the new features must be extensively tested, especially in
situations of interest for cosmological simulations (i.e., gas
dynamics in presence of a Dark Matter and/or stellar components).
Here, we have restricted our analysis to standard hydrodynamical problems. Other
and new tests will be presented in the companion paper by Merlin
et al. (2009, in preparation), dedicated to the inclusion in
\textsc{EvoL} of radiative cooling, chemical
evolution, star formation, energy feedback, and other non-standard
algorithms.

\begin{acknowledgements}
The authors would like to thank D.J. Price, S. Borgani, P.A. Thomas,
L. Secco, S. Pasetto and G. Tormen for the useful discussions and
for their help. All simulations were run using the \textsc{Cineca}
facilities and/or the \textsc{Monster} cluster at the Department of
Astronomy (Padova). The pictures have been produced using
\textsc{MatLab} or \textsc{Splash} by D. Price \citet{Price2007splash}.
\end{acknowledgements}


\bibliographystyle{apj}           

\bibliography{mnemonic,biblio}    


\appendix \label{appendix_A}
\section{$N$-dimensional kernel}

The general form for the $N$-dimensional spline kernel function ($1
\leq N \leq 3$) is

\begin{eqnarray}
&& W_N(r,\epsilon) = \eta_N \times
\begin{cases}
\frac{2}{3} - u^2+\frac{1}{2}u^3 & \text{if $0 \leq u < 1$,} \\
\frac{1}{6}(2-u)^3 &\text{if $1 \leq u < 2$,} \\
0 &\text{if $u \geq 2$,}
\end{cases}
\label{splinegen}
\end{eqnarray}

\noindent where

\begin{eqnarray}
&& \eta_N = \frac{1}{h^N} \times
\begin{cases}
1 & \text{if N = 1,} \\
15 / (7 \pi) &\text{if N = 2,} \\
3 / (2 \pi) &\text{if N = 3.}
\end{cases}
\end{eqnarray}

\noindent Derivatives are straightforwardly obtained from this
expression. The 1-D and 2-D kernels have been used where necessary
in some of the hydrodynamical tests presented in Sect. \ref{test}.

\section{Summary of equations of motion and conservation}

Gravitational acceleration:

\begin{eqnarray}
\frac{d \vec{v}_{i,grav}}{dt} & = & \\
& & -\sum_j m_j \left[\frac{\phi_{ij}'(\epsilon_i)+\phi_{ij}'(\epsilon_j)}{2}\right]\frac{\vec{r}_i-\vec{r}_j}{|\vec{r}_i-\vec{r}_j|} \nonumber\\
& & -\sum_j m_j \frac{1}{2} \left[ \frac{\xi_i}{\Upsilon_i} \frac{\partial W_{ij}(\epsilon_i)}{\partial \vec{r}_i} + \frac{\xi_j}{\Upsilon_j} \frac{\partial W_{ij}(\epsilon_i)}{\partial \vec{r}_j} \right], \nonumber
\end{eqnarray}

\noindent Hydrodynamical acceleration (for SPH particles only):

\begin{eqnarray}
\frac{d\vec{v}_{i,hyd}}{dt} & = & \sum_j m_j \times  \\
& & \left[ \frac{P_i}{\rho_i^2} \left( 1+\frac{\zeta_i/m_j}{\Omega_i^*}\right) \nabla_i W_{ij}(h_i) + \right. \nonumber \\
& & \left. \frac{P_j}{\rho_j^2} \left( 1+\frac{\zeta_j/m_i}{\Omega_j^*}\right) \nabla_i W_{ij}(h_j) + \Pi_{ij} \bar{\nabla_i W_{ij}} \right], \nonumber
\end{eqnarray}

\noindent Specific internal energy evolution (for SPH particles
only):

\begin{eqnarray}
\frac{du_i}{dt} & = &\\ \nonumber
& & \sum_j m_j \left[ \frac{P_i}{\rho_i^2}\left( 1+\frac{\zeta_i/m_j}{\Omega_i^*}\right) (\vec{v}_j-\vec{v}_i) \cdot \nabla_i W_{ij}(h_i) \right. \\ \nonumber
& & + \left. \left( \frac{1}{2}\Pi_{ij} + \Pi_{ij}^u \right) w_{ij} |\bar{\nabla_i W_{ij}}| \right].
\end{eqnarray}

\noindent In these expressions,

\begin{eqnarray}
\Omega_i^* = 1 - \frac{\partial h_i}{\partial n_i} \sum_j \frac{\partial W_{ij}(h_i)}{\partial h_i},
\end{eqnarray}

\begin{eqnarray}
\Upsilon_i = \left[ 1 - \frac{\partial \epsilon_i}{\partial n_i} \sum_j \frac{\partial W_{ij}(\epsilon_i)}{\partial \epsilon_i}\right],
\end{eqnarray}

\begin{eqnarray}
\xi_i = \frac{\partial \epsilon_i}{\partial n_i} \sum_j \frac{\partial \phi_{ij}(\epsilon_i)}{\partial \epsilon_i},
\end{eqnarray}

\noindent and

\begin{eqnarray}
\zeta_i = \frac{\partial h_i}{\partial n_i}\sum_j m_j \frac{\partial W_{ij}(h_i)}{\partial h_i}.
\end{eqnarray}

\noindent The smoothing kernel $W$ is given by Eq. \ref{spline},
whereas the softened gravitational potential $\phi$ is given by Eq.
\ref{softphi}.

\noindent The standard equation of state is that of an ideal gas,

\begin{eqnarray}
P = (\gamma - 1) \rho u,
\end{eqnarray}

\noindent where $\gamma$ is the adiabatic index (5/3 for a monatomic
gas); a more general equation of state will be presented in Merlin
et al. (2009, in preparation).

%

\end{document}